\documentclass[11pt,a4paper,oneside]{article}

\pdfoutput=1
\usepackage{jheppub}

\usepackage{verbatim}
\usepackage{graphicx}
\usepackage{mathrsfs}
\usepackage{hyperref}
\usepackage{appendix}
\usepackage{caption}
\usepackage{float}
\usepackage{subcaption}
\usepackage{array,multirow,tabu}
\usepackage{arydshln}
\usepackage{ulem}
\usepackage{booktabs}
\usepackage{graphicx}
\usepackage{amssymb}
\usepackage{relsize}
\usepackage{slashed}
\usepackage{amsmath}
\usepackage{setspace}
\usepackage{lscape}
\usepackage{alltt}

\definecolor{mmaLabel}{RGB}{70,70,153}            
\definecolor{mmaLink}{RGB}{20,40,153}
\definecolor{mmaUndefined}{RGB}{0,44,195}         
\definecolor{mmaFunctionLocal}{RGB}{60,125,145}   
\definecolor{mmaLocal}{RGB}{67,137,88}           
\definecolor{mmaMessage}{RGB}{129,43,38}
\definecolor{mmaError}{RGB}{255,51,51}
\definecolor{mmaSyntaxError}{RGB}{194,85,204}
\definecolor{mmaEmphasizedError}{RGB}{204,0,0}
\definecolor{mmaEmphasizedErrorBackground}{RGB}{255,225,130}
\definecolor{mmaFormattingError}{RGB}{255,85,85}
\definecolor{mmaFormattingErrorBackground}{RGB}{255,230,230}
\definecolor{mmaString}{gray}{.4}  
\definecolor{mmaComment}{gray}{.6}

\preprint{IPPP/20/86}

\usepackage[dvipsnames]{xcolor}

\title{EFT Diagrammatica: UV Roots of the CP-conserving SMEFT}

\author[1]{Supratim Das Bakshi, Joydeep Chakrabortty, Suraj Prakash, Shakeel Ur Rahaman,}

\author[2]{and Michael Spannowsky}

\emailAdd{sdbakshi@iitk.ac.in, joydeep@iitk.ac.in, surajprk@iitk.ac.in, shakel@iitk.ac.in, michael.spannowsky@durham.ac.uk}

\affiliation[1]{Indian Institute of Technology Kanpur, Kalyanpur, Kanpur 208016, Uttar Pradesh, INDIA}

\affiliation[2]{Institute for Particle Physics Phenomenology, Department of Physics, Durham University, Durham DH1 3LE, U.K}

\abstract{Effective Field Theories are an established framework to bridge the gap between UV and low energy theories. In the context of the Standard Model, the bottom-up approach extends its operator set and thus equips us to astutely probe its observables while encapsulating indirect evidence of unknown high scale theories. While the top-down approach, on the other hand, employs functional techniques to integrate out the heavy fields from a BSM Lagrangian leading to a set of SMEFT operators. An intricate interplay of the two approaches enhances the efficacy of the SMEFT in making meaningful predictions while providing a platform for conducting a coherent comparison of new physics scenarios. However, while the bottom-up approach fails to indicate the origin of the effective operators, the top-down approach is highly dependent on the specific model assumptions of the UV theory. We, for the first time, are proposing a diagrammatic approach to establish selection criteria for the allowed heavy field representations corresponding to each SMEFT operator. This, in turn, paves the way to construct observable driven new physics models. While we take a predominantly minimalistic approach, we also highlight the necessity for non-minimal interactions for certain operators. }

\begin{document}
	
	\maketitle
	
\section{Introduction}\label{sec:intro}
	
The contemporary paradigm for conducting phenomenological analyses in particle physics is based on Effective Field Theories (EFTs). EFT are a natural choice considering the proliferation of scales in subatomic physics. A popular EFT paradigm is the so-called \textit{bottom-up} approach, first introduced in \cite{Weinberg:1980wa,BUCHMULLER1986621,Grzadkowski:2010es}, which not only attempts to encapsulate the effects of unknown high energy theories but also provides useful contributions to the observables defined by the renormalizable Lagrangian. This method involves the construction of higher mass dimension operator bases, and the parameter space is defined by the Wilson coefficients corresponding to these operators. A number of computational tools have been developed that automate the procedure of operator construction \cite{Criado:2019ugp,Fonseca:2017lem,Gripaios:2018zrz,Banerjee:2020bym,Marinissen:2020jmb}. Standard Model Effective Field Theory (SMEFT) is the \textit{bottom-up} extension of the Standard Model which consists of 59 operators at dimension-6 considering only a single flavour of fermions\footnote{This counting does not include hermitian conjugates of the operators, nor does it include operators that violate accidental symmetries of the renormalizable Lagrangian. Taking those into account, we get 2 operators at dimension-5 \cite{WEINBERG1979327,Henning:2015alf}, 84 at dimension-6 \cite{Grzadkowski:2010es}, 30 at dimension-7 \cite{Lehman:2014jma}, 993 at dimension-8 \cite{Murphy:2020rsh,Li:2020gnx} and 560 operators at dimension-9 \cite{Li:2020xlh,Liao:2020jmn} for SMEFT.} and brings to light several interesting features not encountered at the renormalizable level. The only downside of this approach is that it is oblivious to the exact UV roots of each of the operators and the associated phenomenology. Hence, the Wilson coefficients remain without origin and independent \cite{Georgi:1994qn}.

The other face of the coin, popularly known as the \textit{top-down} approach, starts at a higher scale where one constructs the Lagrangian for a particular model and identifies certain degrees of freedom as heavy and systematically integrates them out to obtain a set of higher mass dimension operators composed entirely of the lighter fields \cite{Henning:2014wua,Henning:2016lyp,Haisch:2020ahr,Jiang:2018pbd,Zhang:2016pja,Ellis:2016enq,Ellis:2017jns,Ellis:2020ivx,delAguila:2016zcb,Kramer:2019fwz,Drozd:2015rsp}. To affirm the predictions of a UV theory against the SM observables, the field content of the chosen UV theory must first be brought down to the SM field content and this is where the \textit{top-down} approach finds great utility. This procedure ultimately leads to various subsets of the 59 SMEFT operators, constructed via the \textit{bottom-up} approach. These may or may not overlap for different UV models. Various computational packages have been developed to automate the procedure for certain cases \cite{Bakshi:2018ics,Aebischer:2018bkb,Criado:2017khh,Celis:2017hod}. 

Also, from the point of view of symmetries, there is no restriction on the choice of field as well as symmetry extensions of the SM that can be treated as a UV theory. The only means of eliminating candidate models is through phenomenological analyses. But, even then a large number of models still appear to be viable as they all contribute to some observable or another \cite{Deshpande:1977rw,Bambhaniya:2013yca,deBlas:2014mba,arhrib:hal-00608687,Angelescu:2020yzf,Duka:1999uc,PhysRevD.44.837,Saad:2017pqj}. Conducting a comparative analysis of every single one of them becomes tedious and impractical, more so for non-minimal scenarios with multiple heavy fields. In such cases, we encounter a multitude of scales which implies that a cascade of EFTs may be required to determine the relations among SMEFT Wilson coefficients and BSM parameters, and a systematic procedure must be developed to address this. 

It is desirable to have a more structured method of cataloguing the UV models that lead to specific SMEFT operators, which in turn can be correlated with observables. Clearly, instead of starting with a different BSM Lagrangian each time and comparing the various subsets of SMEFT each of them leads to, it is more economical if we approach this issue in a retrograde manner, where based on the observables under study we first identify the necessary operators and then attempt to enumerate the specific list of heavy fields that can generate the particular operator(s). This allows us to conduct our analysis in a minimal sense and also highlights which combinations of heavy fields may lead to redundant contributions. 

This operator-driven BSM model building is what we have addressed in this work, i.e., our primary aim has been to identify the possible UV roots of each SMEFT operator when considering 1-particle-irreducible (1PI) diagrams up to one-loop-level built of interactions involving the SM as well as heavy fields. We must emphasize that we have only considered extensions to the SM particle content and the internal symmetry has been kept fixed at $SU(3)_C\otimes SU(2)_L\otimes U(1)_Y$. Also, we realise that the SMEFT operators may receive radiative corrections of the SM as well as heavy fields but accounting for such corrections would require us to delve deeper into the incorporation of a renormalization prescription which is beyond the scope of the current work and such analysis will form a part of our future objectives. We have started by delineating a schematic unfolding of the set of independent operator classes of mass dimension-6 in section \ref{sec:lorentz-inv-unfolding}. The external legs of the contact operators have been judiciously segregated and internal lines have been added suitably while ensuring Lorentz invariance at each vertex to obtain a representative set of tree- and one-loop-level diagrams for each operator class and the corresponding subclasses if any. In section \ref{sec:smeft-unfolding}, we have first prepared an exhaustive list of heavy field representations which can appear at various renormalizable vertices when the internal symmetry of the SM is imposed in addition to Lorentz invariance. Then using these interactions as fundamental building blocks we have carefully outlined how each SMEFT operator of mass dimension-6 can be unfolded to reveal these heavy fields within tree-level diagrams with heavy propagators and one-loop-level diagrams involving pure-heavy or light-heavy-mixed loops. We must remark that the analysis here is by no means exhaustive and conforms to a notion of minimality which has been described in the paper. As a matter of fact, in section \ref{sec:non-minimal-cases}, we have addressed a few ways in which our set can be extended by accounting for non-minimal cases. We have also emphasized special cases of certain operators where this non-minimality is unavoidable. In section \ref{sec:section5}, we have validated our results by focussing on a single heavy scalar representation and showing that the various SMEFT operators whose unfolding incorporates it are precisely the same as the ones obtained when the UV theory containing the particular heavy scalar is subjected to the \textit{top-down} analysis \cite{deBlas:2017xtg,Bakshi:2020eyg,Gherardi:2020det,Dawson:2020oco,Ellis:2018gqa}. We have also described how an extension of the SM containing a second Higgs doublet along with a scalar with a non-trivial color quantum number can be seen to account for all the operators that are significant in the context of electroweak precision observables (EWPO) and Higgs signal strength measurements \cite{Brivio:2017btx,Brivio:2017vri,Dawson:2019clf,Baglio:2020oqu,Dawson:2020oco,Ellis:2018gqa,DasBakshi:2020ejz}. In addition, we have emphasized the need for non-minimal extensions, analyses at the level of higher loops, as well as the necessity of developing novel observables to account for the remaining SMEFT operators.
		
\section{Unfolding effective operators into Lorentz invariant vertices}\label{sec:lorentz-inv-unfolding}
	
	We intend to highlight a generalizable procedure using which, provided a complete and independent set of operators, we can trace the origin of each operator from candidate UV theories containing specific heavy fields. This indeed is the opposite of conventional analyses where one starts with a UV theory, identifies certain degrees of freedom as heavy and after suitably integrating out obtains a subset of the effective Lagrangian, i.e., higher mass dimension operator basis, of a low energy theory \cite{Gaillard:1985uh,PhysRevLett.55.2394,Bilenky:1993bt,Manohar:2018aog,Chan1985}. The following points emphasize the salient aspects of this reverse-engineering procedure:
	
	\begin{itemize}
		\item The building blocks of our analysis are Feynman diagrams which in turn are constituted of certain well-defined symbols. For instance, dashed, solid, and wiggly lines indicate scalars, fermions, and gauge bosons respectively. Different colours differentiate between heavy, light, and general fields. These along with additional symbols for $\mathcal{D}_{\mu}$, $\gamma^{\mu}$, and their combinations have been listed in  Table~\ref{table:symbol-legend}.
		
		\begin{table}[!htb]
			\centering
			\renewcommand{\arraystretch}{1.6}
			{\small\begin{tabular}{||c|c||c|c||c|c||}
					\hline
					\textbf{Symbol}&
					\textbf{Stands for}&
					\textbf{Symbol}&
					\textbf{Stands for}&
					\textbf{Symbol}&
					\textbf{Stands for}\\
					\hline
					
					\multirow{2}{*}{\includegraphics[height=0.6cm, width=2cm]{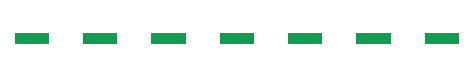}}&
					\multirow{2}{*}{Scalar}&
					\multirow{2}{*}{\includegraphics[height=0.6cm, width=2cm]{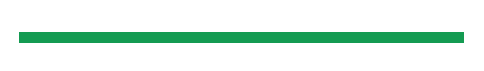}}&
					\multirow{2}{*}{Fermion}&
					\multirow{2}{*}{\includegraphics[height=1cm, width=2cm]{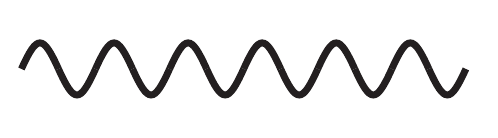}}&
					SM Gauge\\
					
					&
					&
					&
					&
					&
					Boson\\
					\hline
					
					\multirow{2}{*}{\includegraphics[height=0.6cm, width=2cm]{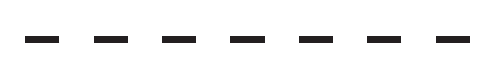}}&
					Light (SM)&
					\multirow{2}{*}{\includegraphics[height=0.6cm, width=2cm]{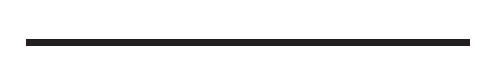}}&
					Light (SM)&
					\multirow{2}{*}{\includegraphics[height=0.7cm, width=1.8cm]{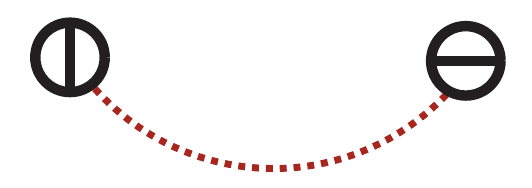}}&
					\multirow{2}{*}{$(\,\cdot\,)\mathcal{D}^{\mu}(\,\cdot\,\cdot\,)\mathcal{D}_{\mu}(\,\cdot\,)$}\\
					
					&
					scalar ($\phi$)&
					&
					fermion ($\psi$)&
					&
					\\
					\hline
					
					\multirow{2}{*}{\includegraphics[height=0.6cm, width=2cm]{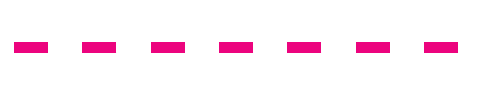}}&
					Heavy&
					\multirow{2}{*}{\includegraphics[height=0.6cm, width=2cm]{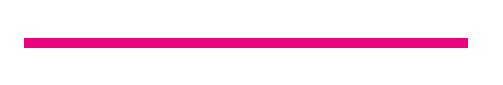}}&
					Heavy&
					\multirow{2}{*}{\includegraphics[height=0.7cm, width=1.8cm]{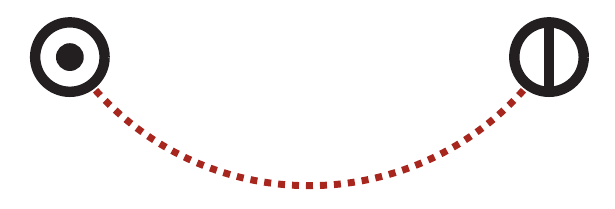}}&
					\multirow{2}{*}{$(\cdot)\gamma^{\mu}(\cdot\,\cdot)\,\mathcal{D}_{\mu}(\,\cdot\,)$}\\
					
					&
					scalar ($\Phi$)&
					&
					fermion ($\Psi$)&
					&
					\\
					\hline
					
					\multirow{2}{*}{\includegraphics[height=0.6cm, width=0.6cm]{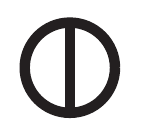}}&
					\multirow{2}{*}{$\mathcal{D}_{\mu}$}&
					\multirow{2}{*}{\includegraphics[height=0.6cm, width=0.6cm]{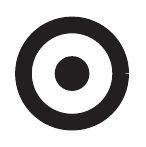}}&
					\multirow{2}{*}{$\gamma^{\mu}$}&
					\multirow{2}{*}{\includegraphics[height=0.6cm, width=0.6cm]{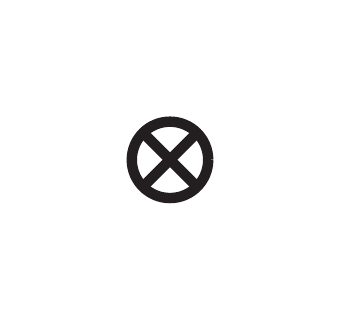}}&
					\multirow{2}{*}{$\square\,\equiv\,\mathcal{D}^2$}\\
					
					&
					&
					&
					&
					&
					\\
					\hline
			\end{tabular}}
			\caption{The symbols used throughout the paper and their meaning.}
			\label{table:symbol-legend}
		\end{table}

		\item At the intermediate step between these symbols and Feynman diagrams lie the vertices obtained from the renormalizable Lagrangian\footnote{It must be kept in mind that we are working in 3+1 space-time dimensions, therefore a renormalizable Lagrangian consists of terms with mass dimension 4 and individual components, i.e., bosons, fermions, derivatives and field strength tensors have mass dimensions of 1, $\frac{3}{2}$, 1 and 2 respectively.}. The most general collection of vertices, which includes trilinear and quartic self-interactions of scalars and gauge bosons, Yukawa terms connecting a pair of fermions with a scalar and the kinetic terms of scalars and fermions, have been depicted in Fig.~\ref{fig:renorm-vertices}. One can see how these vertices incorporate the symbols defined in Table~\ref{table:symbol-legend}.

		\begin{figure}[!htb]
			\centering
			\renewcommand{\thesubfigure}{\roman{subfigure}}
			\hspace{1.3cm}
			\begin{subfigure}[t]{3.7cm}
				\centering
				\includegraphics[width=3.5cm,height=2.8cm]{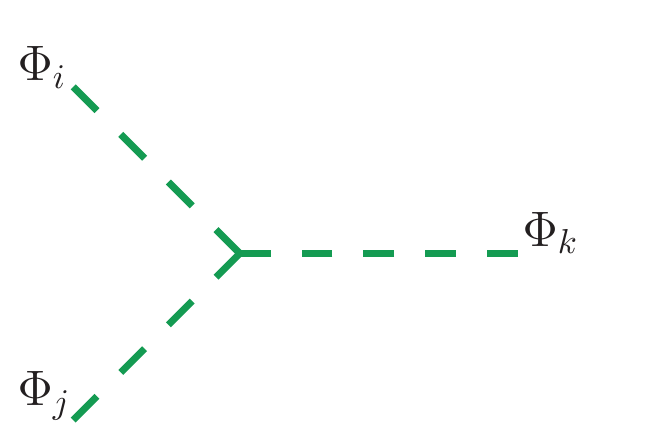}
				\caption{$[\Phi]^3$}\label{subfig:phi3-vertex}
			\end{subfigure}
			\begin{subfigure}[t]{3.7cm}
				\centering
				\includegraphics[width=3.5cm]{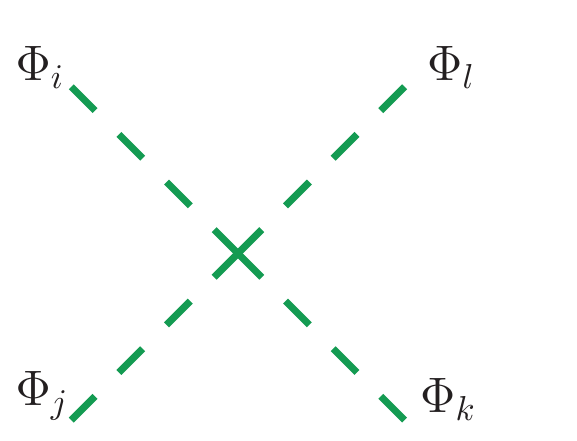}
				\caption{$[\Phi]^4$}\label{subfig:phi4-vertex}
			\end{subfigure}
			\begin{subfigure}[t]{3.7cm}
				\centering
				\includegraphics[width=3.5cm]{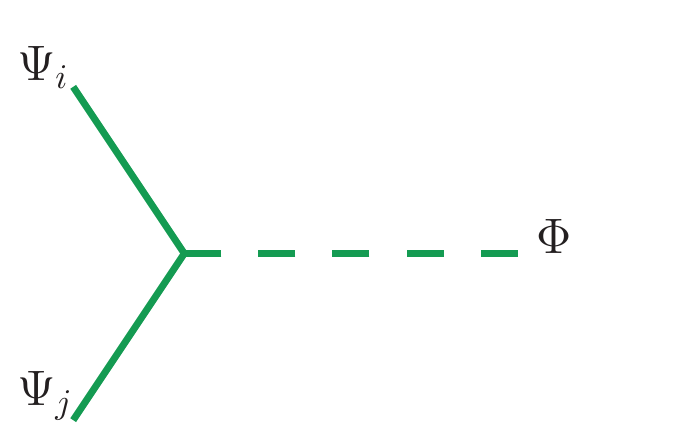}
				\caption{$[\Psi]^2\,\Phi$}\label{subfig:yukawa-vertex}
			\end{subfigure}
			\newline
			\begin{subfigure}[t]{3.7cm}
				\centering
				\includegraphics[width=3.5cm]{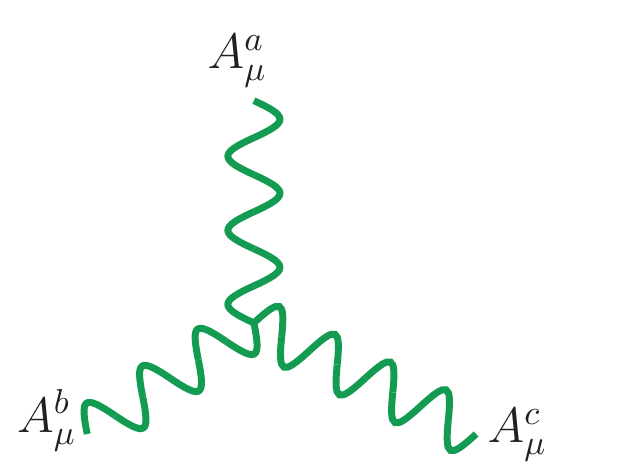}
				\caption{$[A_{\mu}]^3$}\label{subfig:gauge-trilinear-vertex}
			\end{subfigure}
			\begin{subfigure}[t]{3.7cm}
				\centering
				\includegraphics[width=3.5cm]{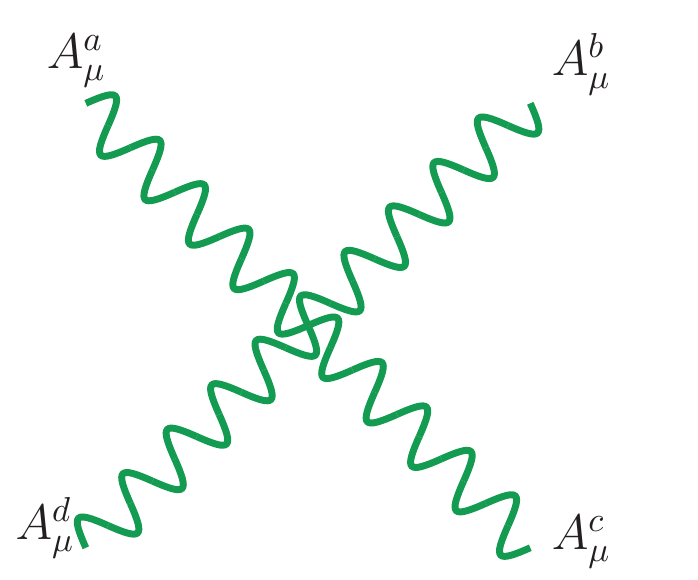}
				\caption{$[A_{\mu}]^4$}\label{subfig:gauge-quartic-vertex}
			\end{subfigure}
			\begin{subfigure}[t]{3.7cm}
				\centering
				\includegraphics[width=3.5cm]{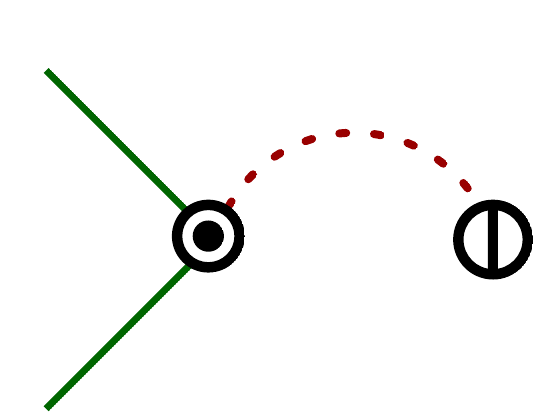}
				\caption{$[\Psi]^2\,\mathcal{D}$}\label{subfig:fermion-kinetic-vertex}
			\end{subfigure}
			\begin{subfigure}[t]{3.7cm}
				\centering
				\includegraphics[width=3.5cm]{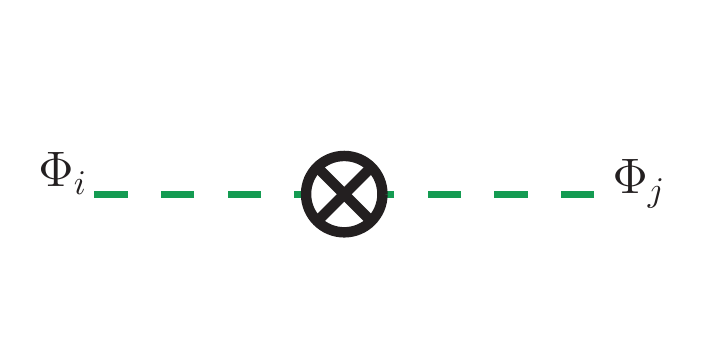}
				\caption{$\Phi\,\square\,\Phi$}\label{subfig:scalar-kinetic-vertex}
			\end{subfigure}
			\caption{Lorentz invariant vertices obtained from terms of the renormalizable Lagrangian.}
			\label{fig:renorm-vertices}
		\end{figure}
		
		\item Though our method is generic and applicable for any effective field theory, in this work, we restrict ourselves to the effective Lagrangians of mass dimension-6. In Fig.~\ref{fig:eff-ops-schematic}, we have collected schematic representations of Lorentz invariant operator classes that constitute a complete and independent set at dimension-6 after certain other classes are found to be redundant based on equations of motion (EoM) of the fields, integration by parts (IBP) and removal of total derivatives. See \cite{Grzadkowski:2010es,Banerjee:2020jun} for detailed discussions.
		
		\begin{figure}[!htb]
			\centering
			\renewcommand{\thesubfigure}{\roman{subfigure}}
			\begin{subfigure}[t]{3cm}
				\centering
				\includegraphics[width=2.5cm,height=2.4cm]{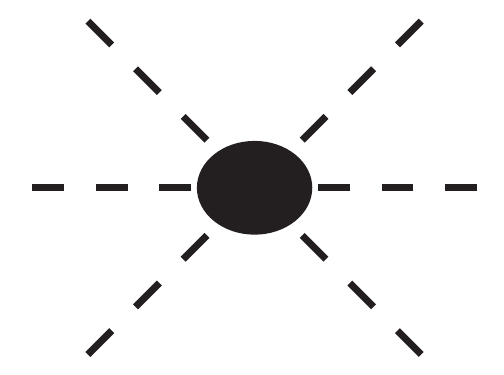}
				\caption{$\phi^6$}\label{subfig:phi6}
			\end{subfigure}
			\begin{subfigure}[t]{3cm}
				\centering
				\includegraphics[width=2.4cm]{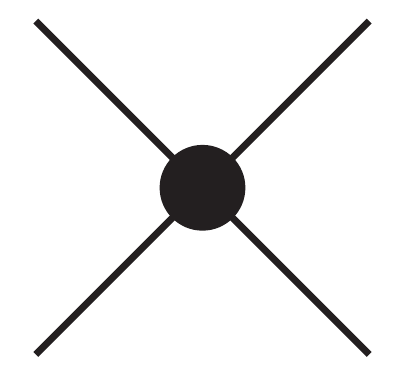}
				\caption{$\psi^4$}\label{subfig:psi4}
			\end{subfigure}
			\begin{subfigure}[t]{3cm}
				\centering
				\includegraphics[width=2.2cm,height=2.4cm]{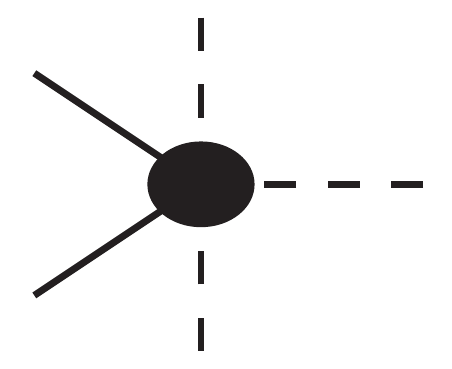}
				\caption{$\psi^2\phi^3$}\label{subfig:psi2phi3}
			\end{subfigure}
			\begin{subfigure}[t]{3cm}
				\centering
				\includegraphics[width=2.6cm,height=2.4cm]{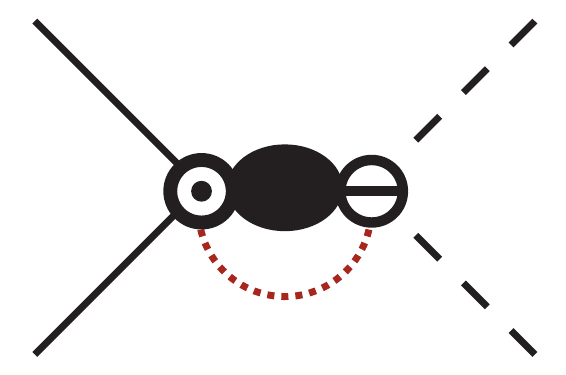}
				\caption{$\psi^2\phi^2 \mathcal{D}$}\label{subfig:psi2phi2D}
			\end{subfigure}
			\newline
			\begin{subfigure}[t]{7cm}
				\centering
				\includegraphics[width=2.2cm,height=2.2cm]{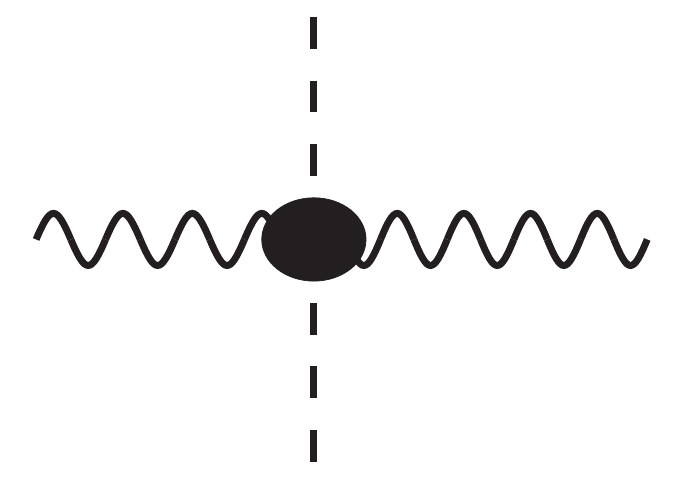}
				\includegraphics[width=2.2cm,height=2.2cm]{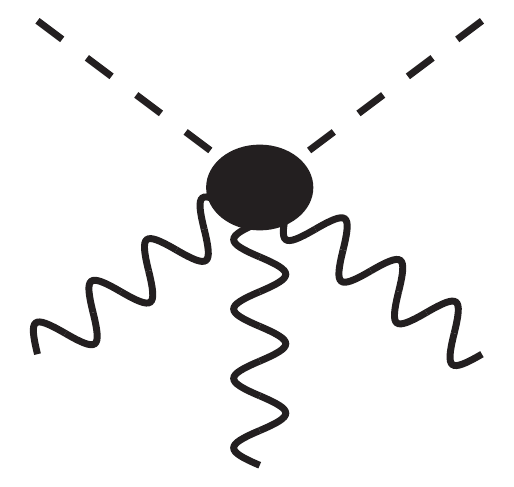}
				\includegraphics[width=2.2cm,height=2.2cm]{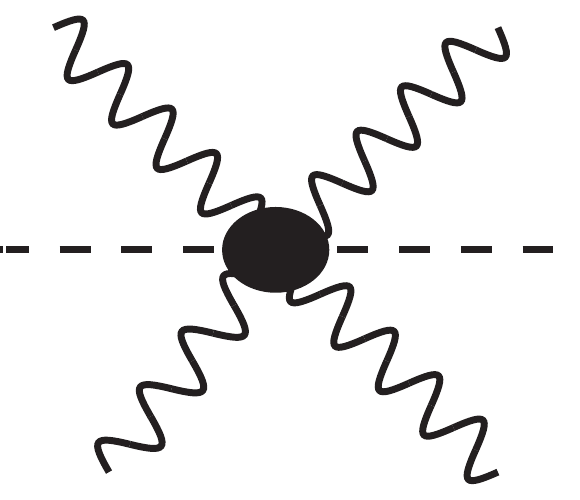}
				\caption{$\phi^2X^2$}\label{subfig:phi2x2}
			\end{subfigure}
			\begin{subfigure}[t]{6cm}
				\centering
				\includegraphics[width=2.4cm,height=2.2cm]{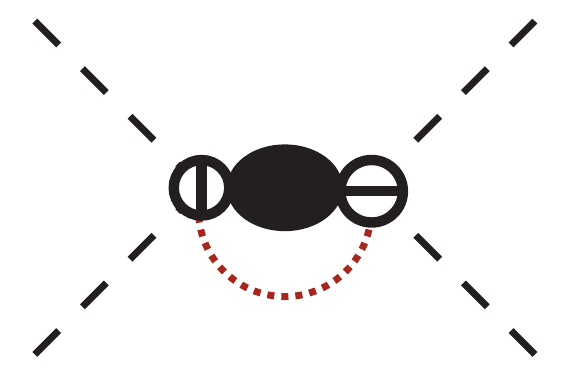}
				\includegraphics[width=2.4cm,height=2.2cm]{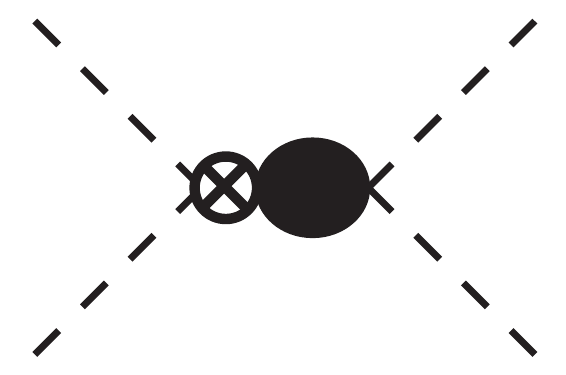}
				\caption{$\phi^4\mathcal{D}^2$}\label{subfig:phi4D2}
			\end{subfigure}
			\newline
			\begin{subfigure}[t]{5cm}
				\centering
				\includegraphics[width=2.3cm,height=2.2cm]{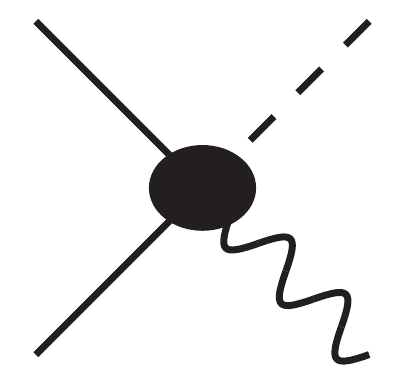}
				\includegraphics[width=2.3cm,height=2.2cm]{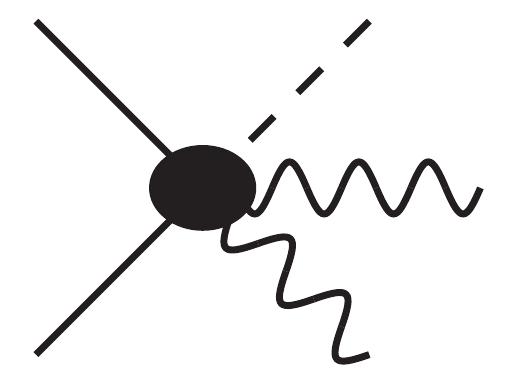}
				\caption{$\psi^2\phi X$}\label{subfig:psi2phix}
			\end{subfigure}
			\begin{subfigure}[t]{10cm}
				\centering
				\includegraphics[width=2.2cm,height=2.2cm]{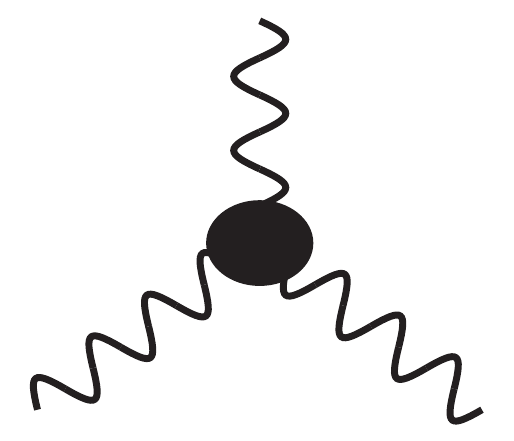}
				\includegraphics[width=2.2cm,height=2.2cm]{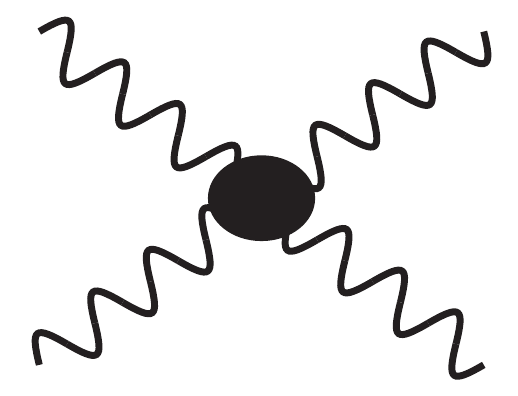}
				\includegraphics[width=2.2cm,height=2.2cm]{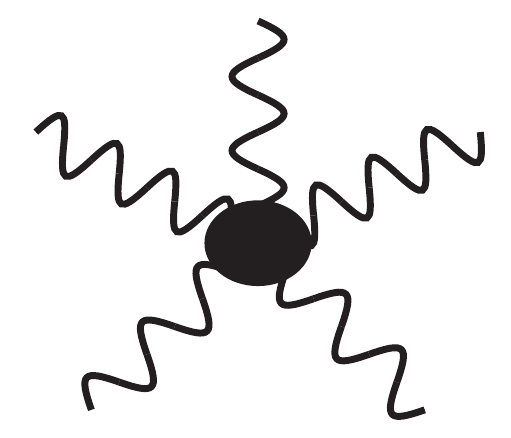}
				\includegraphics[width=2.2cm,height=2.2cm]{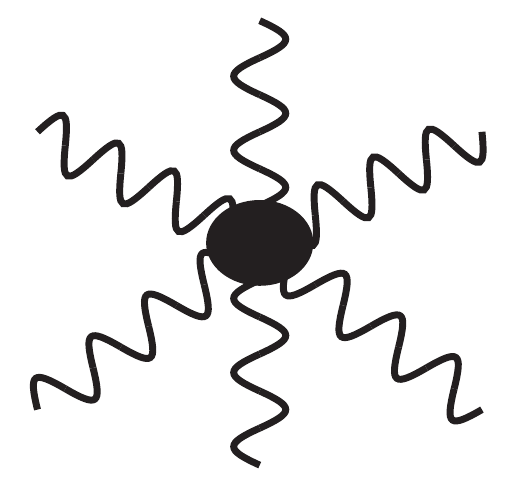}
				\caption{$X^3$}\label{subfig:x3}
			\end{subfigure}
			\caption{Schematic representation of SMEFT operator classes at mass dimension-6 in the Warsaw basis. The dotted red line indicates the contraction of Lorentz indices as mentioned in Table~\ref{table:symbol-legend}.}
			\label{fig:eff-ops-schematic}
		\end{figure}

		\item One can see that multiple diagrams have been drawn corresponding to the classes containing $X$ in Fig.~\ref{fig:eff-ops-schematic}. The reason is that each $X$ which denotes a field strength tensor can contain up to two gauge bosons within it. Thus, the multiple diagrams hint at how several effective vertices can be obtained from the same effective operator. 
		
		\item The next important step is the ``unfolding" of the effective operators, shown in Fig.~\ref{fig:eff-ops-schematic}, using the renormalizable vertices depicted in  Fig.~\ref{fig:renorm-vertices}, through tree- and one-loop-level diagrams. This has been done for each class and the results have been shown in Figs.~\ref{fig:phi6-unfolding}-\ref{fig:psi2phix-unfolding}. 
		
		\item We must comment at this stage that the diagrams for individual classes need not be exhaustive but care has been taken to include the ones which are relevant for the most general analysis.
		
		\item Also, the unfolding at this stage has been done with only Lorentz invariance in mind. So, within the loops, we have kept open the possibility of having light as well as heavy propagators, but it must be understood that there is at least one heavy field.
	\end{itemize}

	\begin{figure}[!htb]
		\centering
		\renewcommand{\thesubfigure}{\roman{subfigure}}
		\begin{subfigure}[t]{2cm}
			\centering
			\includegraphics[width=1.7cm,height=1.7cm]{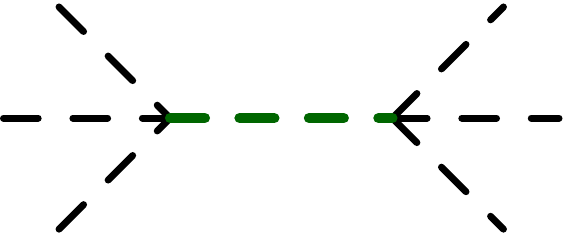}
			\caption{}\label{subfig:phi6-tree1}
		\end{subfigure}
		\begin{subfigure}[t]{1.8cm}
			\centering
			\includegraphics[width=1.5cm,height=2.2cm]{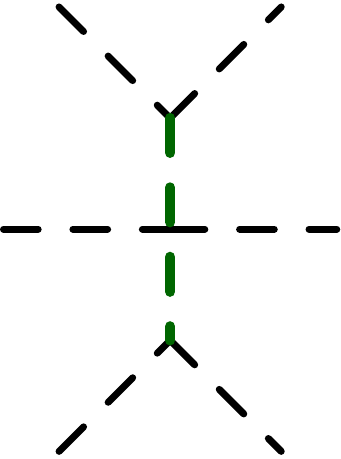}
			\caption{}\label{subfig:phi6-tree2}
		\end{subfigure}
		\begin{subfigure}[t]{1.8cm}
			\centering
			\includegraphics[width=1.5cm,height=2.2cm]{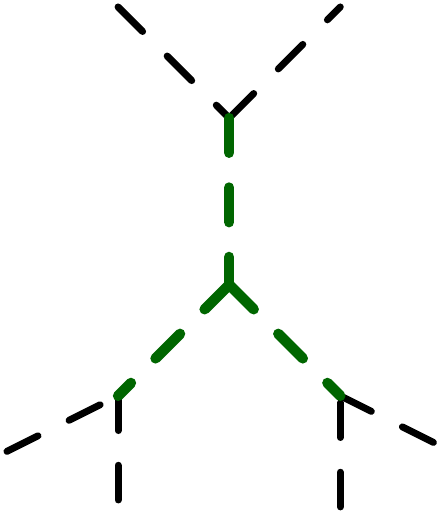}
			\caption{}\label{subfig:phi6-tree3}
		\end{subfigure}
		\begin{subfigure}[t]{2cm}
			\centering
			\includegraphics[width=1.5cm,height=2.2cm]{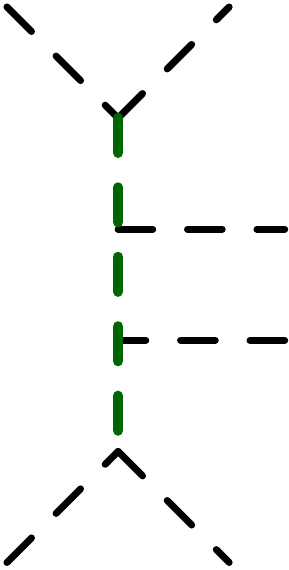}
			\caption{}\label{subfig:phi6-tree4}
		\end{subfigure}
		\begin{subfigure}[t]{2.2cm}
			\centering
			\includegraphics[width=1.8cm,height=1.8cm]{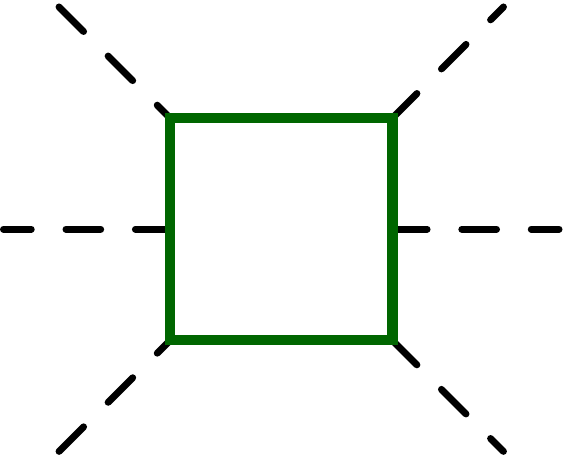}
			\caption{}\label{subfig:phi6-loop1}
		\end{subfigure}
		\begin{subfigure}[t]{2.2cm}
			\centering
			\includegraphics[width=1.8cm,height=1.8cm]{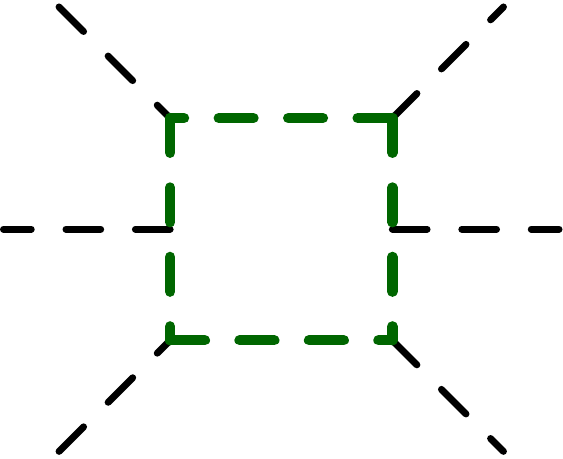}
			\caption{}\label{subfig:phi6-loop2}
		\end{subfigure}
		\begin{subfigure}[t]{2.2cm}
			\centering
			\includegraphics[width=2cm,height=2cm]{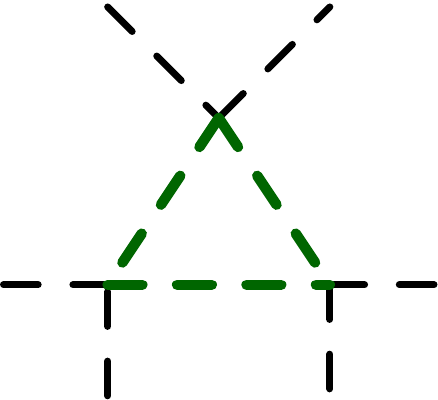}
			\caption{}\label{subfig:phi6-loop3}
		\end{subfigure}
%		\begin{subfigure}[t]{2.2cm}
%			\centering
%			\includegraphics[width=2cm,height=2.5cm]{phi6loop4.pdf}
%			\caption{}\label{subfig:phi6-loop4}
%		\end{subfigure}
		\caption{Tree- (i)-(iv) and one-loop-level (v)-(viii) diagrams built of Lorentz invariant renormalizable interactions of the UV theory that lead to effective operators of $\phi^6$ class.}
		\label{fig:phi6-unfolding}
	\end{figure}
	
	\begin{figure}[!htb]
		\centering
		\renewcommand{\thesubfigure}{\roman{subfigure}}
		\hspace{0.6cm}
		\begin{subfigure}[t]{2.2cm}
			\centering
			\includegraphics[width=2cm,height=1.8cm]{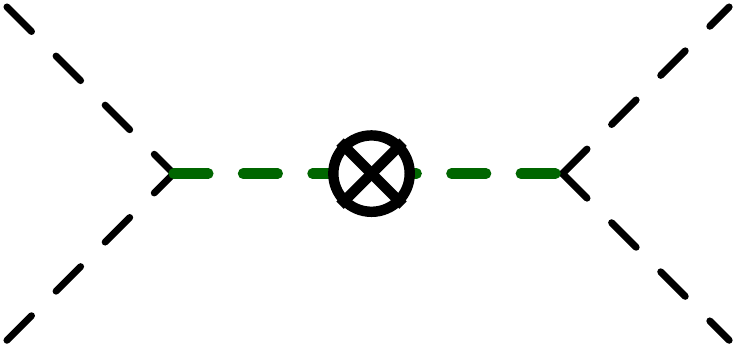}
			\caption{}\label{subfig:phi4D2-tree1}
		\end{subfigure}
		\begin{subfigure}[t]{2.2cm}
			\centering
			\includegraphics[width=2cm,height=1.8cm]{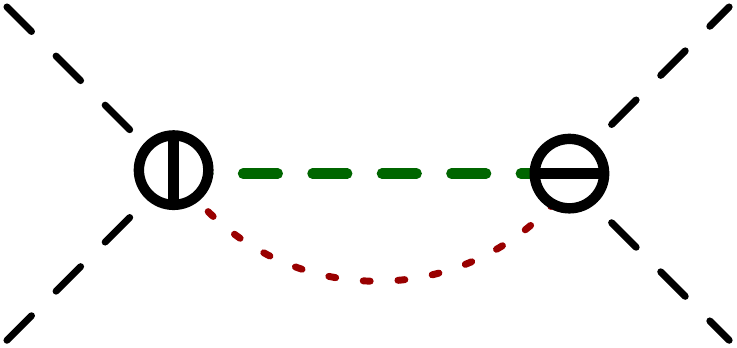}
			\caption{}\label{subfig:phi4D2-tree2}
		\end{subfigure}
		\begin{subfigure}[t]{2.2cm}
			\centering
			\includegraphics[width=2cm,height=1.8cm]{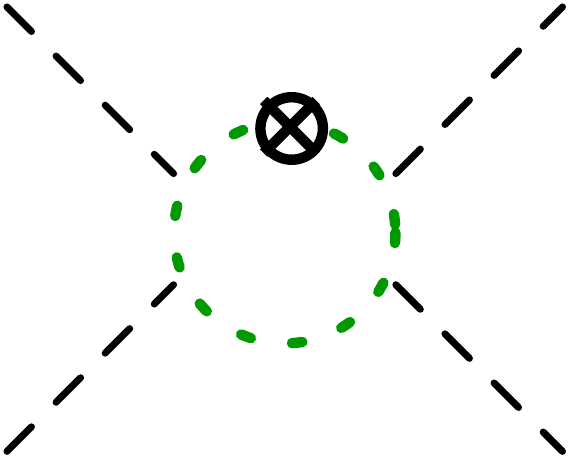}
			\caption{}\label{subfig:phi4D2-loop7}
		\end{subfigure}
		\begin{subfigure}[t]{2.2cm}
			\centering
			\includegraphics[width=2cm,height=1.7cm]{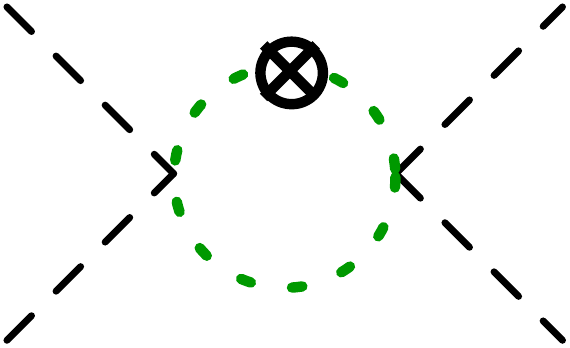}
			\caption{}\label{subfig:phi4D2-loop8}
		\end{subfigure}
		\begin{subfigure}[t]{2.2cm}
			\centering
			\includegraphics[width=2cm,height=1.8cm]{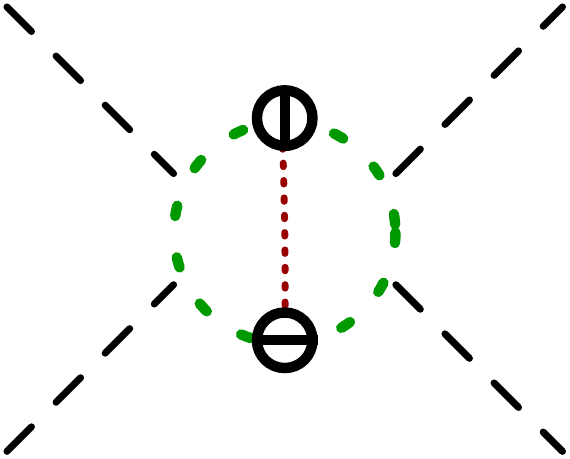}
			\caption{}\label{subfig:phi4D2-loop1}
		\end{subfigure}
		\begin{subfigure}[t]{2.2cm}
			\centering
			\includegraphics[width=2cm,height=1.8cm]{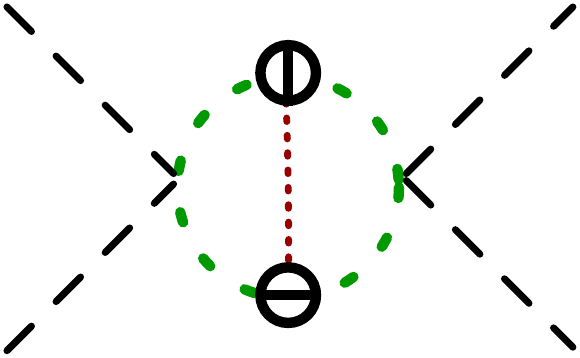}
			\caption{}\label{subfig:phi4D2-loop2}
		\end{subfigure}
		\begin{subfigure}[t]{2.2cm}
			\centering
			\includegraphics[width=2cm,height=1.8cm]{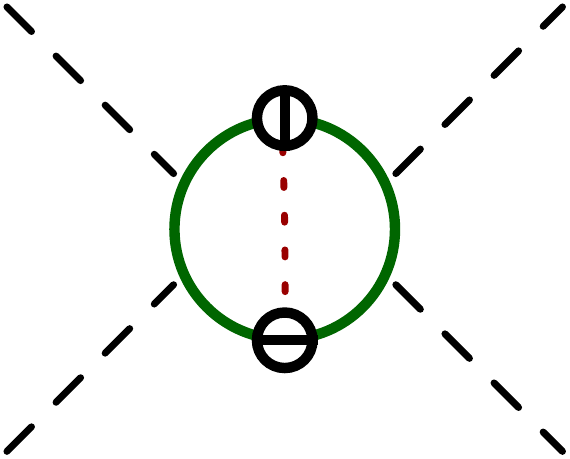}
			\caption{}\label{subfig:phi4D2-loop3}
		\end{subfigure}
		\caption{Tree- (i)-(ii) and one-loop-level (iii)-(v) diagrams built of Lorentz invariant renormalizable interactions of the UV theory that lead to effective operators of $\phi^4\mathcal{D}^2$ class.}
		\label{fig:phi4D2-unfolding}
	\end{figure}

	\begin{figure}[!htb]
		\centering
		\renewcommand{\thesubfigure}{\roman{subfigure}}
		\begin{subfigure}[t]{2.2cm}
			\centering
			\includegraphics[width=2cm,height=1.8cm]{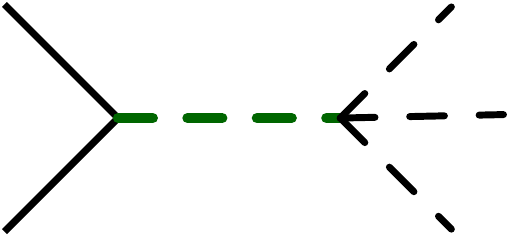}
			\caption{}\label{subfig:psi2phi3-tree1}
		\end{subfigure}
		\begin{subfigure}[t]{2.2cm}
			\centering
			\includegraphics[width=2cm,height=2.2cm]{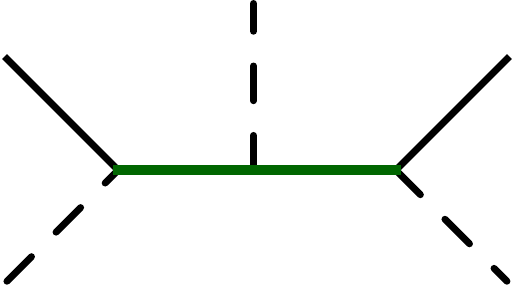}
			\caption{}\label{subfig:psi2phi3-tree2}
		\end{subfigure}
		\begin{subfigure}[t]{2.2cm}
			\centering
			\includegraphics[width=2cm,height=2.2cm]{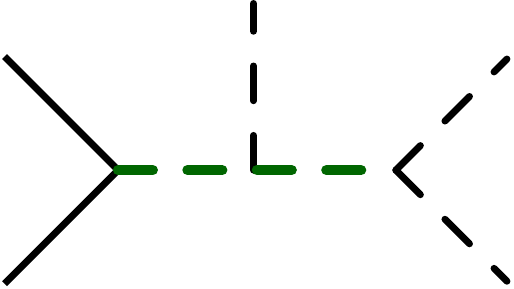}
			\caption{}\label{subfig:psi2phi3-tree3}
		\end{subfigure}
%		\begin{subfigure}[t]{2.5cm}
%			\centering
%			\includegraphics[width=2.2cm,height=2.2cm]{psi2phi3loop1.pdf}
%			\caption{}\label{subfig:psi2phi3-loop1}
%		\end{subfigure}
		\begin{subfigure}[t]{2.5cm}
			\centering
			\includegraphics[width=2.2cm,height=2.2cm]{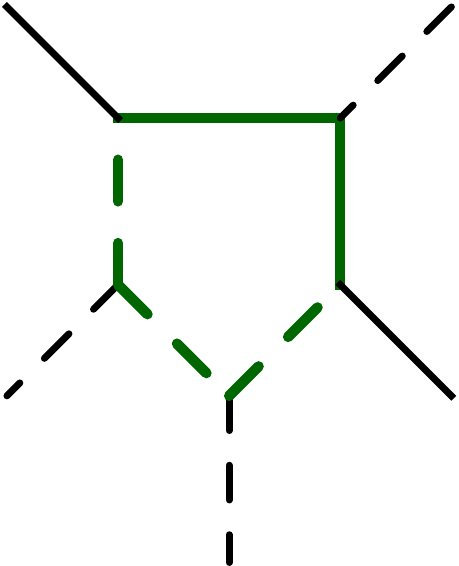}
			\caption{}\label{subfig:psi2phi3-loop2}
		\end{subfigure}
		\begin{subfigure}[t]{2.5cm}
			\centering
			\includegraphics[width=2.2cm,height=2.2cm]{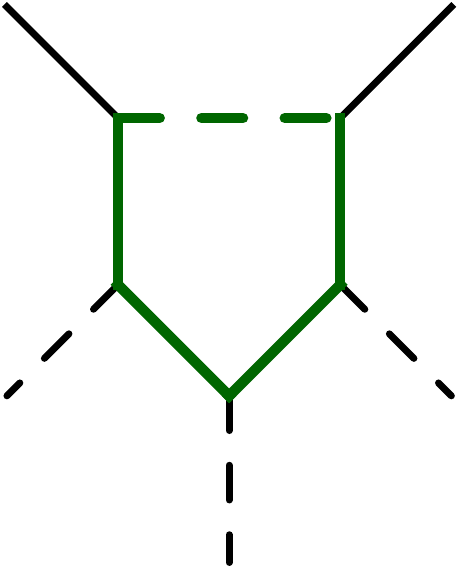}
			\caption{}\label{subfig:psi2phi3-loop3}
		\end{subfigure}
		\begin{subfigure}[t]{2.5cm}
			\centering
			\includegraphics[width=2.2cm,height=2.2cm]{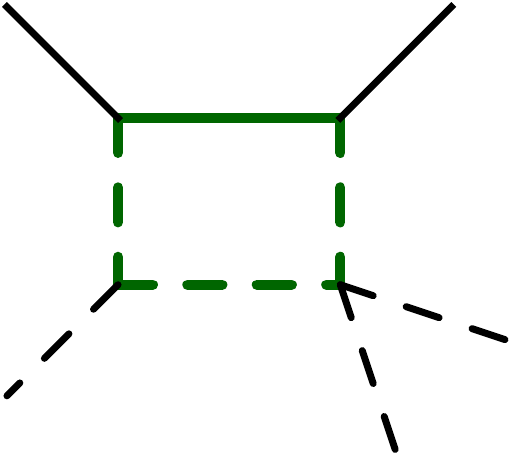}
			\caption{}\label{subfig:psi2phi3-loop4}
		\end{subfigure}
		\caption{Tree- (i)-(iii) and one-loop-level (iv)-(vii) diagrams built of Lorentz invariant renormalizable interactions of the UV theory that lead to effective operators of $\psi^2\phi^3$ class.}
		\label{fig:psi2phi3-unfolding}
	\end{figure}
	
	\begin{figure}[!htb]
		\centering
		\renewcommand{\thesubfigure}{\roman{subfigure}}
		\hspace{0.6cm}
		\begin{subfigure}[t]{5cm}
			\centering
			\includegraphics[width=3.8cm,height=2cm]{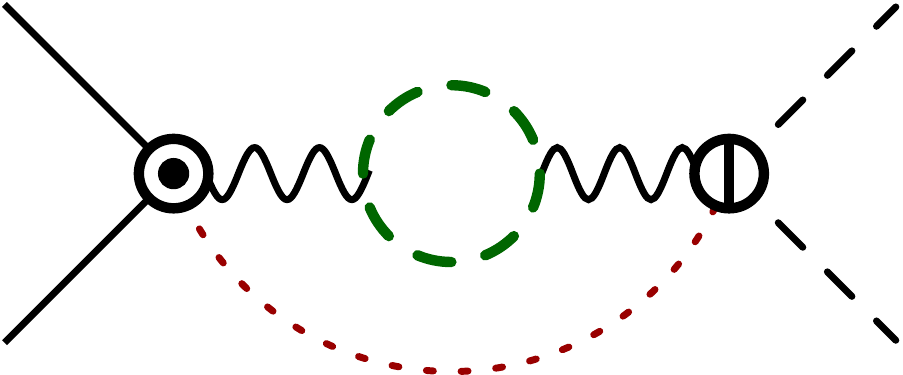}
			\caption{}\label{subfig:psi2phi2D-loop1}
		\end{subfigure}
		\begin{subfigure}[t]{5cm}
			\centering
			\includegraphics[width=3.8cm,height=2cm]{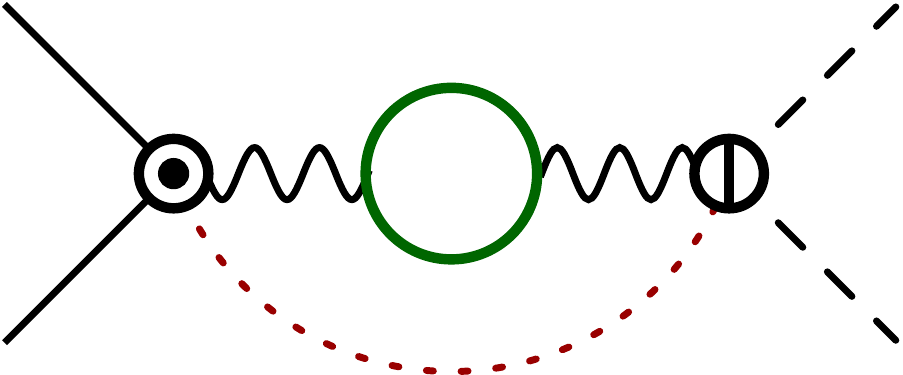}
			\caption{}\label{subfig:psi2phi2D-loop2}
		\end{subfigure}
		\caption{One-loop diagrams built of Lorentz invariant renormalizable interactions of the UV theory that lead to effective operators of $\psi^2\phi^2\mathcal{D}$ class.}
		\label{fig:psi2phi2D-unfolding}
	\end{figure}
	
	\begin{figure}[!htb]
		\centering
		\renewcommand{\thesubfigure}{\roman{subfigure}}
		\begin{subfigure}[t]{2.7cm}
			\centering
			\includegraphics[width=2.3cm,height=2cm]{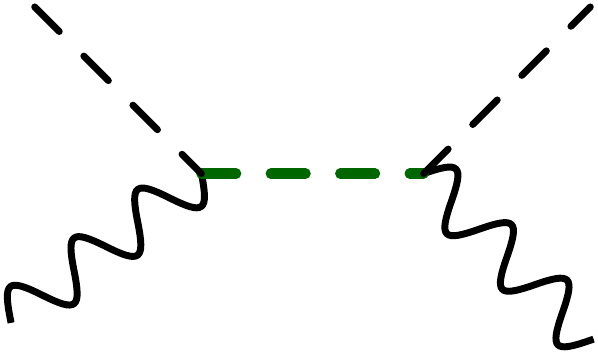}
			\caption{}\label{subfig:phi2x2-tree1}
		\end{subfigure}
		\begin{subfigure}[t]{2.7cm}
			\centering
			\includegraphics[width=2.3cm,height=2cm]{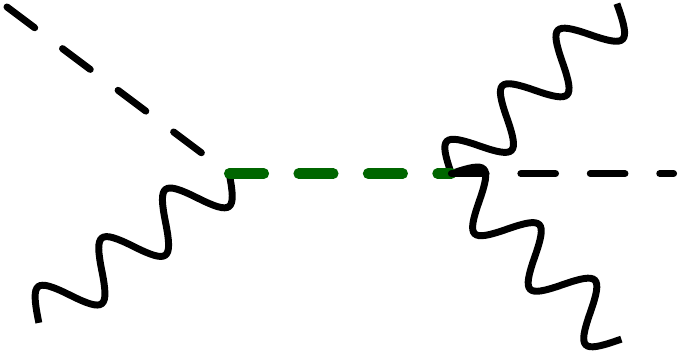}
			\caption{}\label{subfig:phi2x2-tree3}
		\end{subfigure}
		\begin{subfigure}[t]{2.7cm}
			\centering
			\includegraphics[width=2.3cm,height=2cm]{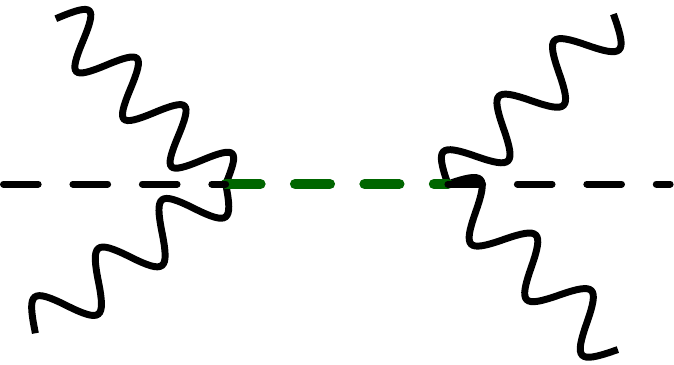}
			\caption{}\label{subfig:phi2x2-tree2}
		\end{subfigure}
		\begin{subfigure}[t]{2.7cm}
			\centering
			\includegraphics[width=2.1cm,height=2cm]{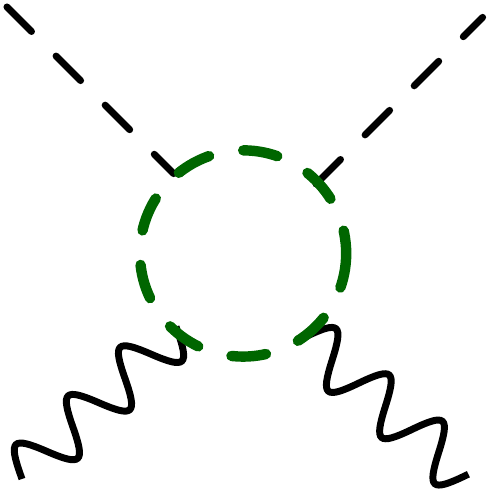}
			\caption{}\label{subfig:phi2x2-loop1}
		\end{subfigure}
		\begin{subfigure}[t]{2.7cm}
			\centering
			\includegraphics[width=2.1cm,height=2cm]{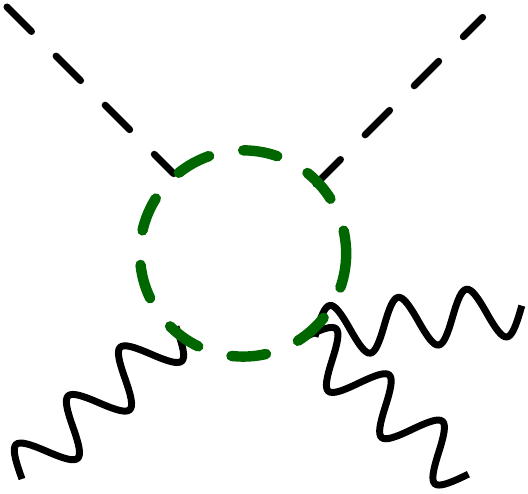}
			\caption{}\label{subfig:phi2x2-loop2}
		\end{subfigure}
		\begin{subfigure}[t]{2.7cm}
			\centering
			\includegraphics[width=2.3cm,height=2cm]{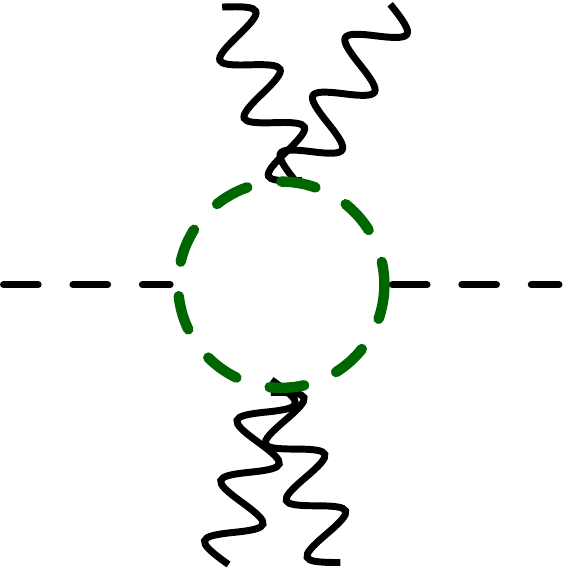}
			\caption{}\label{subfig:phi2x2-loop3}
		\end{subfigure}
		\newline
		\begin{subfigure}[t]{2.7cm}
			\centering
			\includegraphics[width=2.2cm,height=2cm]{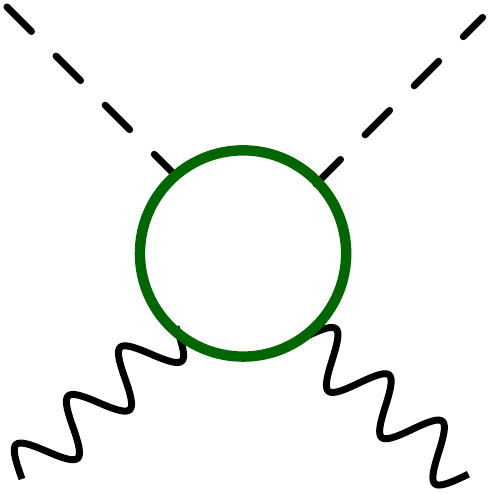}
			\caption{}\label{subfig:phi2x2-loop4}
		\end{subfigure}
		\begin{subfigure}[t]{2.7cm}
			\centering
			\includegraphics[width=2.3cm,height=2cm]{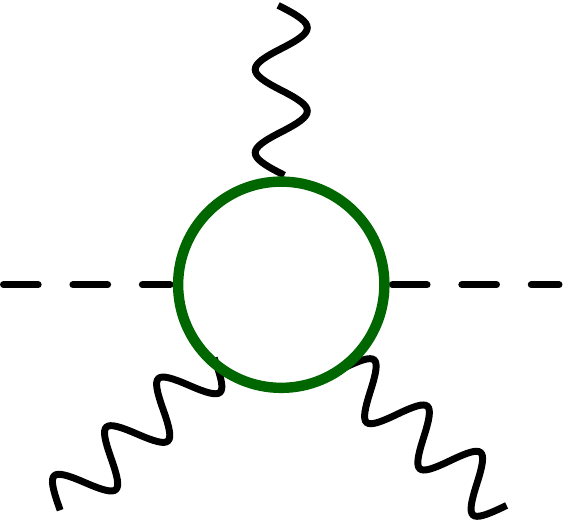}
			\caption{}\label{subfig:phi2x2-loop5}
		\end{subfigure}
		\begin{subfigure}[t]{2.7cm}
			\centering
			\includegraphics[width=2.3cm,height=2cm]{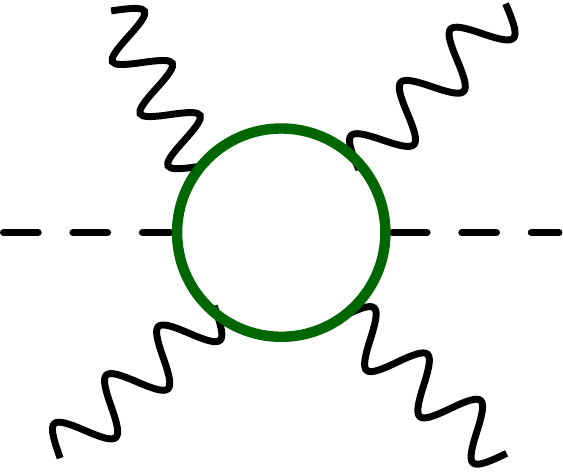}
			\caption{}\label{subfig:phi2x2-loop6}
		\end{subfigure}
		\begin{subfigure}[t]{2.7cm}
			\centering
			\includegraphics[width=2.3cm,height=1.7cm]{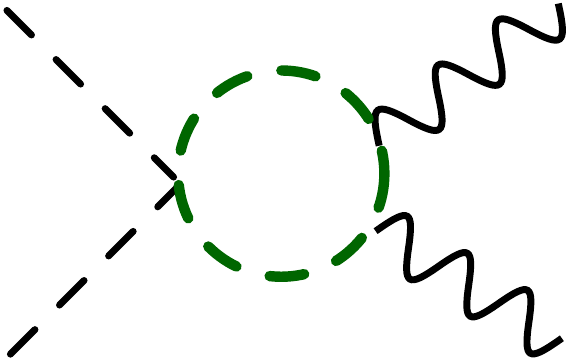}
			\caption{}\label{subfig:phi2x2-loop7}
		\end{subfigure}
		\begin{subfigure}[t]{2.7cm}
			\centering
			\includegraphics[width=2.3cm,height=1.7cm]{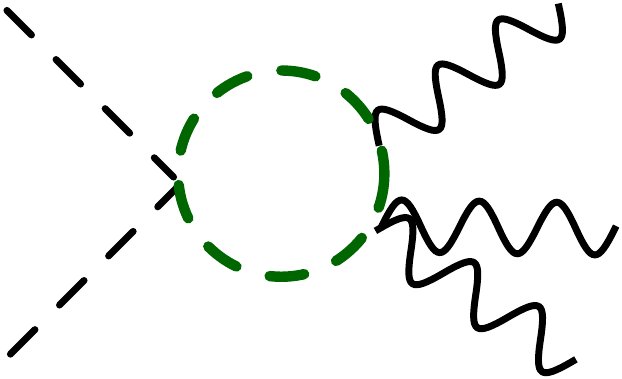}
			\caption{}\label{subfig:phi2x2-loop8}
		\end{subfigure}
		\begin{subfigure}[t]{2.7cm}
			\centering
			\includegraphics[width=2.3cm,height=2cm]{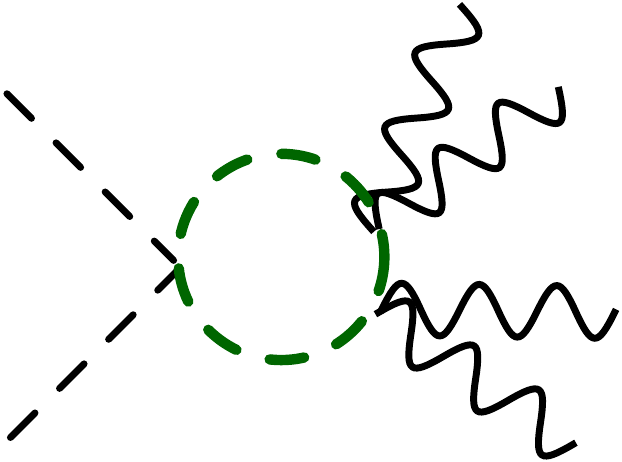}
			\caption{}\label{subfig:phi2x2-loop9}
		\end{subfigure}
		\caption{Tree- (i)-(ii) and one-loop-level (iii)-(viii) diagrams built of Lorentz invariant renormalizable interactions of the UV theory that lead to effective operators of $\phi^2X^2$ class.}
		\label{fig:phi2x2-unfolding}
	\end{figure}

	\begin{figure}[!htb]
		\centering
		\renewcommand{\thesubfigure}{\roman{subfigure}}
		\begin{subfigure}[t]{3cm}
			\centering
			\includegraphics[width=2.2cm,height=1.8cm]{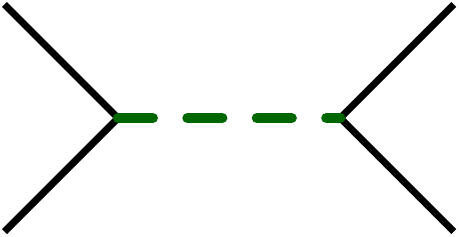}
			\caption{}\label{subfig:psi4-tree1}
		\end{subfigure}
		\begin{subfigure}[t]{3cm}
			\centering
			\includegraphics[width=2.2cm,height=2.2cm]{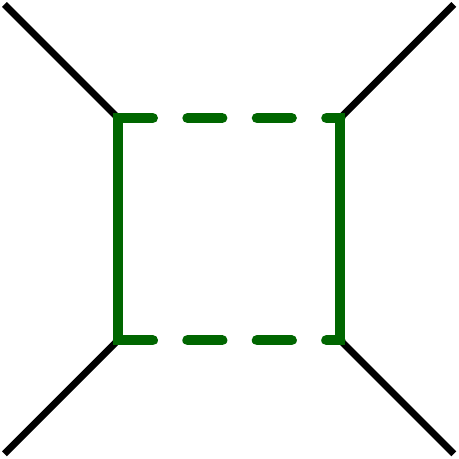}
			\caption{}\label{subfig:psi4-loop1}
		\end{subfigure}
	
		\caption{(i) Tree- and (ii) one-loop-level diagrams built of Lorentz invariant renormalizable interactions of the UV theory that lead to effective operators of $\psi^4$ class.}
		\label{fig:psi4-unfolding}
	\end{figure}

	\begin{figure}[!htb]
		\centering
		\renewcommand{\thesubfigure}{\roman{subfigure}}
		\hspace{0.6cm}
		\begin{subfigure}[t]{2cm}
			\centering
			\includegraphics[width=1.8cm,height=2cm]{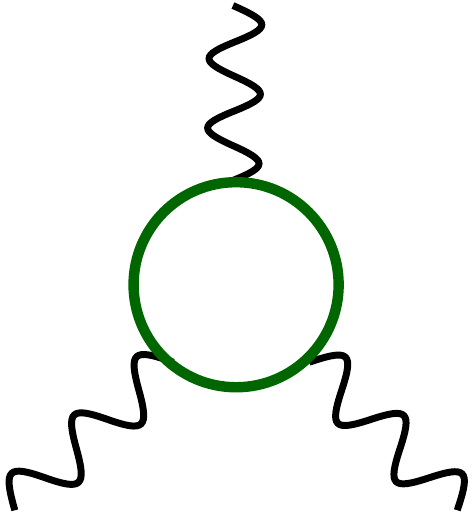}
			\caption{}\label{subfig:x3-loop1}
		\end{subfigure}
		\begin{subfigure}[t]{2cm}
			\centering
			\includegraphics[width=1.8cm,height=2cm]{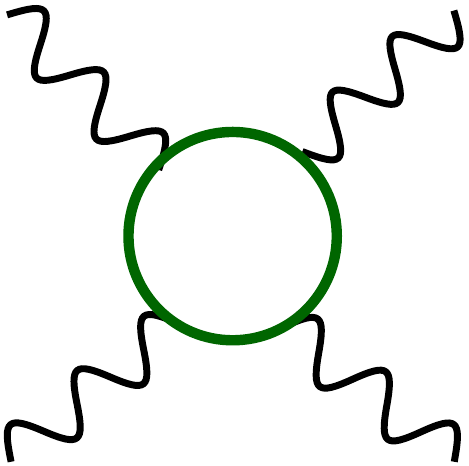}
			\caption{}\label{subfig:x3-loop2}
		\end{subfigure}
		\begin{subfigure}[t]{2cm}
			\centering
			\includegraphics[width=1.8cm,height=2cm]{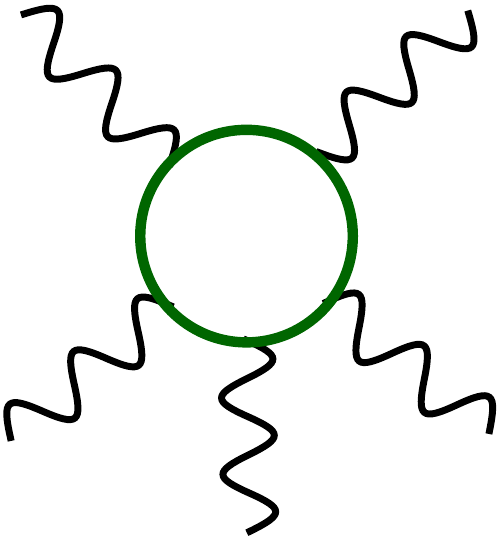}
			\caption{}\label{subfig:x3-loop3}
		\end{subfigure}
		\begin{subfigure}[t]{2cm}
			\centering
			\includegraphics[width=1.8cm,height=2cm]{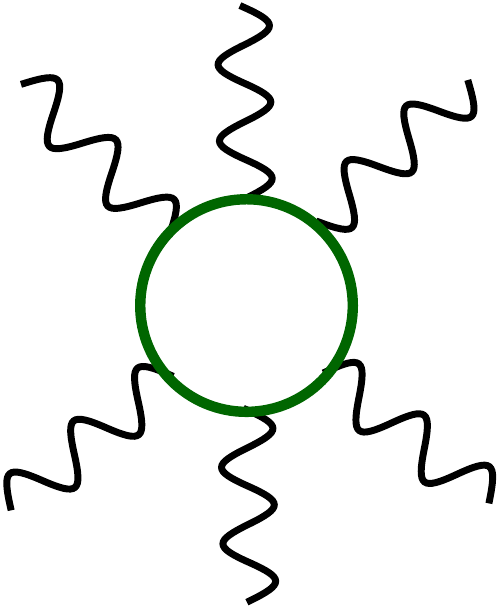}
			\caption{}\label{subfig:x3-loop4}
		\end{subfigure}
		\begin{subfigure}[t]{2cm}
			\centering
			\includegraphics[width=1.8cm,height=2cm]{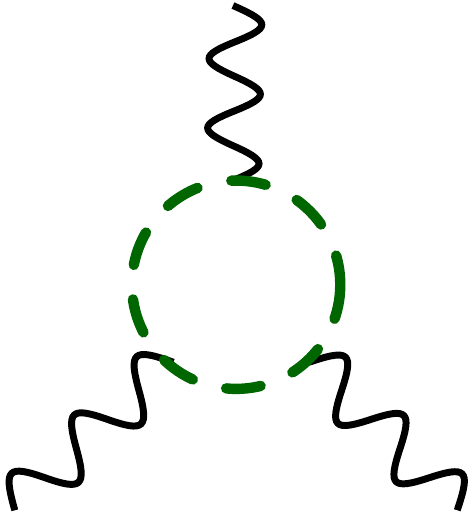}
			\caption{}\label{subfig:x3-loop5}
		\end{subfigure}
		\begin{subfigure}[t]{2cm}
			\centering
			\includegraphics[width=1.8cm,height=2cm]{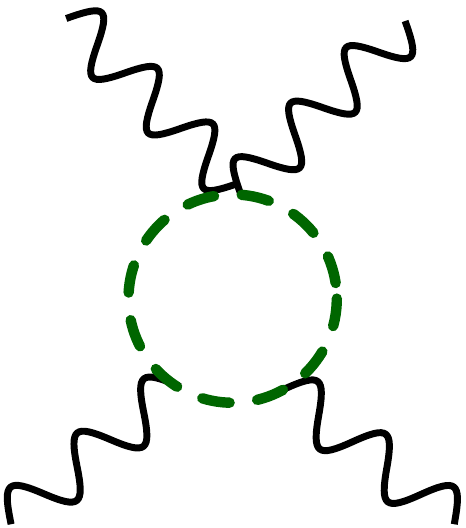}
			\caption{}\label{subfig:x3-loop6}
		\end{subfigure}
		\begin{subfigure}[t]{2cm}
			\centering
			\includegraphics[width=1.8cm,height=2cm]{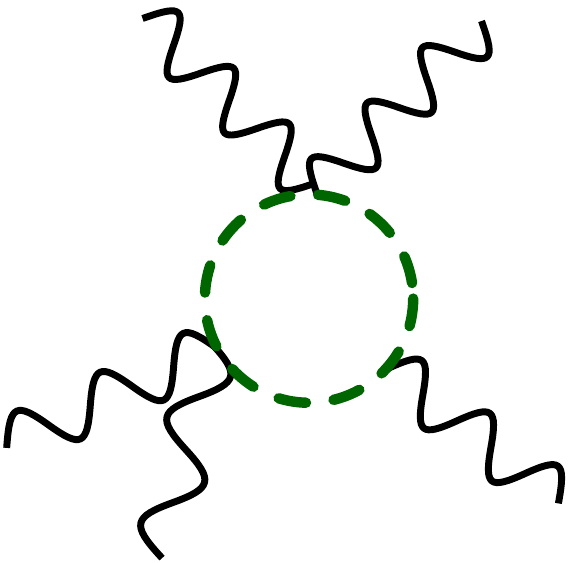}
			\caption{}\label{subfig:x3-loop7}
		\end{subfigure}
		\begin{subfigure}[t]{2cm}
			\centering
			\includegraphics[width=1.8cm,height=2cm]{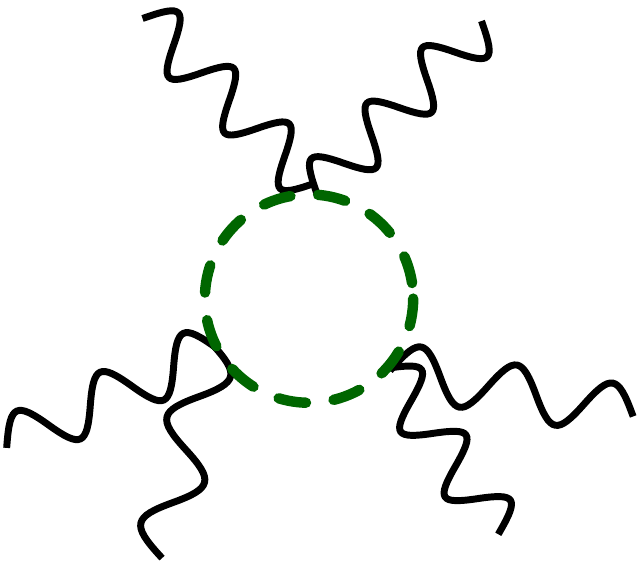}
			\caption{}\label{subfig:x3-loop8}
		\end{subfigure}
		\caption{One-loop diagrams built of Lorentz invariant renormalizable interactions of the UV theory that lead to effective operators of $X^3$ class.}
		\label{fig:x3-unfolding}
	\end{figure}
	
	\begin{figure}[!htb]
		\centering
		\renewcommand{\thesubfigure}{\roman{subfigure}}
		\hspace{0.6cm}
		\begin{subfigure}[t]{2.5cm}
			\centering
			\includegraphics[width=2.2cm,height=1.8cm]{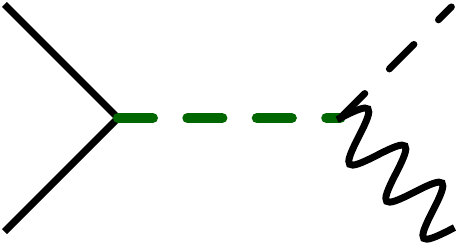}
			\caption{}\label{subfig:psi2phix-tree1}
		\end{subfigure}
		\begin{subfigure}[t]{2.5cm}
			\centering
			\includegraphics[width=2.2cm,height=1.8cm]{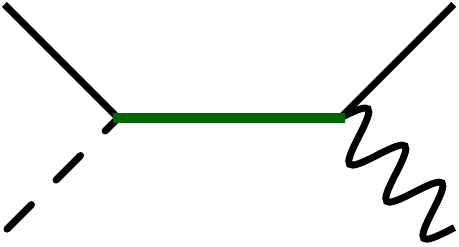}
			\caption{}\label{subfig:psi2phix-tree3}
		\end{subfigure}
		\begin{subfigure}[t]{2.5cm}
			\centering
			\includegraphics[width=2.2cm,height=2.2cm]{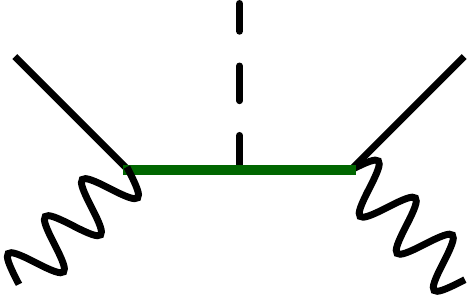}
			\caption{}\label{subfig:psi2phix-tree4}
		\end{subfigure}
		\begin{subfigure}[t]{2.5cm}
			\centering
			\includegraphics[width=2.2cm,height=2.2cm]{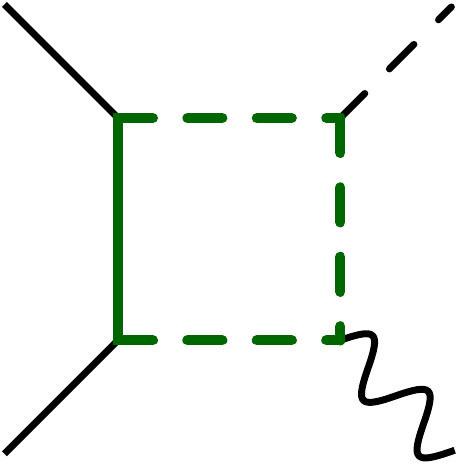}
			\caption{}\label{subfig:psi2phix-loop1}
		\end{subfigure}
		\begin{subfigure}[t]{2.5cm}
			\centering
			\includegraphics[width=2.2cm,height=2.2cm]{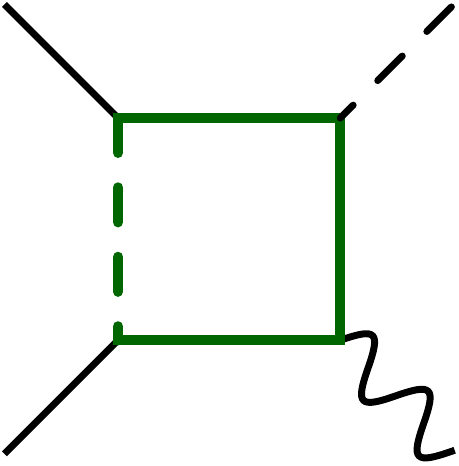}
			\caption{}\label{subfig:psi2phix-loop2}
		\end{subfigure}
		\begin{subfigure}[t]{2.5cm}
			\centering
			\includegraphics[width=2.2cm,height=2.2cm]{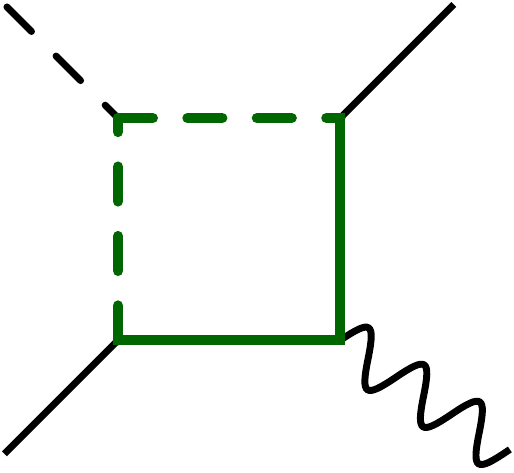}
			\caption{}\label{subfig:psi2phix-loop5}
		\end{subfigure}
		\caption{Tree- (i)-(iv) and one-loop-level (v)-(ix) diagrams built of Lorentz invariant renormalizable interactions of the UV theory that lead to effective operators of $\psi^2\phi X$ class.}
		\label{fig:psi2phix-unfolding}
	\end{figure}
	
	\clearpage
	\section{Identifying heavy fields corresponding to dimension-6 SMEFT operators}\label{sec:smeft-unfolding}
	
	In this section, we have substantiated the general ideas presented in the previous one using the CP even dimension-6 SMEFT operators as the backdrop, i.e., henceforth the light fields have been restricted to be Standard Model fields, described in appendix \ref{sec:app}. The sequence of steps involved can be described under two main headings:
	
	\subsubsection*{$(i)$ Cataloguing an exhaustive list of heavy field representations}
	
	\begin{itemize}
		\item Based on Fig.~\ref{fig:renorm-vertices}, we have listed all possible combinations of light and heavy fields that lead to a particular vertex. In the rest of this paper, these have been referred to as ``fundamental vertices".  
		
		\item For each vertex and its sub-cases, based on the fixed light fields, we have identified all possible representations of the heavy fields. We have presented the results for vertices composed of scalars, fermions, gauge bosons and their combinations in Tables~\ref{table:sm-vertices-1}-\ref{table:sm-vertices-3}. 
	\end{itemize}
	\begin{table}[h]
		\centering
		\renewcommand{\arraystretch}{1.8}
		{\small\begin{tabular}{|c|c|c|c|}
				\hline
				\textbf{Vertex}&
				\textbf{S. No.}&
				\textbf{Light fields}&
				\textbf{Heavy field(s)}\\
				\hline

				\multirow{4}{*}{\includegraphics[width=3.6cm, height=2.8cm]{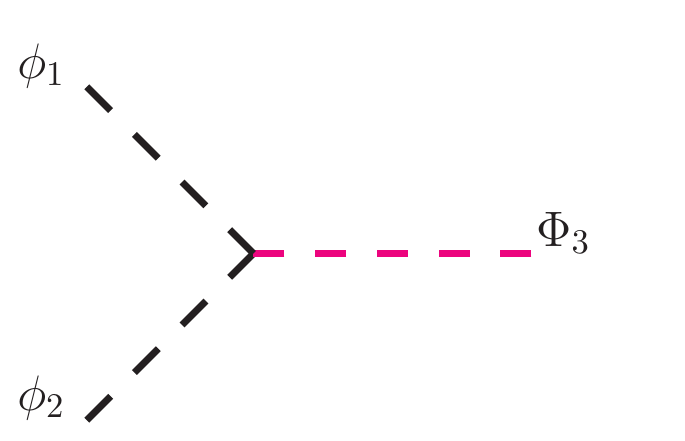}}&
				\multirow{2}{*}{\hypertarget{vertex-1-i}{V1-(i)}}&	
				\multirow{2}{*}{$\phi_1 = \phi_2 = H_{(1,2,\frac{1}{2})}$ or $H^{\dagger}_{(1,2,-\frac{1}{2})}$}&
				\multirow{2}{*}{$\Phi_3 \in \{(1,3,\pm1),\,\, (1,1,\pm1)\}$}\\
				
				&
				&
				&
				\\
				\cdashline{2-4}
				
				&
				\multirow{2}{*}{\hypertarget{vertex-1-ii}{V1-(ii)}}&
				\multirow{2}{*}{$\phi_1 = H,\,\,\,\, \phi_2 = H^{\dagger}$}&
				\multirow{2}{*}{$\Phi_3 \in \{(1,3,0),\,\, (1,1,0)\}$}\\

				&
				&
				&
				\\
				
				\hline
				\multirow{3}{*}{\includegraphics[width=3.6cm, height=2.2cm]{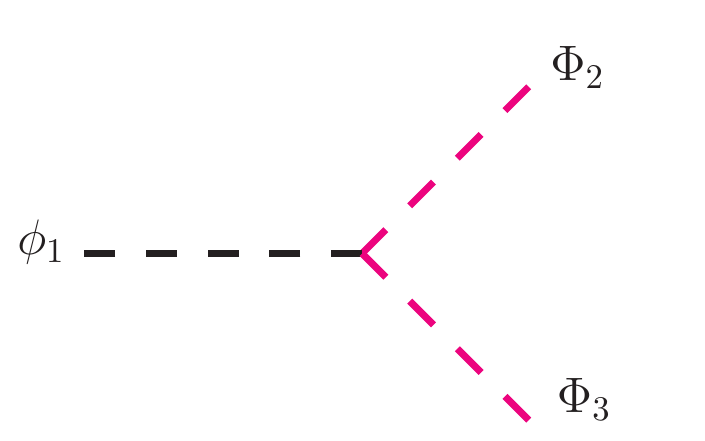}\label{johncena}}&
				\multirow{3}{*}{\hypertarget{vertex-2}{V2}}&
				\multirow{3}{*}{$\phi_1 = H$ or $H^{\dagger}$}&
				$\Phi_2 \in (R_{C_2}, R_{L_2}, Y_2)$, $\Phi_3 \in (R_{C_3}, R_{L_3}, Y_3)$  \\
				
				&
				&
				&
				with $R_{C_2} \otimes R_{C_3} \equiv 1$, $R_{L_2} \otimes R_{L_3} \equiv 2$\\
				
				&
				&
				&
				 and $Y_2 + Y_3 = \pm\frac{1}{2}$.\\
				
				\hline
				
				\multirow{4}{*}{\includegraphics[width=3.6cm, height=2.8cm]{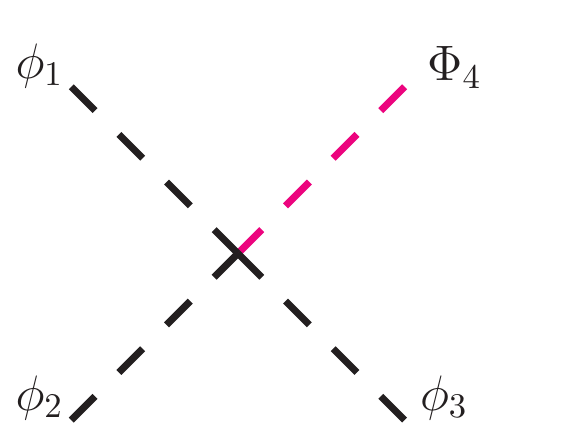}}&
				\multirow{2}{*}{\hypertarget{vertex-3-i}{V3-(i)}}&
				\multirow{2}{*}{$\phi_1 = \phi_2 = \phi_3 = H$ or $H^{\dagger}$}&
				\multirow{2}{*}{$\Phi_4 \in \{(1,4,\pm\frac{3}{2}),\,\, (1,2,\pm\frac{3}{2})\}$}\\
				
				&
				&
				&
				\\
				\cdashline{2-4}
				
				&
				\multirow{2}{*}{\hypertarget{vertex-3-ii}{V3-(ii)}}&
				\multirow{2}{*}{$\phi_1 = \phi_2 = H,\,\,\,\, \phi_3 = H^{\dagger}$}&
				\multirow{2}{*}{$\Phi_4 \in \{(1,4,\pm\frac{1}{2}),\,\, (1,2,\pm\frac{1}{2})\}$}\\
				
				&
				&
				&
				\\
				
				\hline

				\multirow{5}{*}{\includegraphics[width=3.6cm, height=2.8cm]{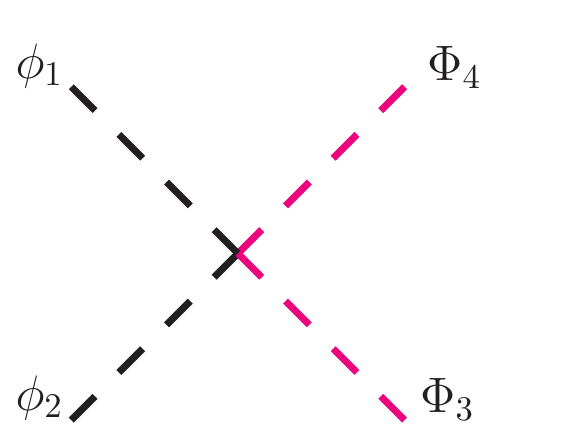}}&
				\multirow{2}{*}{\hypertarget{vertex-4-i}{V4-(i)}}&
				\multirow{2}{*}{$\phi_1 = H,\,\,\,\, \phi_2 = H^{\dagger}$}&
				\multirow{2}{*}{$\Phi_3 \in (\{1,R_C\},\{1,R_L\},\{0,Y\}),\,\,\,\, \Phi_4 = \Phi_3^{\dagger}$}.\\
				
				&
				&
				&
				\\
				\cdashline{2-4}
				
				&
				\multirow{3}{*}{\hypertarget{vertex-4-ii}{V4-(ii)}}&
				\multirow{3}{*}{$\phi_1 = \phi_2 = H$ or $H^{\dagger}$}&
				$\Phi_3 \in (R_{C_3}, R_{L_3}, Y_3)$, $\Phi_4 \in (R_{C_4}, R_{L_4}, Y_4)$\\

				&
				&
				&
				 with $R_{C_3} \otimes R_{C_4} \equiv 1$, $R_{L_3} \otimes R_{L_4} \equiv 1$ or $3$\\

				&
				&
				&
				 and $Y_3 + Y_4 = \pm1$.\\
				\hline
		\end{tabular}}
		\caption{Allowed heavy field representations when the light degrees of freedom are the Standard Model ones. Here,  $\phi_i$ denote the SM Higgs and $\Phi_i$ the various heavy scalars. $R_C$ and $R_L$ denote representations under $SU(3)_C$ and $SU(2)_L$ gauge groups and $Y$ refers to the $U(1)_Y$ hypercharge of the field. Their appearance describes the cases where the vertex is not constituted by a unique heavy field representation but can involve a plethora of them. }
		\label{table:sm-vertices-1}
	\end{table}

	\begin{table}[h]
		\centering
		\renewcommand{\arraystretch}{1.7}
		{\begin{tabular}{|c|c|c|c|}
				\hline
				\textbf{Vertex}&
				\textbf{S. No.}&	
				\textbf{Light fields}&
				\textbf{Heavy field(s)}\\
				\hline
				
				\multirow{15}{*}{\includegraphics[width=3.6cm, height=3.2cm]{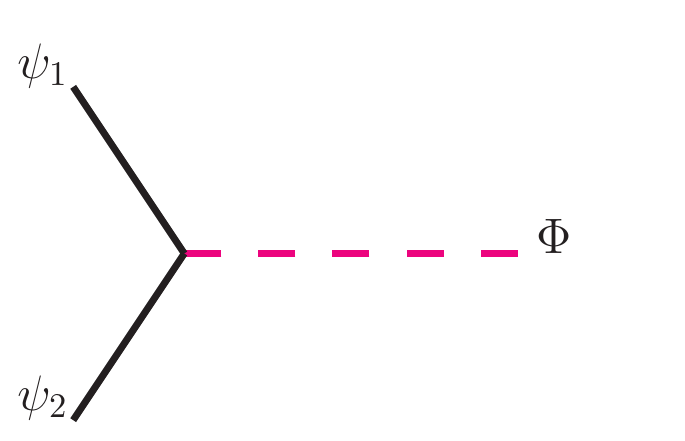}}&
				\hypertarget{vertex-5-i}{V5-(i)}&
				$\psi_1=\psi_2 = e_{(1,1,-1)}$&
				$\Phi \in (1,1,2)$\\
				\cdashline{2-4}
				
				&
				\hypertarget{vertex-5-ii}{V5-(ii)}&
				$\psi_1=\psi_2 = l_{(1,2,-\frac{1}{2})}$&
				$\Phi \in \{(1,1,1),\,\, (1,3,1)\}$\\
				\cdashline{2-4}
				
				&
				\hypertarget{vertex-5-iii}{V5-(iii)}&
				$\psi_1=\psi_2 = d_{(3,1,-\frac{1}{3})}$&
				$\Phi \in \{(3,1,\frac{2}{3}),\,\, (\bar{6},1,\frac{2}{3})\}$\\
				\cdashline{2-4}
				
				&
				\hypertarget{vertex-5-iv}{V5-(iv)}&
				$\psi_1=\psi_2 = u_{(3,1,\frac{2}{3})}$&
				$\Phi \in \{(3,1,-\frac{4}{3}),\,\, (\bar{6},1,-\frac{4}{3})\}$\\
				\cdashline{2-4}
				
				&
				\hypertarget{vertex-5-v}{V5-(v)}&
				$\psi_1=\psi_2 = q_{(3,2,\frac{1}{6})}$&
				$\Phi \in \{(3,1,-\frac{1}{3}),\,\, (3,3,-\frac{1}{3}), (\bar{6},1,-\frac{1}{3}),\,\,(\bar{6},3,-\frac{1}{3})\}$\\
				
				\cdashline{2-4}
				
				&
				\hypertarget{vertex-5-vi}{V5-(vi)}&
				$(\psi_1, \psi_2) = (\bar{l}, e)$&
				$\Phi \in (1,2,\frac{1}{2})$\\
				\cdashline{2-4}
				
				&
				\hypertarget{vertex-5-vii}{V5-(vii)}&
				$(\psi_1, \psi_2) = (\bar{q}, d)$&
				$\Phi \in \{(1,2,\frac{1}{2}), (8,2,\frac{1}{2})\}$\\
				\cdashline{2-4}
				
				&
				\hypertarget{vertex-5-viii}{V5-(viii)}&
				$(\psi_1, \psi_2) = (\bar{u}, q)$&
				$\Phi \in \{(1,2,\frac{1}{2}), (8,2,\frac{1}{2})\}$\\
				\cdashline{2-4}
				
				&
				\hypertarget{vertex-5-ix}{V5-(ix)}&
				$(\psi_1, \psi_2) = (q, l)$&
				$\Phi \in \{(\bar{3},1,\frac{1}{3}),\,\,(\bar{3},3,\frac{1}{3})\}$\\
				\cdashline{2-4}
				
				&
				\hypertarget{vertex-5-x}{V5-(x)}&
				$(\psi_1, \psi_2) = (u, d)$&
				$\Phi \in \{(3,1,-\frac{1}{3}),\,\,(\bar{6},1,-\frac{1}{3})\}$\\
				\cdashline{2-4}
				
				&
				\hypertarget{vertex-5-xii}{V5-(xi)}&
				$(\psi_1, \psi_2) = (u, e)$&
				$\Phi \in (\bar{3},1,\frac{1}{3})$\\
				\cdashline{2-4}
				
				&
				\hypertarget{vertex-5-xiii}{V5-(xii)}&
				$(\psi_1, \psi_2) = (d, e)$&
				$\Phi \in (\bar{3},1,\frac{4}{3})$\\
				\cdashline{2-4}
				
				&
				\hypertarget{vertex-5-xiv}{V5-(xiii)}&
				$(\psi_1, \psi_2) = (\bar{q}, e)$&
				$\Phi \in (3,2,\frac{7}{6})$\\
				\cdashline{2-4}
				
				&
				\hypertarget{vertex-5-xv}{V5-(xiv)}&
				$(\psi_1, \psi_2) = (\bar{l}, u)$&
				$\Phi \in (\bar{3},2,-\frac{7}{6})$\\
				\cdashline{2-4}
				
				&
				\hypertarget{vertex-5-xvi}{V5-(xv)}&
				$(\psi_1, \psi_2) = (\bar{l}, d)$&
				$\Phi \in (\bar{3},2,-\frac{1}{6})$\\
				\hline
				
				\multirow{10}{*}{\includegraphics[width=3.6cm, height=3.2cm]{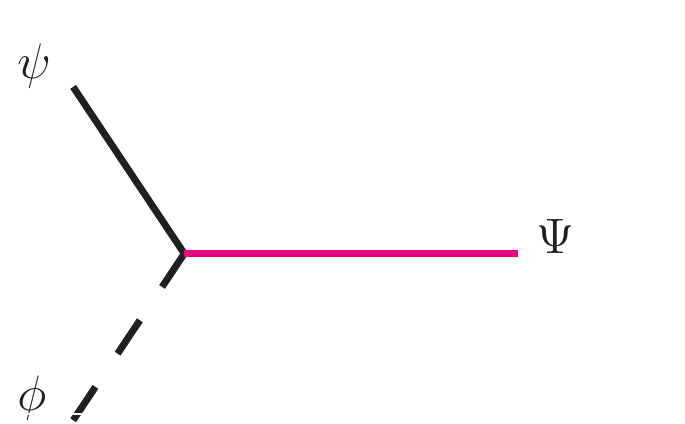}}&
				\hypertarget{vertex-6-i}{V6-(i)}&
				$\phi = H_{(1,2,\frac{1}{2})}, \,\,\,\, \psi = l$&
				$\Psi \in \{(1,1,0), (1,3,0)\}$\\
				\cdashline{2-4}
				
				&
				\hypertarget{vertex-6-ii}{V6-(ii)}&
				$\phi = H, \,\,\,\, \psi = e$&
				$\Psi \in (1,2,\frac{1}{2})$\\
				\cdashline{2-4}

				&
				\hypertarget{vertex-6-iii}{V6-(iii)}&
				$\phi = H, \,\,\,\, \psi = q$&
				$\Psi \in \{(\bar{3},1,-\frac{2}{3}), (\bar{3},3,-\frac{2}{3})\}$\\
				\cdashline{2-4}
				
				&
				\hypertarget{vertex-6-iv}{V6-(iv)}&
				$\phi = H, \,\,\,\, \psi = u$&
				$\Psi \in (\bar{3},2,-\frac{7}{6})$\\
				\cdashline{2-4}
				
				&
				\hypertarget{vertex-6-v}{V6-(v)}&
				$\phi = H, \,\,\,\, \psi = d$&
				$\Psi \in (\bar{3},2,-\frac{1}{6})$\\
				\cdashline{2-4}
				
				&
				\hypertarget{vertex-6-vi}{V6-(vi)}&
				$\phi = H^{\dagger}, \,\,\,\, \psi = l$&
				$\Psi \in \{(1,1,1), (1,3,1)\}$\\
				\cdashline{2-4}
				
				&
				\hypertarget{vertex-6-vii}{V6-(vii)}&
				$\phi = H^{\dagger}, \,\,\,\, \psi = e$&
				$\Psi \in (1,2,\frac{3}{2})$\\
				\cdashline{2-4}
				
				&
				\hypertarget{vertex-6-viii}{V6-(viii)}&
				$\phi = H^{\dagger}, \,\,\,\, \psi = q$&
				$\Psi \in \{(\bar{3},1,\frac{1}{3}), (\bar{3},3,\frac{1}{3})\}$\\
				\cdashline{2-4}
				
				&
				\hypertarget{vertex-6-ix}{V6-(ix)}&
				$\phi = H^{\dagger}, \,\,\,\, \psi = u$&
				$\Psi \in (\bar{3},2,-\frac{1}{6})$\\
				\cdashline{2-4}
				
				&
				\hypertarget{vertex-6-x}{V6-(x)}&
				$\phi = H^{\dagger}, \,\,\,\, \psi = d$&
				$\Psi \in (\bar{3},2,\frac{5}{6})$\\
				\hline
		\end{tabular}}
		\caption{Table \ref{table:sm-vertices-1} continued. Here,  $\phi_i$ and $\psi_i$ denote the SM Higgs and the SM fermions respectively, whereas $\Phi_i$ and $\Psi_i$ denote the various heavy scalars and heavy fermions.}
		\label{table:sm-vertices-2}
	\end{table}
	
	\clearpage
	\begin{table}[h]
		\centering
		\renewcommand{\arraystretch}{1.8}
		{\begin{tabular}{|c|c|c|c|}
				\hline
				\textbf{Vertex}&
				\textbf{S. No.}&
				\textbf{Light fields}&
				\textbf{Heavy field(s)}\\
				\hline
				
				\multirow{10}{*}{\includegraphics[width=3.6cm, height=3.2cm]{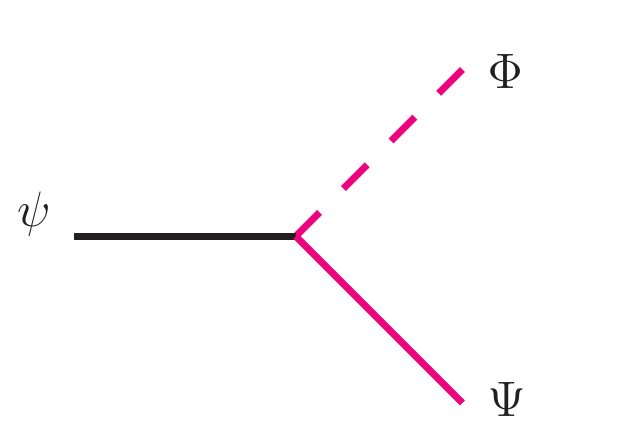}}&
				\multirow{2}{*}{\hypertarget{vertex-7-i}{V7-(i)}}&
				\multirow{2}{*}{$\psi = e_{(1,1,-1)}$}&
				$\Phi \in (R_{C_1}, R_{L_1}, Y_1)$, $\Psi \in (R_{C_2}, R_{L_2}, Y_2)$ with\\

				&
				&
				&
				$R_{C_1} \otimes R_{C_2} \equiv 1$, $R_{L_1} \otimes R_{L_2} \equiv 1$ and $Y_1 + Y_2 = 1$,\\
				\cdashline{2-4}
				
				&
				\multirow{2}{*}{\hypertarget{vertex-7-ii}{V7-(ii)}}&
				\multirow{2}{*}{$\psi = l_{(1,2,-\frac{1}{2})}$}&
				$\Phi \in (R_{C_1}, R_{L_1}, Y_1)$, $\Psi \in (R_{C_2}, R_{L_2}, Y_2)$ with \\

				&
				&
				&
				$R_{C_1} \otimes R_{C_2} \equiv 1$, $R_{L_1} \otimes R_{L_2} \equiv 2$ and $Y_1 + Y_2 = \frac{1}{2}$,\\
				\cdashline{2-4}
				
				&
				\multirow{2}{*}{\hypertarget{vertex-7-iii}{V7-(iii)}}&
				\multirow{2}{*}{$\psi = d_{(3,1,-\frac{1}{3})}$}&
				$\Phi \in (R_{C_1}, R_{L_1}, Y_1)$, $\Psi \in (R_{C_2}, R_{L_2}, Y_2)$ with \\

				&
				&
				&
				$R_{C_1} \otimes R_{C_2} \equiv \bar{3}$, $R_{L_1} \otimes R_{L_2} \equiv 1$ and $Y_1 + Y_2 = \frac{1}{3}$,\\
				\cdashline{2-4}
				
				&
				\multirow{2}{*}{\hypertarget{vertex-7-iv}{V7-(iv)}}&
				\multirow{2}{*}{$\psi = u_{(3,1,\frac{2}{3})}$}&
				$\Phi \in (R_{C_1}, R_{L_1}, Y_1)$, $\Psi \in (R_{C_2}, R_{L_2}, Y_2)$ with \\

				&
				&
				&
				$R_{C_1} \otimes R_{C_2} \equiv \bar{3}$, $R_{L_1} \otimes R_{L_2} \equiv 1$ and $Y_1 + Y_2 = -\frac{2}{3}$,\\
				
				\cdashline{2-4}
				
				&
				\multirow{2}{*}{\hypertarget{vertex-7-v}{V7-(v)}}&
				\multirow{2}{*}{$\psi = q_{(3,2,\frac{1}{6})}$}&
				$\Phi \in (R_{C_1}, R_{L_1}, Y_1)$, $\Psi \in (R_{C_2}, R_{L_2}, Y_2)$ with \\

				&
				&
				&
				$R_{C_1} \otimes R_{C_2} \equiv \bar{3}$, $R_{L_1} \otimes R_{L_2} \equiv 2$ and $Y_1 + Y_2 = -\frac{1}{6}$,\\
				
				\hline

				\multirow{3}{*}{\includegraphics[width=3.6cm, height=2.4cm]{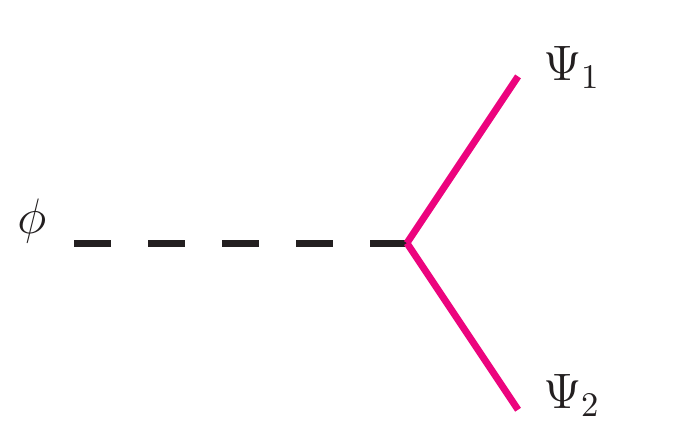}}&
				\multirow{3}{*}{\hypertarget{vertex-8}{V8}}&
				\multirow{3}{*}{$\phi = H$ or $H^{\dagger}$}&
				$\Psi_1 \in (R_{C_1}, R_{L_1}, Y_1)$, $\Psi_2 \in (R_{C_2}, R_{L_2}, Y_2)$ \\

				&
				&
				&
				with $R_{C_1} \otimes R_{C_2} \equiv 1$, $R_{L_1} \otimes R_{L_2} \equiv 2$ \\
				
				&
				&
				&
				and $Y_1 + Y_2 = \pm\frac{1}{2}$,\\
				
				\hline
				
				\multirow{3}{*}{\includegraphics[width=3.6cm, height=1.8cm]{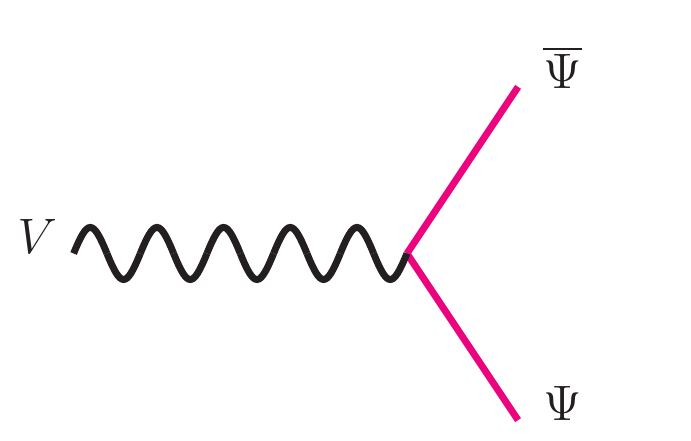}}&
				\hypertarget{vertex-9-i}{V9-(i)}&
				$V = B_{(1,1,0)}$&
				$\Psi \in (\{1,R_C\}, \{1,R_L\}, Y)$\\
				\cdashline{2-4}
				
				&
				\hypertarget{vertex-9-ii}{V9-(ii)}&
				$V = W_{(1,3,0)}$&
				$\Psi \in (\{1,R_C\}, R_L, \{0,Y\})$\\
				\cdashline{2-4}
				
				&
				\hypertarget{vertex-9-iii}{V9-(iii)}&
				$V = G_{(8,1,0)}$&
				$\Psi \in (R_C, \{1,R_L\}, \{0,Y\})$\\
				\hline
				
				\multirow{3}{*}{\includegraphics[width=3.6cm, height=1.8cm]{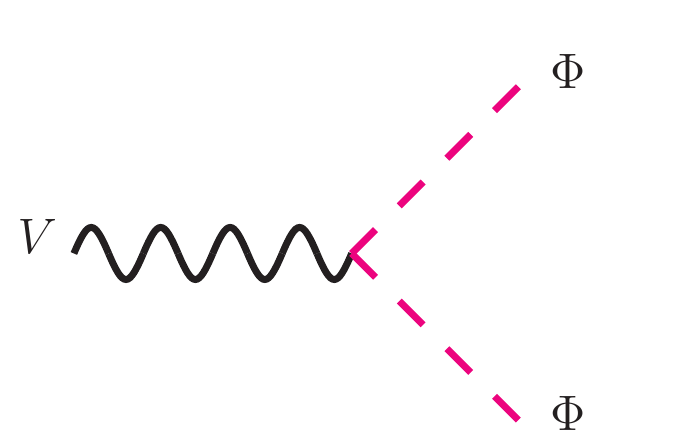}}&
				\hypertarget{vertex-10-i}{V10-(i)}&
				$V = B$&
				$\Phi \in (\{1,R_C\}, \{1,R_L\}, Y)$\\
				\cdashline{2-4}
				
				&
				\hypertarget{vertex-10-ii}{V10-(ii)}&
				$V = W$&
				$\Phi \in (\{1,R_C\}, R_L, \{0,Y\})$\\
				\cdashline{2-4}
				
				&
				\hypertarget{vertex-10-iii}{V10-(iii)}&
				$V = G$&
				$\Phi \in (R_C, \{1,R_L\}, \{0,Y\})$\\
				
				\hline
				
				\multirow{3}{*}{\includegraphics[width=3.6cm, height=1.8cm]{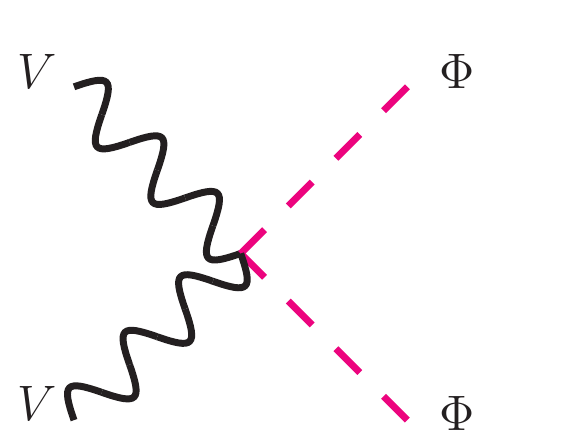}}&
				\hypertarget{vertex-11-i}{V11-(i)}&
				$V = B$&
				$\Phi \in (\{1,R_C\}, \{1,R_L\}, Y)$\\
				\cdashline{2-4}
				
				&
				\hypertarget{vertex-11-ii}{V11-(ii)}&
				$V = W$&
				$\Phi \in (\{1,R_C\}, R_L, \{0,Y\})$\\
				\cdashline{2-4}
				
				&
				\hypertarget{vertex-11-iii}{V11-(iii)}&
				$V = G$&
				$\Phi \in (R_C, \{1,R_L\}, \{0,Y\})$\\
				\hline
		\end{tabular}}
		\caption{Table \ref{table:sm-vertices-2} continued. Here,  $\phi_i$, $\psi_i$ and $V$ denote the SM Higgs, the SM fermions and the SM gauge bosons respectively, whereas $\Phi_i$ and $\Psi_i$ denote the various heavy scalars and heavy fermions. $R_C$ and $R_L$ denote $SU(3)_C$ and $SU(2)_L$ quantum numbers respectively and $Y$ refers to the $U(1)_Y$ hypercharge of the field.}
		\label{table:sm-vertices-3}
	\end{table}

	\newpage
	\subsubsection*{$(ii)$ Unfolding effective operators in terms of UV diagrams}
	
	Having obtained an exhaustive list of heavy field representations based on the renormalizable vertices, we can focus our attention on ascertaining the correspondence between these heavy fields and the dimension-6 $CP$ as well as $B$-, $L$- conserving SMEFT operators.  We have accomplished this through the following steps.
	
	\begin{itemize}
		
	\item  We have unfolded the operators, on a case-by-case basis, into tree- and one-loop-level diagrams describing processes of the UV theory and revealing various vertices of Tables~\ref{table:sm-vertices-1}-\ref{table:sm-vertices-3} as the constituents of those diagrams for different cases.
	
	\item However, we must emphasize that instead of conducting the unfolding exhaustively by drawing the whole myriad of diagrams shown in Figs.~\ref{fig:phi6-unfolding}-\ref{fig:psi2phi2D-unfolding} relevant for each operator and then tracing the possible heavy fields within those diagrams, we have followed a well defined and economical approach. We have searched for the most minimal way in which it can be shown that a particular operator receives non-zero contributions from a single heavy field.
	
	\item Noting the fact that a particular heavy field can very well contribute to the same operator through multiple channels, we have included only one relevant diagram\footnote{Here, relevant refers to diagrams that yield non-zero contributions. Also, it must be noted that since we have not discussed the relative strengths of the coupling constants that may appear at particular vertices, all diagrams at a certain order have been treated on an equal footing.} exhibiting the connection between the two for every combination of heavy field and effective operator.  
	
	\item  For individual operators, we have included diagrams built of only one kind of fundamental vertex and containing a single type of heavy field propagator among the internal lines.   
	
	\item While we have found this prescription to work for operator classes which are composed of one type of fields, e.g., the $\phi^6$, $X^3$ classes and certain operators of the $\psi^4$ class, we have been forced to relax the strict minimality criteria once we start examining operator classes of non-trivial constitution. The shift from our notion of minimality is necessary for the following scenarios:
	
	\begin{itemize}
		\item In situations where the number of heavy fields is still one but due to the appearance of different SM fields as the external legs and the variety of ways in which these legs can be permuted, the number of fundamental vertices involved in the diagram is more than one. This has also been observed to lead to light-heavy mixing in the loop.
		
		\item Situations where in a particular diagram a single vertex demands multiple heavy field representations simultaneously. This mostly corresponds to those entries of Tables~\ref{table:sm-vertices-1}-\ref{table:sm-vertices-3} that lead to $\Phi_i, \Psi_i \,\in\, (R_{C_i}, R_{L_i},Y_i)$ with strict conditions imposed on these quantum numbers depending on the vertex.
		
		\item More dramatic departures where not only multiple heavy fields but sometimes multi-loop diagrams also become necessary. Such cases have been addressed in section~\ref{sec:non-minimal-cases}.
	\end{itemize} 
		
	\item In passing, it must be mentioned that we have excluded the cases involving heavy gauge bosons. This is because our focus is on minimal extensions of the SM and we are not including scenarios that involve the breaking of some higher symmetry.
		
	\item We must emphasize that the results of Tables~\ref{table:sm-vertices-1}-\ref{table:sm-vertices-3} are necessary in order to understand the results collected in Tables~\ref{table:smeft-phi6}-\ref{table:smeft-psi4-tensor-op}. 
	
	\begin{itemize}
		\item Each of these tables contains a list of heavy fields that can be back-traced from a particular operator.
		
		\item Next to each operator we have attached the diagram it originates from. The diagrams can denote pure heavy propagators at tree- or one-loop-level, as well as light-heavy mixing in the loops
		
		\item Corresponding to each heavy field appearing from a diagram we have also listed the vertex or the group of vertices constituting the diagram that involves the particular heavy field.
		
		\item For ease of readability we have established hyperlinks to the entries of Tables~\ref{table:sm-vertices-1}-\ref{table:sm-vertices-3}, so that one can suitably verify the appearance of the heavy field. 
	\end{itemize}
		
	\end{itemize}

	\subsection{\Large$\phi^6$}	
	
	Being consistent with our idea of minimality, we have tried to identify heavy fields that can lead to $\mathcal{Q}_H  = (H^{\dagger}H)^3$ through tree- and one-loop-level diagrams built of the smallest number of fundamental vertices listed in Table~\ref{table:sm-vertices-1}. We have collected our results in Table~\ref{table:smeft-phi6} and the following are salient points related to the results:
	
	\begin{itemize}
		\item Our notion of minimality restricts us to the following diagrams:
		\begin{enumerate}
			\item Fig.~\ref{subfig:phi6-tree1} - a tree-level diagram containing only quartic scalar vertices joined by a heavy propagator.
			
			\item Fig.~\ref{subfig:phi6-loop2} - one-loop-level diagram containing only trilinear scalar vertices. This presents two distinct sub-cases involving
			
			\begin{enumerate}
				\item light-heavy mixing in the loop between the SM scalar and the heavy scalar.
				
				\item mixing between two distinct heavy fields in the loop. This corresponds to the case where the UV theory has a degenerate spectrum. We have mostly avoided such cases in the remainder of this paper.  
			\end{enumerate}
		
			\item Fig.~\ref{subfig:phi6-loop3} - one-loop-level diagram containing only quartic scalar vertices. This contains both heavy-heavy mixing as well as a single heavy field in the loop.
		\end{enumerate}

		\item  A simple inspection reveals that the heavy field representations obtained from Fig.~\ref{subfig:phi6-loop2} do in fact appear at tree-level through Fig.~\ref{subfig:phi6-tree2} but this has not been included here as it contains both trilinear as well as quartic scalar vertices, therefore not being a minimal option.
		
		\item Specific heavy field representations are fixed using the contents of Table~\ref{table:sm-vertices-1} as well as by noting the permutations of the external legs of the effective operator in the unfolded diagrams. For instance, if we consider the tree-level diagram in Table~\ref{table:smeft-phi6}, then different heavy fields emerge based on whether we have $H^3$ or $H^2H^{\dagger}$ on one of the vertices. To highlight these differences we have separated the multiple heavy fields embedded in the same diagram through a dashed line in the Tables and we have also referenced the specific vertices involved next to them.
		
		\item It must be noted that in each diagram in Tables~\ref{table:smeft-phi6}-\ref{table:smeft-psi4-tensor-op}, the black and pink lines represent light and heavy fields respectively.
	\end{itemize}
	
	\begin{table}[!htb]
		\centering
		\renewcommand{\arraystretch}{2.4}
		{\scriptsize\begin{tabular}{||c|c|c||c|c|c||}
				\hline
				\multicolumn{6}{|c|}{$\mathcal{Q}_H:\,(H^{\dagger}H)^3$}\\
				\hline
				
				\textbf{Heavy fields}&
				\textbf{Diagram}&
				\textbf{Vertices}&
				\textbf{Heavy fields}&
				\textbf{Diagram}&
				\textbf{Vertices}\\
				\hline

				\multirow{2}{*}{$(1,3,1),\,\, (1,1,1)$}&
				\multirow{4}{*}{\includegraphics[width=3.2cm, height=2.5cm]{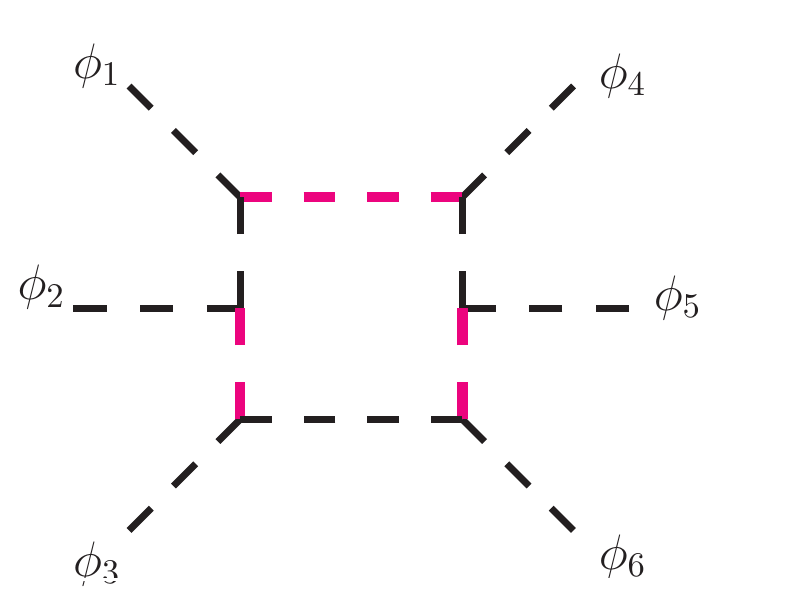}}&
				\multirow{2}{*}{\hyperlink{vertex-1-i}{V1-(i)}}&
				\multirow{2}{*}{$(1,4,\frac{3}{2}),\,\, (1,2,\frac{3}{2})$}&
				\multirow{4}{*}{\includegraphics[width=3.2cm, height=2.5cm]{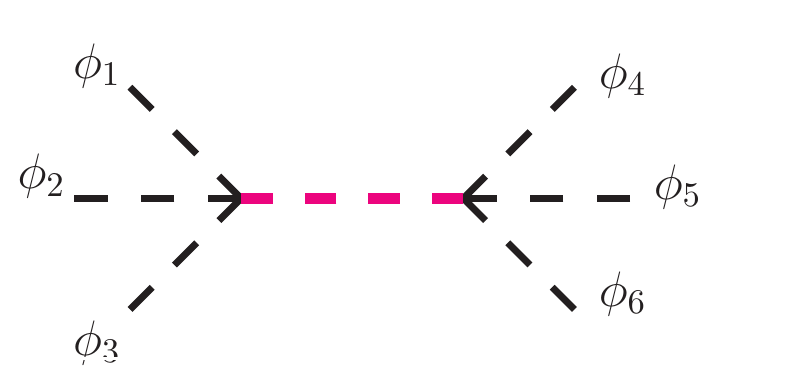}}&
				\multirow{2}{*}{\hyperlink{vertex-3-i}{V3-(i)}}\\
				
				&
				&
				&
				&
				&
				\\
				\cdashline{1-1}\cdashline{3-3}\cdashline{4-4}\cdashline{6-6}

				\multirow{2}{*}{$(1,3,0),\,\, (1,1,0)$}&
				&
				\multirow{2}{*}{\hyperlink{vertex-1-ii}{V1-(ii)}}&
				\multirow{2}{*}{$(1,4,\frac{1}{2}),\,\, (1,2,\frac{1}{2})$}&
				&
				\multirow{2}{*}{\hyperlink{vertex-3-ii}{V3-(ii)}}\\
				
				&
				&
				&
				&
				&
				\\
				\hline
				
				\multirow{2}{*}{$(R_{C_2},R_{L_2},Y_2)\,\oplus$}&
				\multirow{4}{*}{\includegraphics[width=3.2cm, height=2.3cm]{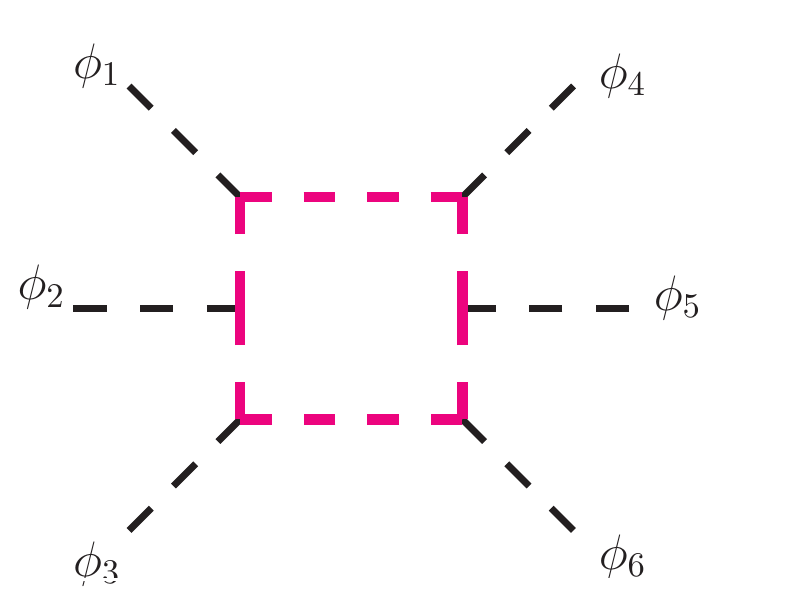}}&
				\multirow{4}{*}{\hyperlink{vertex-2}{V2}}&
				\multirow{2}{*}{$(\{1,R_C\},\{1,R_L\},\{0,Y\})$}&
				\multirow{4}{*}{\includegraphics[width=3.2cm, height=2.3cm]{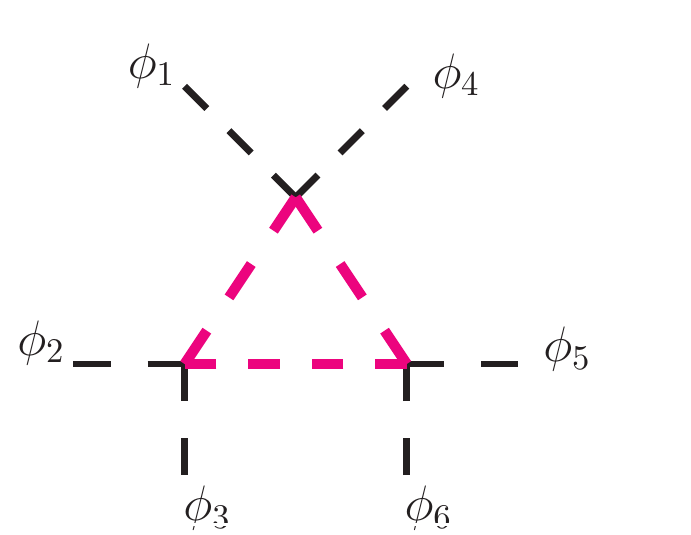}}&
				\multirow{2}{*}{\hyperlink{vertex-4-i}{V4-(i)}}\\
				
				\multirow{2}{*}{$ (R_{C_3},R_{L_3},Y_3)$}&
				&
				&
				&
				&
				\\
				\cdashline{4-4}\cdashline{6-6}
				&
				&
				&
				$(R_{C_3},R_{L_3},Y_3)\,\oplus$&
				&
				\multirow{2}{*}{\hyperlink{vertex-4-ii}{V4-(ii)}}\\
				
				&
				&
				&
				$(R_{C_4},R_{L_4},Y_4)$&
				&
				\\
				\hline
		\end{tabular}}
		\caption{Heavy field representations that are obtained by unfolding the $\phi^6$ operator into non-trivial tree- and(or) one-loop-level diagrams and the corresponding vertices.}
		\label{table:smeft-phi6}
	\end{table}
	\newpage
	\subsection{\Large$\phi^4\mathcal{D}^2$}	
	
	This class contains two operators $\mathcal{Q}_{H\square}$ and $\mathcal{Q}_{H\mathcal{D}}$. In both cases, the external states contain 4 scalars and 2 derivatives. The difference is that in the first case the derivatives are Lorentz contracted together to form the $\mathcal{D}^2$ operator, whereas in the second case the derivatives act on separate fields. Due to this, we get an extra diagram for $\mathcal{Q}_{H\mathcal{D}}$ with heavy-heavy mixing of fermions in the loop. Tree level diagrams with a heavy scalar propagator and one-loop-level diagrams with heavy scalar as well as heavy-heavy mixing are common to both operators. Specifically, looking at Fig.~\ref{fig:phi4D2-unfolding}, the following diagrams appear:
	\begin{enumerate}
		\item Tree-level diagrams: Figs. \ref{subfig:phi4D2-tree1} and \ref{subfig:phi4D2-tree2} - Both diagrams contain trilinear scalar vertices and a scalar propagator but in the first case the derivatives are contracted together whereas in the second case they act on separate fields.

		\item One-loop diagrams
		\begin{enumerate}
			\item Fig. \ref{subfig:phi4D2-loop7} - This diagram contains trilinear scalar vertices and a scalar loop with the derivatives contracted together.
			
			\item Fig. \ref{subfig:phi4D2-loop8} - This diagram contains quartic scalar vertices and a scalar loop with the derivatives contracted together.

			\item Fig. \ref{subfig:phi4D2-loop1} - This diagram contains trilinear scalar vertices and a scalar loop with the derivatives acting on separate fields.
			
			\item Fig. \ref{subfig:phi4D2-loop2} - This diagram contains quartic scalar vertices and a scalar loop with the derivatives acting on separate fields.
			
			\item Fig. \ref{subfig:phi4D2-loop3} - This diagram contains Yukawa-like vertices and a fermion loop with the derivatives acting on separate fields.
		\end{enumerate}
	\end{enumerate}	
	Figs. \ref{subfig:phi4D2-tree1}, \ref{subfig:phi4D2-loop7} and \ref{subfig:phi4D2-loop8} contribute to $\mathcal{Q}_{H\square}$, while Figs.~\ref{subfig:phi4D2-tree2}, \ref{subfig:phi4D2-loop1}, \ref{subfig:phi4D2-loop2} and \ref{subfig:phi4D2-loop3} contribute to $\mathcal{Q}_{H\mathcal{D}}$.
	Representations of the heavy fields are once again fixed using the contents of Tables~\ref{table:sm-vertices-1} and \ref{table:sm-vertices-3}. The detailed results have been collected in Table~\ref{table:smeft-phi4D2-1}.
	
	\begin{table}[!htb]
		\centering
		\renewcommand{\arraystretch}{1.8}
		{\scriptsize\begin{tabular}{||c|c|c||c|c|c||}
				
				\hline
				\multicolumn{3}{||c||}{$\mathcal{Q}_{H\square}:\,(H^{\dagger}H)\square (H^{\dagger}H)$}&
				\multicolumn{3}{c||}{$\mathcal{Q}_{H\mathcal{D}}:\,(H^{\dagger}\mathcal{D}_{\mu} H) (H^{\dagger} \mathcal{D}^{\mu} H)$}\\
				\hline
				
				\textbf{Heavy fields}&
				\textbf{Diagram}&
				\textbf{Vertices}&
				\textbf{Heavy fields}&
				\textbf{Diagram}&
				\textbf{Vertices}\\
				\hline

				\multirow{2}{*}{$(1,3,1),\,\,(1,1,1)$}&
				\multirow{4}{*}{\includegraphics[width=3cm,height=2cm]{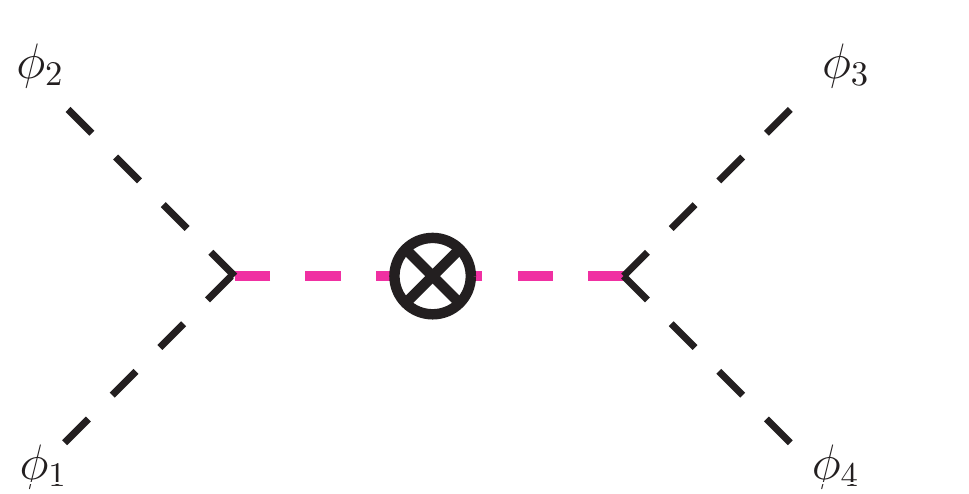}}&
				\multirow{2}{*}{\hyperlink{vertex-1-i}{V1-(i)}}&
				\multirow{2}{*}{$(1,3,1),\,\,(1,1,1)$}&
				\multirow{4}{*}{\includegraphics[width=3cm,height=2cm]{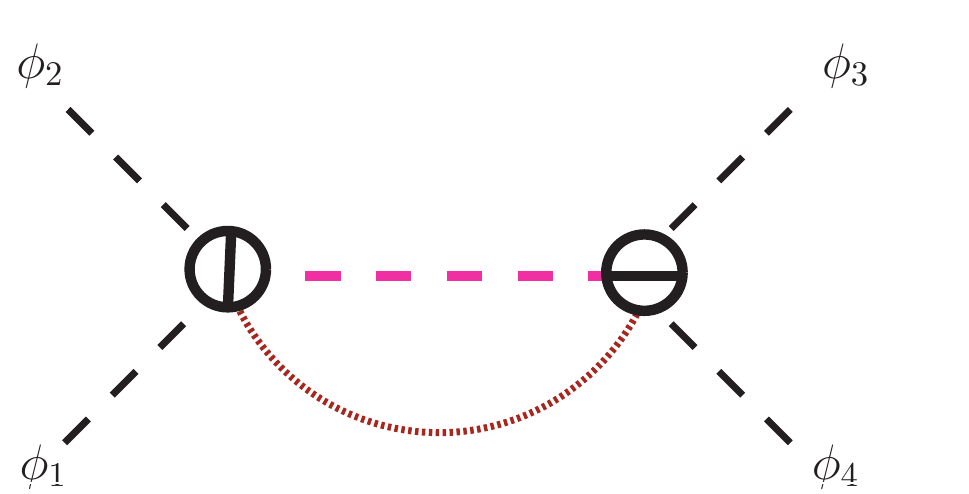}}&
				\multirow{2}{*}{\hyperlink{vertex-1-i}{V1-(i)}}\\

				&
				&
				&
				&
				&
				\\
				\cdashline{1-1}\cdashline{3-3}\cdashline{4-4}\cdashline{6-6}

				\multirow{2}{*}{$(1,3,0),\,\,(1,1,0)$}&
				&
				\multirow{2}{*}{\hyperlink{vertex-1-ii}{V1-(ii)}}&
				\multirow{2}{*}{$(1,3,0),\,\,(1,1,0)$}&
				&
				\multirow{2}{*}{\hyperlink{vertex-1-ii}{V1-(ii)}}\\

				&
				&
				&
				&
				&
				\\
				\hline

				\multirow{2}{*}{$(R_{C_2},R_{L_2},Y_2)\,\oplus $}&
				\multirow{2}{*}{\multirow{2}{*}{\includegraphics[width=3cm,height=2cm]{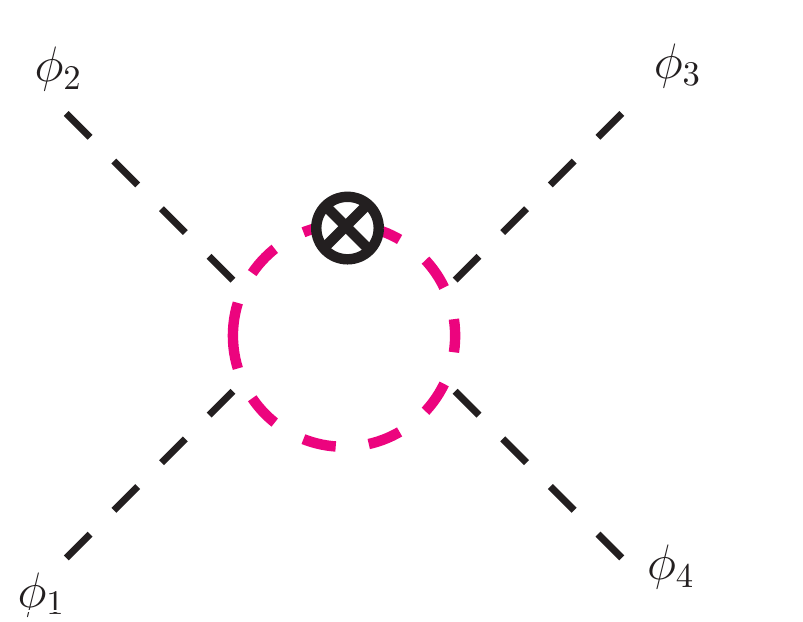}}}&
				\multirow{4}{*}{\hyperlink{vertex-2}{V2}}&
				\multirow{2}{*}{$(R_{C_2},R_{L_2},Y_2)\,\oplus $}&
				\multirow{4}{*}{\includegraphics[width=3cm,height=2cm]{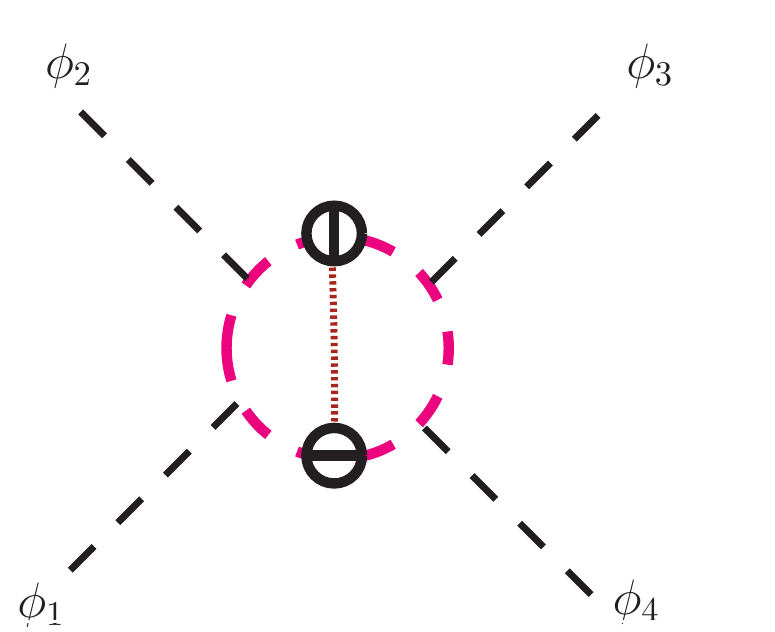}}&
				\multirow{4}{*}{\hyperlink{vertex-2}{V2}}\\			
				
				\multirow{2}{*}{$ (R_{C_3},R_{L_3},Y_3)$}&
				&
				&
				\multirow{2}{*}{$ (R_{C_3},R_{L_3},Y_3)$}&
				&
				\\
				
				&
				&
				&
				&
				&
				\\

				&
				&
				&
				&
				&
				\\
				\hline

				\multirow{2}{*}{$(\{1,R_C\},\{1,R_L\},\{0,Y\})$}&
				\multirow{4}{*}{\includegraphics[width=3cm,height=2cm]{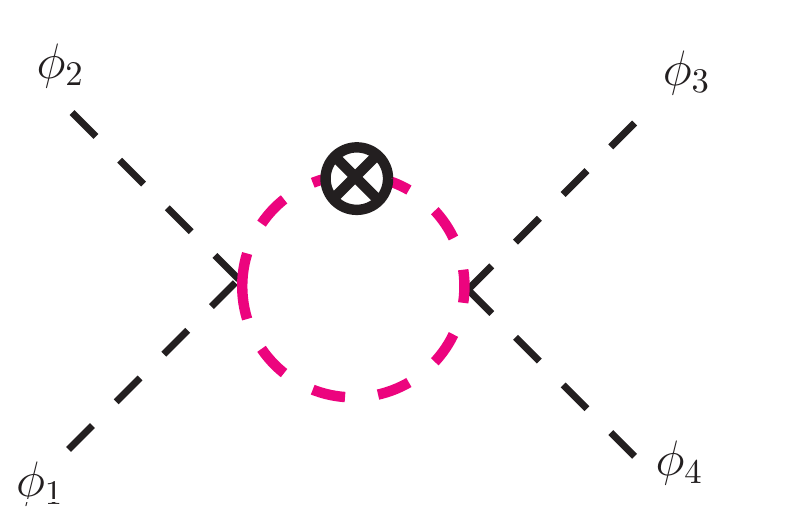}}&
				\multirow{2}{*}{\hyperlink{vertex-4-i}{V4-(i)}}&
				\multirow{2}{*}{$(\{1,R_C\},\{1,R_L\},\{0,Y\})$}&
				\multirow{4}{*}{\includegraphics[width=3cm,height=2cm]{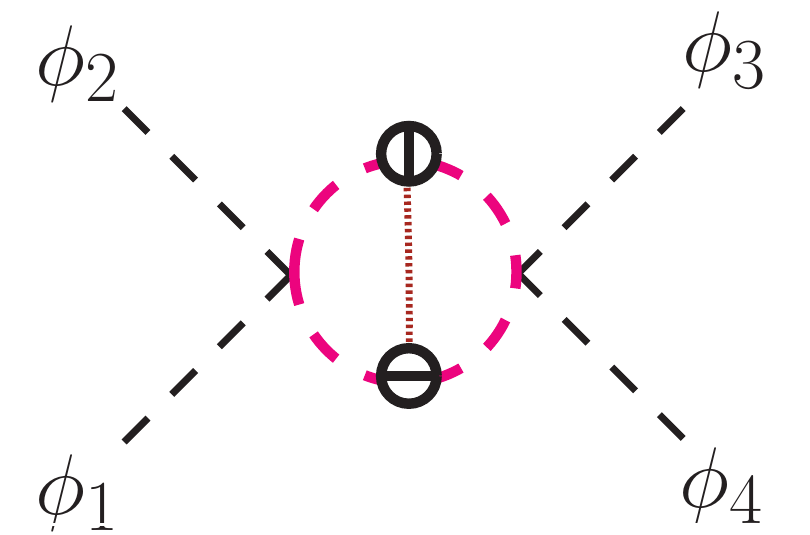}}&
				\multirow{2}{*}{\hyperlink{vertex-4-i}{V4-(i)}}\\

				&
				&
				&
				&
				&
				\\
				\cdashline{1-1}\cdashline{3-3}\cdashline{4-4}\cdashline{6-6}

				$(R_{C_3},R_{L_3},Y_3)\,\oplus$&
				&
				\multirow{2}{*}{\hyperlink{vertex-4-ii}{V4-(ii)}}&
				$(R_{C_3},R_{L_3},Y_3)\,\oplus$&
				&
				\multirow{2}{*}{\hyperlink{vertex-4-ii}{V4-(ii)}}\\

				$(R_{C_4},R_{L_4},Y_4)$&
				&
				&
				$ (R_{C_4},R_{L_4},Y_4)$&
				&
				\\
				\hline
				
				\multicolumn{3}{||c||}{}&
				&
				\multirow{4}{*}{\includegraphics[width=3cm,height=2cm]{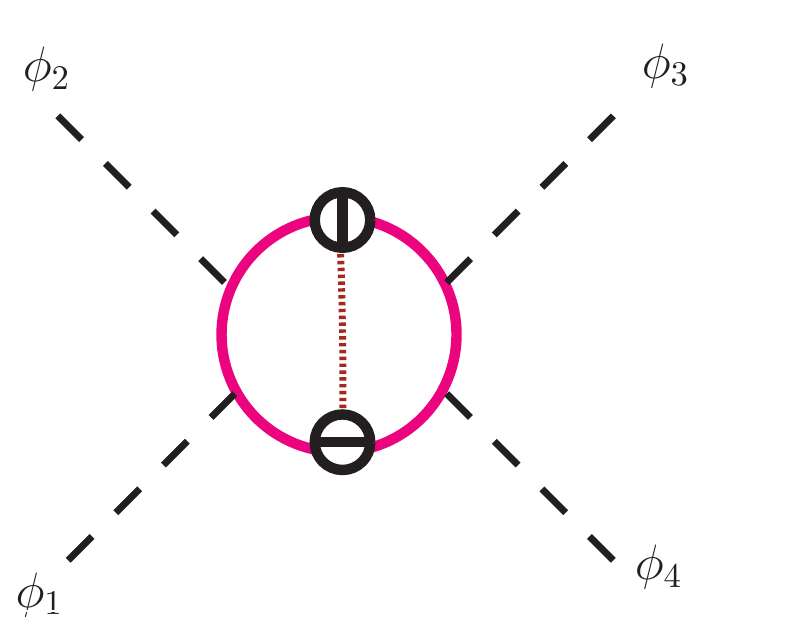}}&
				\multirow{4}{*}{\hyperlink{vertex-8}{V8}}\\
				
				\multicolumn{3}{||c||}{ \textbf{No allowed heavy fermion representation(s)}  }&
				\multirow{2}{*}{$(R_{C_1},R_{L_1},Y_1)\,\oplus $}&
				&
				\\
				
				\multicolumn{3}{||c||}{ \textbf{in this case, unlike $\mathcal{Q}_{H\mathcal{D}}$} }&
				\multirow{2}{*}{$ (R_{C_2},R_{L_2},Y_2)$}&
				&
				\\
				
				\multicolumn{3}{||c||}{}&
				&
				&
				\\
				
				\hline	
		\end{tabular}}
		\caption{Heavy field representations that are obtained by unfolding the $\phi^4 \mathcal{D}^2$ operators into non-trivial tree- and(or) one-loop-level diagrams and the corresponding vertices. }
		\label{table:smeft-phi4D2-1}
	\end{table}
	
%	\newpage
	
	\subsection{\Large$\psi^2\phi^3$}	
	This class contains 3 operators - $\mathcal{Q}_{dH}$, $\mathcal{Q}_{uH}$ and $\mathcal{Q}_{eH}$. Following the scheme of unfolding operators into tree and(or) loop-level diagrams constituted of invariant renormalizable, the emergence of heavy fields is minimally described by the following diagrams:
	
	\begin{enumerate}
		\item Fig.~\ref{subfig:psi2phi3-tree1} - It appears in all three operators and in all 3 cases furnishes a heavy scalar with identical quantum numbers as the SM Higgs, or in other words, the heavy scalar corresponds to the two-Higgs-doublet Model \cite{Branco:2011iw,Bhattacharyya:2015nca}. The vertices involved are the Yukawa vertex and the quartic scalar vertex.
		
		\item Fig.~\ref{subfig:psi2phi3-loop3} - This loop diagram appears in 3 different variations:
		\begin{enumerate}
			\item With a heavy scalar coupling with the SM fermions,
			
			\item With a single heavy fermion propagator in the loop coupling with the SM fermions as well as the SM scalar, and
			
			\item With 2 heavy fermion propagators in the loop, both having the same representations.  
		\end{enumerate}
	\end{enumerate}  
	
	Specific results have been described in Table~\ref{table:smeft-psi2phi3}.	
	\begin{table}[!htb]
		\centering
		\renewcommand{\arraystretch}{2.0}
		{\scriptsize\begin{tabular}{||c|c|c||c|c|c||}
				\hline
				\multicolumn{3}{||c||}{$\mathcal{Q}_{dH}:\,(H^{\dagger}\,H)\,(\overline{q}_p\,d_r\,H)$}&
				\multicolumn{3}{c||}{$\mathcal{Q}_{uH}:\,(H^{\dagger}\,H)\,(\overline{q}_p\,u_r\,\tilde{H})$}\\
				\hline

				\textbf{Heavy fields}&
				\textbf{Diagram}&
				\textbf{Vertices}&
				\textbf{Heavy fields}&
				\textbf{Diagram}&
				\textbf{Vertices}\\
				\hline

				\multirow{4}{*}{$(1,2,\frac{1}{2})$}&
				\multirow{4}{*}{\includegraphics[width=3cm, height=1.8cm]{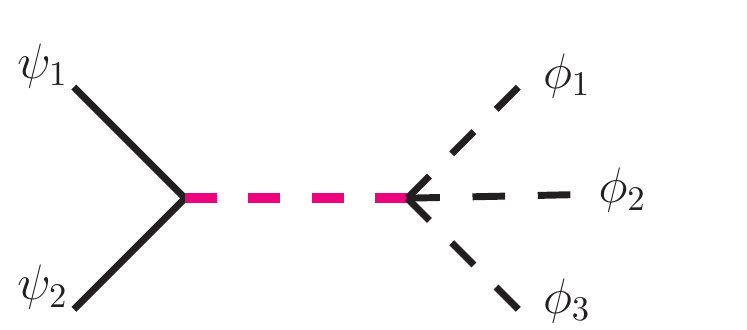}}&
				\multirow{4}{*}{\hyperlink{vertex-5-vii}{V5-(vii)},  \hyperlink{vertex-3-ii}{V3-(ii)}}&
				\multirow{4}{*}{$(1,2,\frac{1}{2})$}&
				\multirow{4}{*}{\includegraphics[width=3cm, height=1.8cm]{smeft-psi2phi3-1.pdf}}&
				\multirow{4}{*}{\hyperlink{vertex-5-viii}{V5-(viii)},  \hyperlink{vertex-3-ii}{V3-(ii)}}\\

				&
				&
				&
				&
				&
				\\

				&
				&
				&
				&
				&
				\\

				&
				&
				&
				&
				&
				\\
				
				\hline

				\multirow{3}{*}{$ (6,1,\frac{1}{3}),$}&
				\multirow{4}{*}{\includegraphics[width=2.8cm, height=1.8cm]{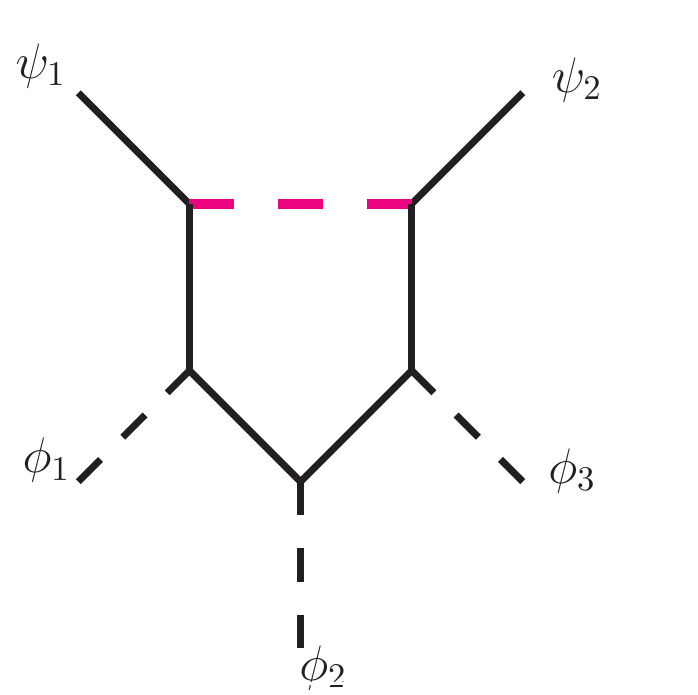}}&
				\multirow{4}{*}{\hyperlink{vertex-5-v}{V5-(v)},  \hyperlink{vertex-5-x}{V5-(x)}}&
				\multirow{2}{*}{$ (3,2,\frac{7}{6}) $}&
				\multirow{4}{*}{\includegraphics[width=2.8cm, height=1.8cm]{smeft-psi2phi3-4.pdf}}&
				\multirow{2}{*}{\hyperlink{vertex-5-xiii}{V5-(xiii)}, \hyperlink{vertex-5-xiv}{V5-(xiv)}}\\

				&
				&
				&
				&
				&
				\\
				\cdashline{4-4}\cdashline{6-6}

				$(3,1,-\frac{1}{3}) $&
				&
				&
				$ (3,1,-\frac{1}{3}),$&
				&
				\multirow{2}{*}{\hyperlink{vertex-5-v}{V5-(v)},  \hyperlink{vertex-5-x}{V5-(x)}}\\

				&
				&
				&
				$ (6,1,\frac{1}{3}) $&
				&
				\\
				\hline

				\multirow{2}{*}{$ (3,2,\frac{7}{6}) $}&
				\multirow{4}{*}{\includegraphics[width=2.8cm, height=1.8cm]{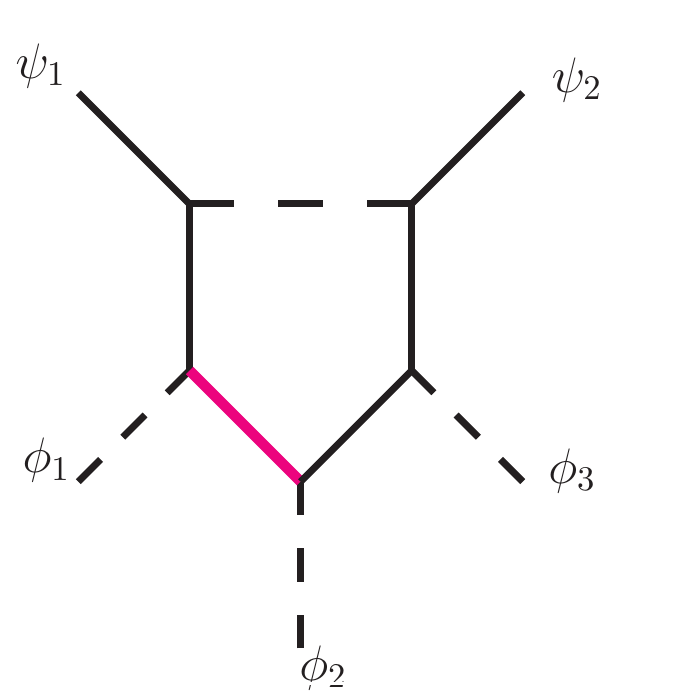}}&
				\multirow{2}{*}{\hyperlink{vertex-6-iv}{V6-(iv)}}&
				\multirow{4}{*}{$ (3,2,-\frac{5}{6}) $}&
				\multirow{4}{*}{\includegraphics[width=2.8cm, height=1.8cm]{smeft-psi2phi3-5.pdf}}&
				\multirow{4}{*}{\hyperlink{vertex-6-x}{V6-(x)}}\\

				&
				&
				&
				&
				&
				\\
				\cdashline{1-1}\cdashline{3-3}

				\multirow{2}{*}{$ (3,3,-\frac{1}{3}) $}&
				&
				\multirow{2}{*}{\hyperlink{vertex-6-viii}{V6-(viii)}}&
				&
				&
				\\

				&
				&
				&
				&
				&
				\\
				\hline

				\multirow{2}{*}{$ (3,2,-\frac{5}{6}) $}&
				\multirow{4}{*}{\includegraphics[width=2.8cm, height=1.8cm]{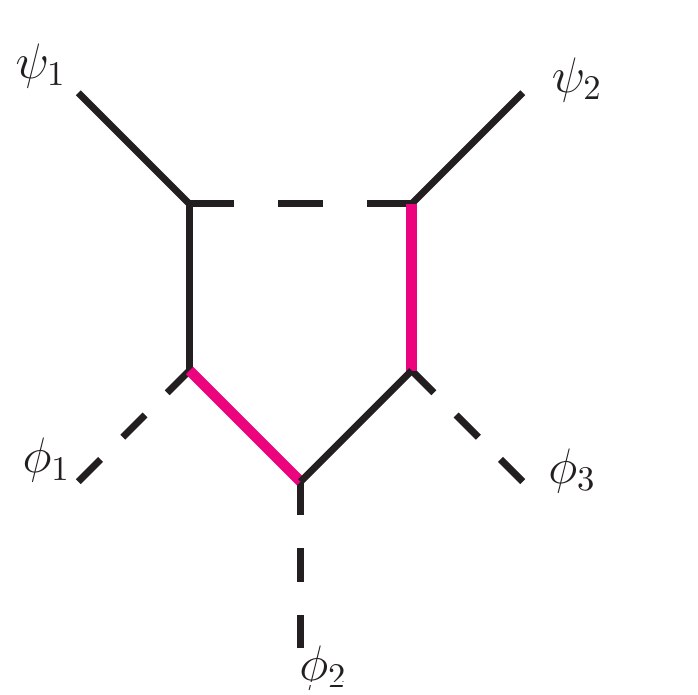}}&
				\multirow{2}{*}{\hyperlink{vertex-6-x}{V6-(x)}}&
				\multirow{3}{*}{$ (3,1,-\frac{1}{3}),$}&
				\multirow{4}{*}{\includegraphics[width=2.8cm, height=1.8cm]{smeft-psi2phi3-6.pdf}}&
				\multirow{4}{*}{\hyperlink{vertex-6-viii}{V6-(viii)}}\\

				&
				&
				&
				&
				&
				\\
				\cdashline{1-1}\cdashline{3-3}

				$ (3,1,-\frac{2}{3}),$&
				&
				\multirow{2}{*}{\hyperlink{vertex-6-iii}{V6-(iii)}}&
				$(3,3,-\frac{1}{3}) $&
				&
				\\

				$(3,3,-\frac{2}{3}) $&
				&
				&
				&
				&
				\\
				\hline
		\end{tabular}}
	{\scriptsize\begin{tabular}{|c|c|c|}
				
				\hline
				\multicolumn{3}{|c|}{$\mathcal{Q}_{eH}:\,(H^{\dagger}\,H)\,(\overline{l}_p\,e_r\,H)$}\\
				\hline

				\textbf{Heavy fields}&
				\textbf{Diagram}&
				\textbf{Vertices}\\
				\hline
				
				\multirow{4}{*}{$(1,2,\frac{1}{2})$}&
				\multirow{4}{*}{\includegraphics[width=3cm, height=2cm]{smeft-psi2phi3-1.pdf}}&
				\multirow{4}{*}{\hyperlink{vertex-5-vi}{V5-(vi)}, \hyperlink{vertex-3-ii}{V3-(ii)}} \\

				&
				&
				\\
				
				&
				&
				\\

				&
				&
				\\
				
				\hline

				\multirow{4}{*}{$ (3,2,\frac{7}{6}) $}&
				\multirow{4}{*}{\includegraphics[width=2.8cm, height=2cm]{smeft-psi2phi3-4.pdf}}&
				\multirow{4}{*}{\hyperlink{vertex-5-xiii}{V5-(xiii)},  \hyperlink{vertex-5-xiv}{V5-(xiv)}}\\

				&
				&
				\\

				&
				&
				\\

				&
				&
				\\
				\hline

				\multirow{4}{*}{$ (1,1,0),\, (1,3,0) $} &
				\multirow{4}{*}{\includegraphics[width=2.8cm, height=2cm]{smeft-psi2phi3-6.pdf}}&
				\multirow{4}{*}{\hyperlink{vertex-6-i}{V6-(i)}}\\

				&
				&
				\\

				&
				&
				\\

				&
				&
				\\
				\hline
		\end{tabular}}
		\caption{Heavy field representations that are obtained by unfolding the $\psi^2\phi^3$ operators into non-trivial tree- and(or) one-loop-level diagrams and the corresponding vertices.}
		\label{table:smeft-psi2phi3}
	\end{table}
	\clearpage
	\subsubsection*{\underline{Taking Equations of Motion into account}}
	
	The schematic effective operators for $\phi^4\mathcal{D}^2$ shown in Fig.~\ref{subfig:phi4D2} can be interpreted in different ways. We have already discussed the cases where the individual derivatives, as well as the $\mathcal{D}^2$ operator, act on the heavy field. If we focus on the case where $\mathcal{D}^2$ acts on an external SM scalar, i.e., the operator $(H^{\dagger}H)\,(H^{\dagger}\mathcal{D}^2H)$, then based on the equation of motion
	
\begin{figure}[!htb]
		\centering
		\includegraphics[height=3cm,width=7.7cm]{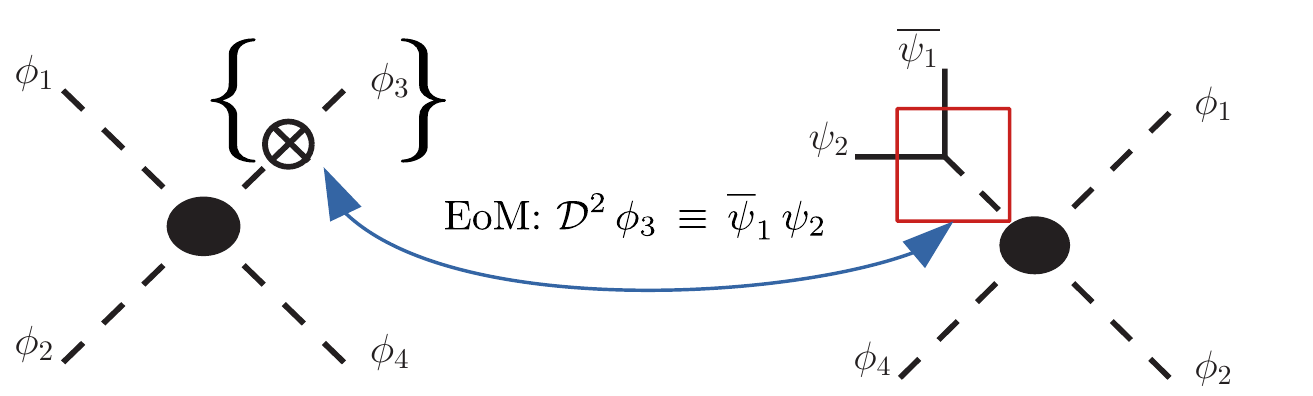}
		\caption{Implementing the equation of motion of the scalar field to relate $(H^{\dagger}H)\,(H^{\dagger}\mathcal{D}^2H)$ to operators of the $\psi^2\,\phi^3$ class.}
		\label{fig:eom-schematic}
	\end{figure}

	\vspace{-0.6cm}
	{\small\begin{eqnarray}
	\mathcal{D}^2 H \,\,\,\supset\,\,\, y^{pr}_e\, \overline{l}_p\, e_r \,\,+\,\, y^{pr}_d\, \overline{q}_p\, d_r \,\,+\,\, y^{pr}_u\, \overline{u}_p\, q_r, 
	\end{eqnarray}} 
	we can establish a connection between this operator and operators of the $\psi^2\phi^3$ class. Fig.~\ref{fig:eom-schematic} shows this connection schematically.
	
	\noindent Due to this EOM, the heavy fields emerging from the unfolding of $(H^{\dagger}H)\,(H^{\dagger}\mathcal{D}^2H)$ into tree and loop level diagrams can, in fact, be connected to $\mathcal{Q}_{eH}$, $\mathcal{Q}_{dH}$ and $\mathcal{Q}_{uH}$. There are a couple of noteworthy points:
	
	\begin{itemize}
		\item The heavy fields which have now been connected to the  $\psi^2\phi^3$ class are different from the ones obtained when individual operators of this class were unfolded into diagrams using renormalizable vertices.
	
		\item Heavy field representations discussed previously in Table~\ref{table:smeft-psi2phi3} involve certain cases which correspond to only one of the three operators $\mathcal{Q}_{eH}$, $\mathcal{Q}_{dH}$ or $\mathcal{Q}_{uH}$ but heavy fields obtained from the unfolding of $(H^{\dagger}H)\,(H^{\dagger}\mathcal{D}^2H)$ relate simultaneously to all three $\psi^2\phi^3$ operators.
	\end{itemize}
	
	\noindent In Table~\ref{table:smeft-eom}, we have listed the heavy fields appearing through the following diagrams:
	
	\begin{enumerate}
		\item Tree level diagram with a single heavy propagator between trilinear scalar vertices.
		
		\item One-loop diagram comprised only of trilinear scalar vertices and involving light-heavy mixing in the loop.
		
		\item One-loop diagram comprised only of Yukawa vertices and involving light-heavy mixing among fermions in the loop.
		
		\item One-loop diagram comprised of quartic scalar vertices and a heavy loop. 
	\end{enumerate}
	
	\begin{table}[!htb]
		\centering
		\renewcommand{\arraystretch}{2.2}
		{\scriptsize\begin{tabular}{|c|ccc|c|}
				
				\hline
				\multicolumn{5}{|c|}{$\mathcal{Q}_{eH}:\,(H^{\dagger}H)(\overline{l}_p\,e_{r}\,H),\,\,\, \mathcal{Q}_{uH}:\,(H^{\dagger}H)(\overline{q}_p\,u_{r}\,\tilde{H}),\,\,\, \mathcal{Q}_{dH}:\,(H^{\dagger}H)(\overline{q}_p\,d_{r}\,H)$}\\
				\hline
				
				\textbf{Heavy fields}&
				\multicolumn{3}{c|}{\textbf{Diagram }}&
				\textbf{Vertices}\\
				\hline

				\multirow{2}{*}{(1,3,1)}&
				\multirow{4}{*}{\includegraphics[width=3.5cm,height=2cm]{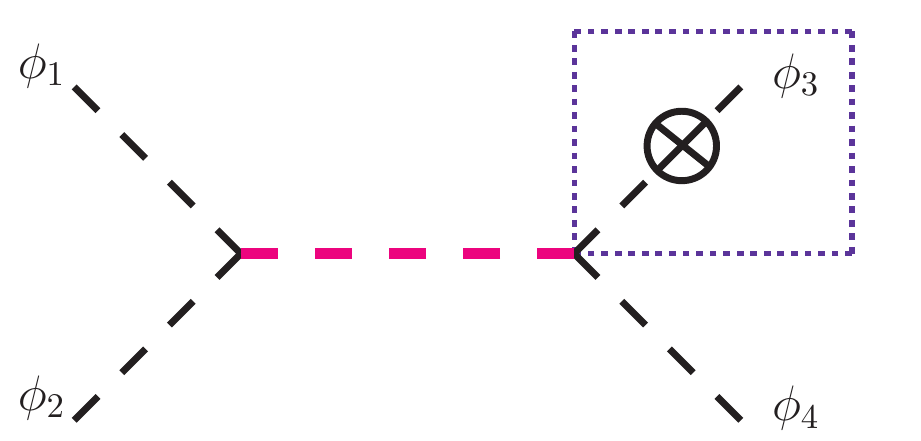}}&
				\multirow{4}{*}{$\equiv$}&
				\multirow{4}{*}{\includegraphics[width=3.5cm,height=2cm]{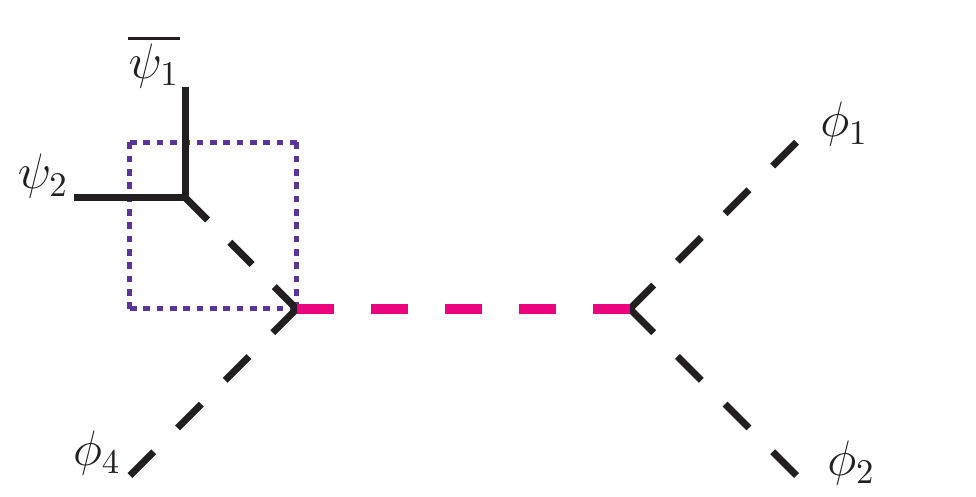}}&
				\multirow{2}{*}{\hyperlink{vertex-1-i}{V1-(i)}}\\

				&
				&
				&
				&
				\\
				\cdashline{1-1}\cdashline{5-5}
				
				\multirow{2}{*}{(1,3,0)}&
				&
				&
				&
				\multirow{2}{*}{\hyperlink{vertex-1-ii}{V1-(ii)}}\\
				
				&
				&
				&
				&
				\\
				\hline
				
				\multirow{2}{*}{(1,1,1)}&
				\multirow{4}{*}{\includegraphics[width=3.5cm,height=2cm]{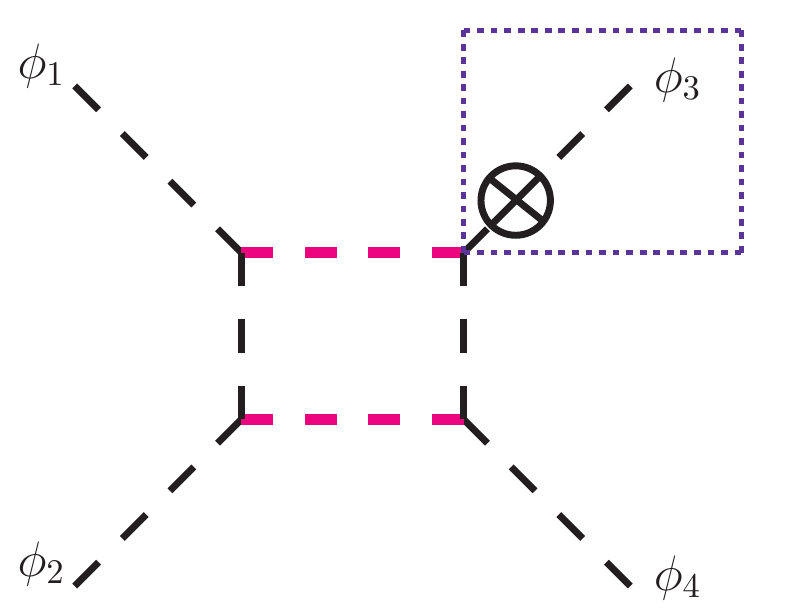}}&
				\multirow{4}{*}{$\equiv$}&
				\multirow{4}{*}{\includegraphics[width=3.5cm,height=2cm]{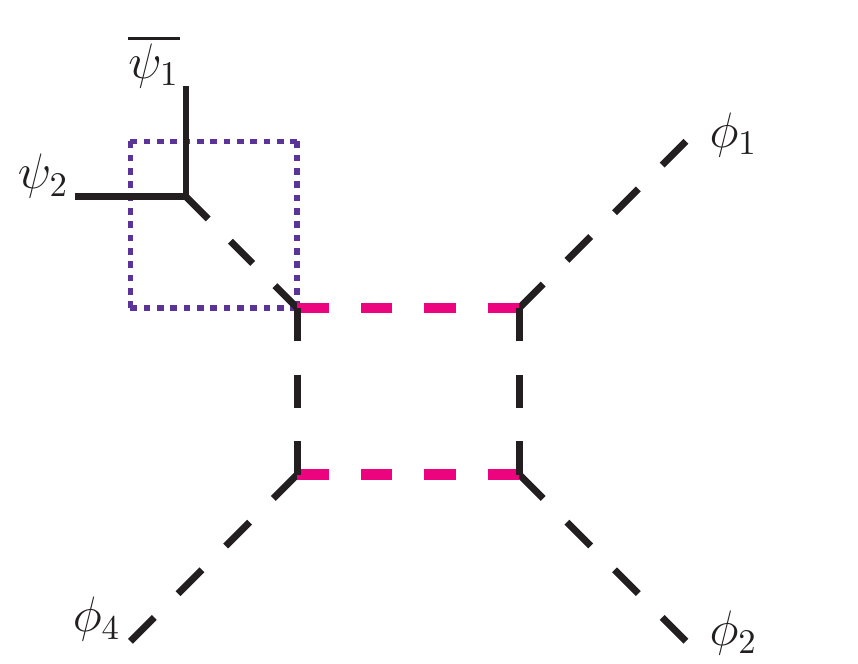}}&
				\multirow{2}{*}{\hyperlink{vertex-1-i}{V1-(i)}}\\

				&
				&
				&
				&
				\\
				\cdashline{1-1}\cdashline{5-5}
				
				\multirow{2}{*}{(1,1,0)}&
				&
				&
				&
				\multirow{2}{*}{\hyperlink{vertex-1-ii}{V1-(ii)}}\\
				
				&
				&
				&
				&
				\\
				\hline
				
				\multirow{2}{*}{(1,1,0), (1,3,0)}&
				\multirow{4}{*}{\includegraphics[width=3.5cm,height=2cm]{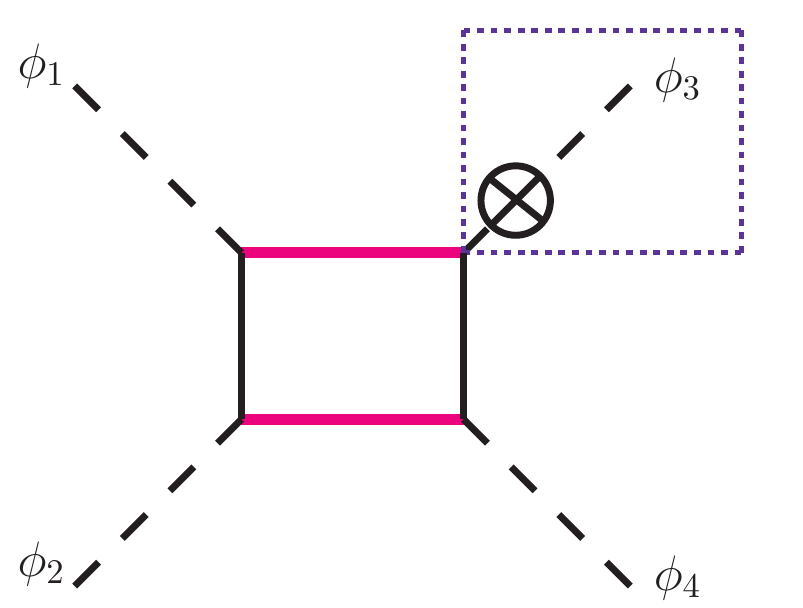}}&
				\multirow{4}{*}{$\equiv$}&
				\multirow{4}{*}{\includegraphics[width=3.5cm,height=2cm]{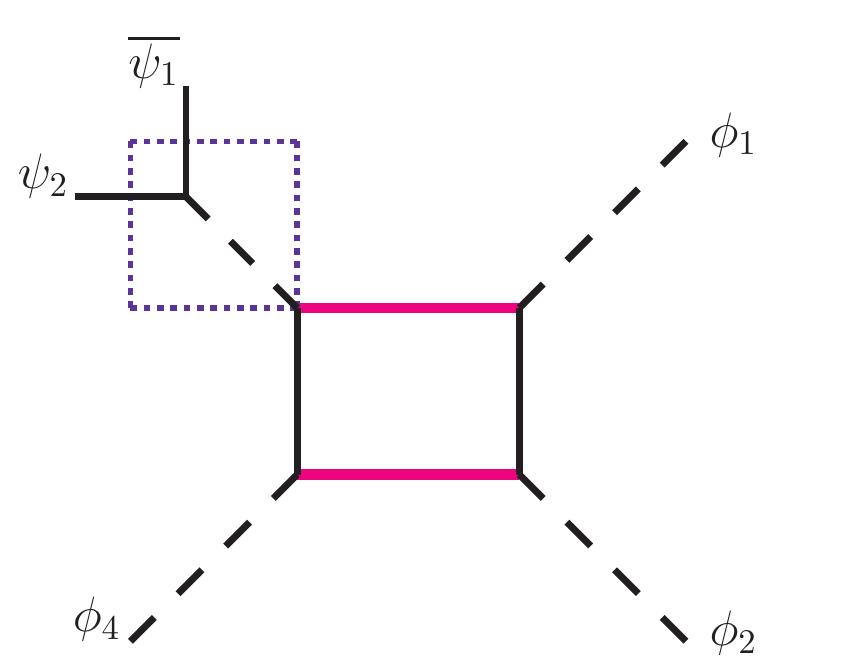}}&
				\multirow{2}{*}{\hyperlink{vertex-6-i}{V6-(i)}}\\

				&
				&
				&
				&
				\\
				\cdashline{1-1}\cdashline{5-5}
				
				\multirow{2}{*}{(1,1,1), (1,3,1)}&
				&
				&
				&
				\multirow{2}{*}{\hyperlink{vertex-6-vi}{V6-(vi)}}\\
				
				&
				&
				&
				&
				\\
				\hline

				\multirow{4}{*}{$ (\{1,R_C\},R_L,\{0,Y\})$ }&
				\multirow{4}{*}{\includegraphics[width=3.5cm,height=2cm]{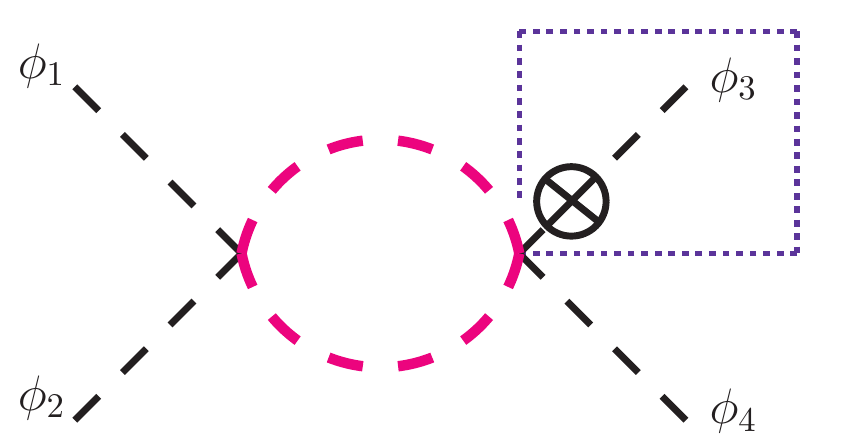}}&
				\multirow{4}{*}{$\equiv$}&
				\multirow{4}{*}{\includegraphics[width=3.5cm,height=2cm]{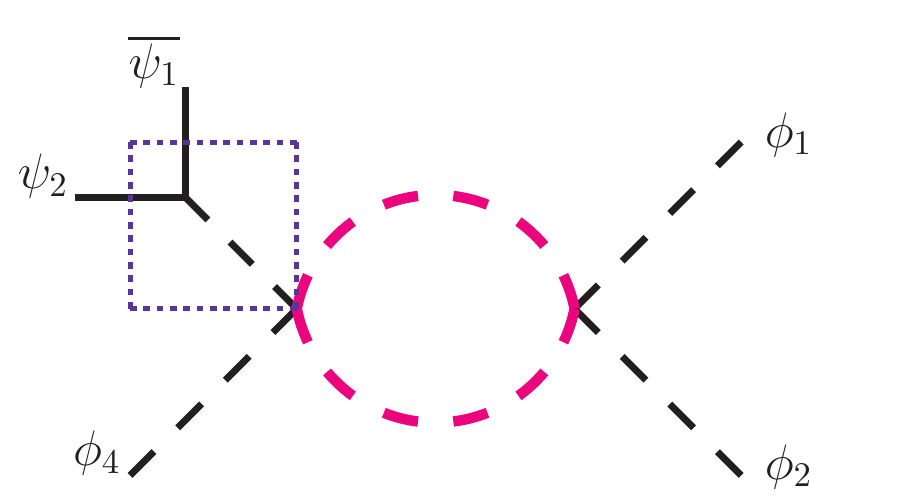}}&
				\multirow{4}{*}{\hyperlink{vertex-4-i}{V4-(i)}}\\

				&
				&
				&
				&
				\\
				
				&
				&
				&
				&
				\\
				
				&
				&
				&
				&
				\\
				\hline
		\end{tabular}}
		\caption{Heavy field representations that indirectly provide non-zero contributions to the $\psi^2\phi^3$ operators when the equation of motion of the SM Higgs is implemented, in the process of unfolding $\phi^4 \mathcal{D}^2$ class of operators. Also, listed are the vertices corresponding to the diagrams.}
		\label{table:smeft-eom}
	\end{table}

	\newpage
	\subsection{\Large$\psi^2\phi^2\mathcal{D}$}
	
	The presence of $\mathcal{D}_{\mu}$ and $\gamma^{\mu}$ within the effective operators of this class, see Fig.~\ref{subfig:psi2phi2D}, indicates that the diagrams leading to these operators must involve vector bosons. Since we are not taking into account scenarios involving heavy gauge bosons, the heavy scalar or fermion appears in a loop and couples to the SM gauge bosons which in turn couple with the external states, see Figs.~\ref{subfig:psi2phi2D-loop1} and \ref{subfig:psi2phi2D-loop2}. Representations of the heavy field are fixed as follows:
	\begin{itemize}
		\item Choice of the external fermion in the effective operator determines the intermediate vector boson. $SU(2)$ singlet fermions couple only with $B_{\mu}$, whereas $SU(2)$ doublets couple with both $B_{\mu}$ and $W^I_{\mu}$. Since the SM scalar is $SU(3)$ singlet, $G^A_{\mu}$ does not appear.
		
		\item Fixing the gauge boson ultimately sets the heavy field representations in accordance with the results of Table~\ref{table:sm-vertices-3}.
	\end{itemize} 	
	Results have been collected in Table~\ref{table:smeft-psi2phi2D}.	
	
	\begin{table}[!htb]
		\centering
		\renewcommand{\arraystretch}{2.4}
		{\scriptsize\begin{tabular}{||>{\centering}p{0.3cm}|c|c|c||>{\centering}p{0.3cm}|c|c|c||}
				
				\hline
				\multicolumn{8}{|c|}{$\mathcal{Q}_{H\psi}/\mathcal{Q}_{H\psi}^{(1)}:\,(H^{\dagger}\,i\, \overleftrightarrow{\mathcal{D}_{\mu}}\,H) (\overline{\psi}\,\gamma^{\mu}\,\psi)$}\\
				\hline
				
				{\large$\mathbf{\psi}$}&
				\textbf{Heavy fields}&
				\textbf{Diagram}&
				\textbf{Vertices}&
				{\large$\mathbf{\psi}$}&
				\textbf{Heavy fields}&
				\textbf{Diagram}&
				\textbf{Vertices}\\
				
				\hline
				$q$&
				\multirow{5}{*}{$(\{1,R_C\},\{1,R_L\},Y)$}&
				\multirow{5}{*}{\includegraphics[width=3cm,height=2cm]{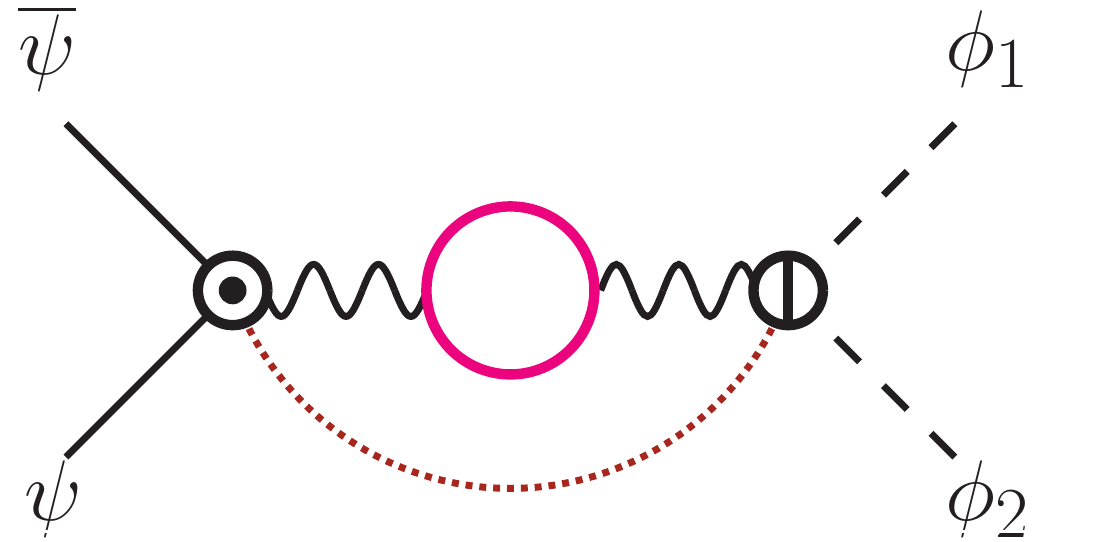}}&
				\multirow{5}{*}{\hyperlink{vertex-9-i}{V9-(i)}}&
				$q$&
				\multirow{5}{*}{$(\{1,R_C\},\{1,R_L\},Y)$}&
				\multirow{5}{*}{\includegraphics[width=3cm,height=2cm]{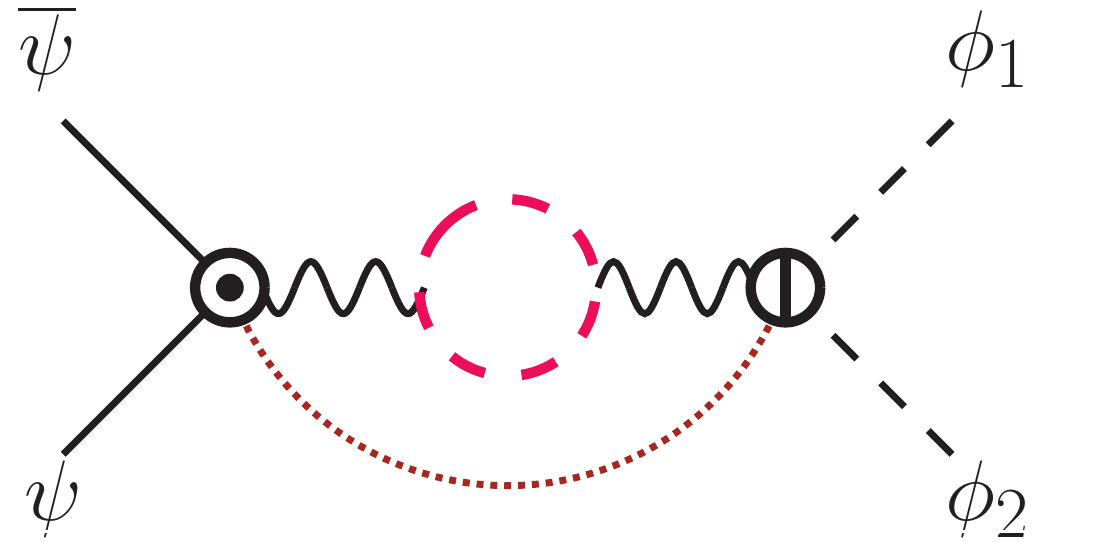}}&
				\multirow{5}{*}{\hyperlink{vertex-10-i}{V10-(i)}}\\
				
				\cdashline{1-1}	\cdashline{5-5}
				$u$&&&&$u$&&&\\
				\cdashline{1-1} \cdashline{5-5}
				$d$&&&&$d$&&&\\
				\cdashline{1-1} \cdashline{5-5}
				$l$&&&&$l$&&&\\
				\cdashline{1-1} \cdashline{5-5}
				$e$&&&&$e$&&&\\				
				
				\hline
				\hline
				\multicolumn{8}{|c|}{$\mathcal{Q}_{H\psi}^{(3)}:\,(H^{\dagger} i\, \overleftrightarrow{\mathcal{D}^I_{\mu}} H) (\overline{\psi}\,\gamma^{\mu}\,\tau^I\,\psi)$}\\
				\hline
				
				{\large$\mathbf{\psi}$}&
				\textbf{Heavy fields}&
				\textbf{Diagram}&
				\textbf{Vertices}&
				{\large$\mathbf{\psi}$}&
				\textbf{Heavy fields}&
				\textbf{Diagram}&
				\textbf{Vertices}\\
				\hline

				\multirow{2}{*}{$q$}&
				\multirow{4}{*}{$(\{1,R_C\},R_L,\{0,Y\})$}&
				\multirow{4}{*}{\includegraphics[width=3cm,height=2cm]{psi2phi2dloop1-table-fermion.pdf}}&
				\multirow{4}{*}{\hyperlink{vertex-9-i}{V9-(ii)}}&
				\multirow{2}{*}{$q$}&
				\multirow{4}{*}{$(\{1,R_C\},R_L,\{0,Y\})$}&
				\multirow{4}{*}{\includegraphics[width=3cm,height=2cm]{psi2phi2dloop1-table.pdf}}&
				\multirow{4}{*}{\hyperlink{vertex-10-ii}{V10-(ii)}}\\
				
				&&&&&&&\\
				\cdashline{1-1}\cdashline{5-5}
				\multirow{2}{*}{$l$}&&&&\multirow{2}{*}{$l$}&&&\\
				
				&&&&&&&\\
				\hline	
		\end{tabular}}
		\caption{Heavy field representations that are obtained by unfolding the $\psi^2\phi^2\mathcal{D}$ operators into non-trivial one-loop-level diagrams and the corresponding vertices.}
		\label{table:smeft-psi2phi2D}
	\end{table}	
	
\newpage

\subsection{\Large$\phi^2X^2$}

Given the external legs of the effective operator, see Fig.~\ref{subfig:phi2x2}, and keeping in mind that there is no mixing in the kinetic sector of the renormalizable Lagrangian, heavy fields emerge only through one-loop-level diagrams, more specifically:

\begin{itemize}
	\item Fig.~\ref{subfig:phi2x2-loop1}, which involves trilinear scalar vertices and vertices that couple scalars with gauge bosons. Heavy fields appear through light-heavy mixing within the loop, and the vector bosons coupling with the light field. 
	
	\item Fig.~\ref{subfig:phi2x2-loop4} involves Yukawa like vertices connecting the SM scalar with two fermions and vertices emerging from fermion kinetic terms. Heavy fields once again emerge through light-heavy mixing but we encounter two distinct cases:
	\begin{enumerate}
		\item Vector bosons coupling with the light (SM) fermion.
		
		\item Vector bosons coupling with the heavy fermion.
	\end{enumerate}
	
	\item Fig.~\ref{subfig:phi2x2-loop7} contains quartic scalar interaction as well as coupling between scalars and gauge bosons. The heavy field appears in the loop (with no mixing).
\end{itemize} 
In each case, the choice of the vector bosons, as well as the light SM fields in the loop, dictates the allowed representations of the heavy fields. We have collected  our results in Tables~\ref{table:smeft-phi2X2-1} and \ref{table:smeft-phi2X2-2}.

\begin{table}[!htb]
	\centering
	\renewcommand{\arraystretch}{2.2}
	{\scriptsize\begin{tabular}{||c|c|c||c|c|c||}
			
			\hline
			\multicolumn{6}{||c||}{$\mathcal{Q}_{HG}:\,(H^{\dagger}\,H)\,(G^{A}_{\mu\nu}\,G^{A\mu\nu})$}\\
			\hline
			
			\textbf{Heavy fields}&
			\textbf{Diagram}&
			\textbf{Vertices}&
			\textbf{Heavy fields}&
			\textbf{Diagram}&
			\textbf{Vertices}\\
			\hline
			
			$(3,1,\frac{2}{3}),\,(3,3,\frac{2}{3})$&
			\multirow{5}{*}{\includegraphics[width=3.2cm, height=2.4cm]{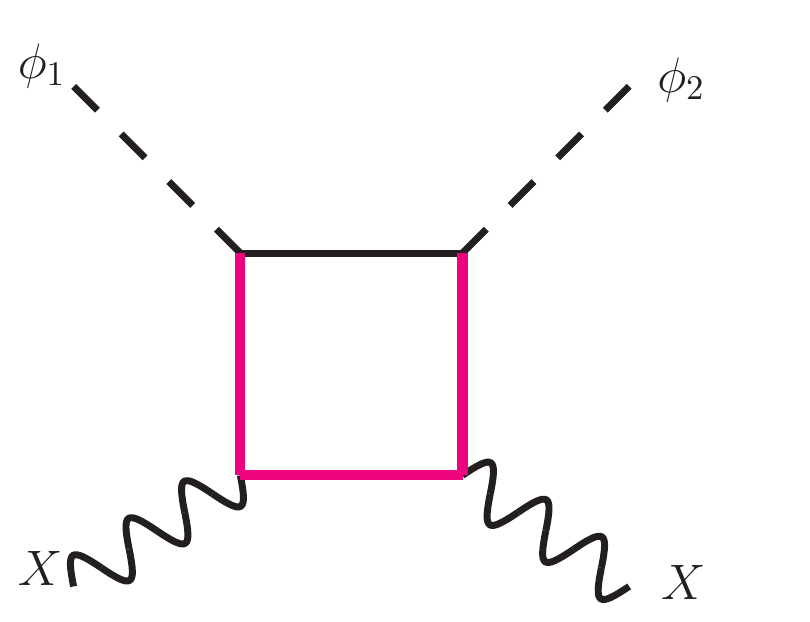}}&
			\hyperlink{vertex-6-iii}{V6-(iii)}&
			\multirow{5}{*}{$(R_C,\{1,R_L\},\{0,Y\})$}&
			\multirow{5}{*}{\includegraphics[width=3.2cm, height=2.4cm]{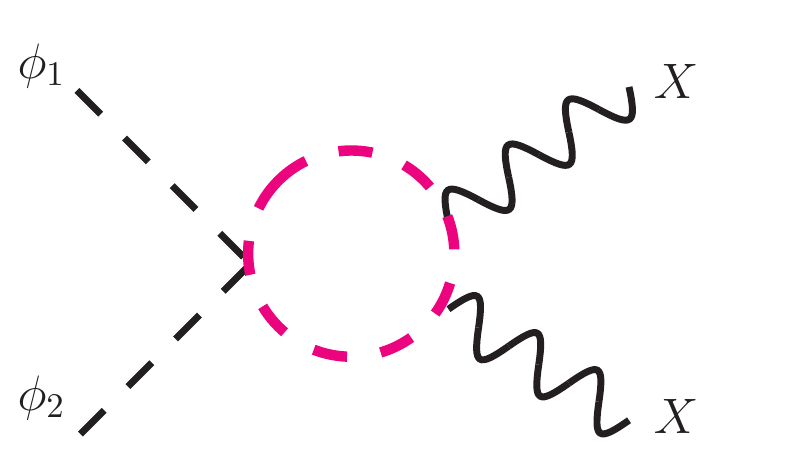}}&
			\multirow{4}{*}{\hyperlink{vertex-1-ii}{V1-(ii)},}\\
			
			\cdashline{1-1}\cdashline{3-3}

			$(3,1,-\frac{1}{3}),\,(3,3,-\frac{1}{3})$&
			&
			\hyperlink{vertex-6-viii}{V6-(viii)}&
			&
			&
			\multirow{4}{*}{\hyperlink{vertex-10-iii}{V10-(iii)}}\\
			\cdashline{1-1}\cdashline{3-3}
			
			$(3,2,\frac{7}{6})$&
			&
			\hyperlink{vertex-6-iv}{V6-(iv)}&
			&
			&
			\\
			
			\cdashline{1-1}\cdashline{3-3}
			
			$(3,2,\frac{1}{6})$&
			&
			\hyperlink{vertex-6-v}{V6-(v)}, \hyperlink{vertex-6-ix}{V6-(ix)}&
			&
			&
			\\
			
			\cdashline{1-1}\cdashline{3-3}

			$(3,2,-\frac{5}{6})$&
			&
			\hyperlink{vertex-6-x}{V6-(x)}&
			&
			&
			\\

			\hline
			\hline
			\multicolumn{6}{||c||}{$\mathcal{Q}_{HW}:\,(H^{\dagger}\,H)\,(W^{I}_{\mu\nu}\,W^{I\mu\nu})$}\\
			\hline
			
			\textbf{Heavy fields}&
			\textbf{Diagram}&
			\textbf{Vertices}&
			\textbf{Heavy fields}&
			\textbf{Diagram}&
			\textbf{Vertices}\\
			\hline

			\multirow{2}{*}{$(1,3,1),\,(1,1,1)$}&
			\multirow{4}{*}{\includegraphics[width=3.2cm, height=2.2cm]{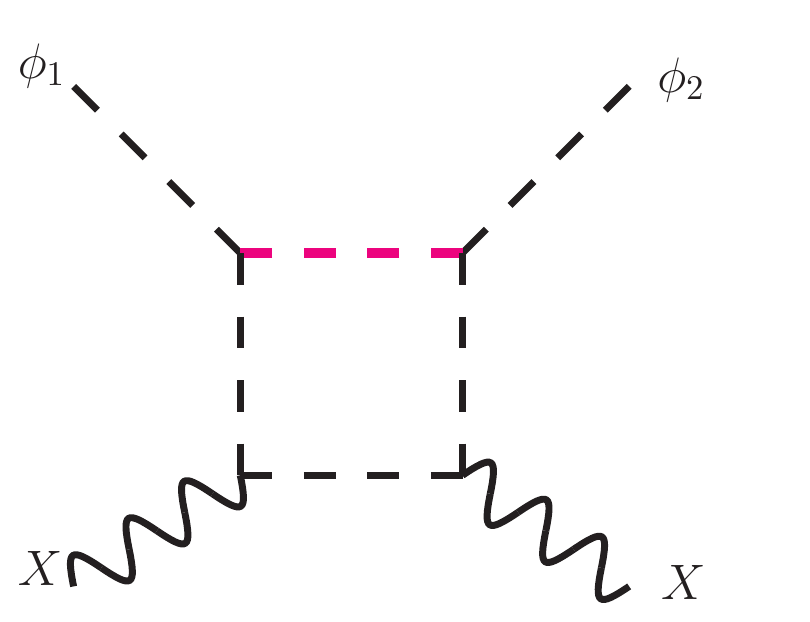}}&
			\multirow{2}{*}{\hyperlink{vertex-1-i}{V1-(i)}}&
			$(3,1,\frac{2}{3}),\,(3,3,\frac{2}{3})$&
			\multirow{4}{*}{\includegraphics[width=3.2cm, height=2.2cm]{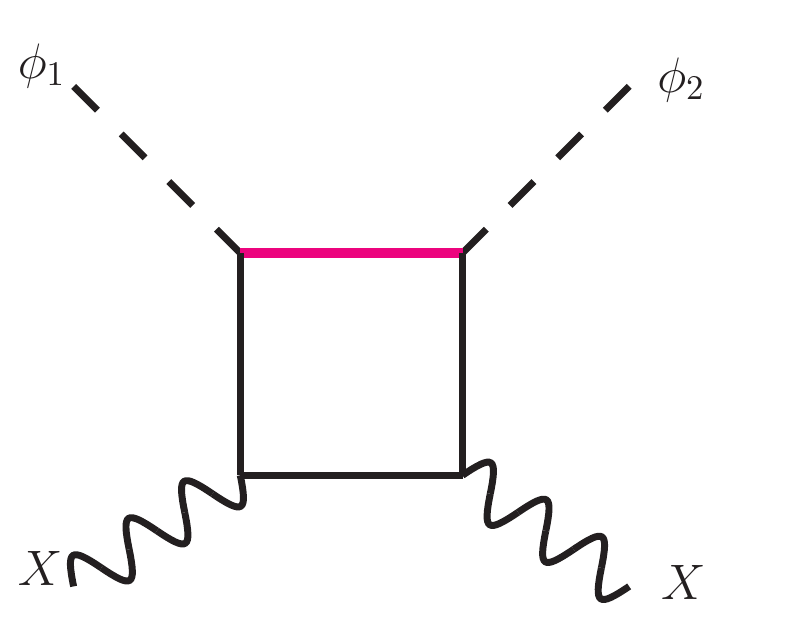}}&
			\hyperlink{vertex-6-iii}{V6-(iii)}\\
			\cdashline{4-4}\cdashline{6-6}
			
			&
			&
			&
			$(3,1,-\frac{1}{3}),\,(3,3,-\frac{1}{3})$&
			&
			\hyperlink{vertex-6-viii}{V6-(viii)}\\
			\cdashline{1-1}\cdashline{3-3}\cdashline{4-4}\cdashline{6-6}

			\multirow{2}{*}{$(1,3,0),\,(1,1,0)$}&
			&
			\multirow{2}{*}{\hyperlink{vertex-1-ii}{V1-(ii)}}&
			$(1,3,0), (1,1,0)$&
			&
			\hyperlink{vertex-6-i}{V6-(i)}\\
			\cdashline{4-4}\cdashline{6-6}

			&
			&
			&
			$(1,3,1), (1,1,1)$&
			&
			\hyperlink{vertex-6-vi}{V6-(vi)}\\
			\hline

			$(3,2,\frac{7}{6})$&
			\multirow{5}{*}{\includegraphics[width=3.2cm, height=2.2cm]{smeft-phi2x2-5.pdf}}&
			\hyperlink{vertex-6-iv}{V6-(iv)}&
			\multirow{5}{*}{$(\{1,R_C\},R_L,\{0,Y\})$}&
			\multirow{5}{*}{\includegraphics[width=3.2cm, height=2.2cm]{smeft-phi2x2-7.pdf}}&
			\multirow{4}{*}{\hyperlink{vertex-1-ii}{V1-(ii)},}\\
			\cdashline{1-1}\cdashline{3-3}
			
			$(3,2,\frac{1}{6})$&
			&
			\hyperlink{vertex-6-v}{V6-(v)}, \hyperlink{vertex-6-ix}{V6-(ix)}&
			&
			&
			\multirow{4}{*}{\hyperlink{vertex-10-ii}{V10-(ii)}}\\
			
			\cdashline{1-1}\cdashline{3-3}
			
			$(3,2,-\frac{5}{6})$&
			&
			\hyperlink{vertex-6-x}{V6-(x)}&
			&
			&
			\\
			
			\cdashline{1-1}\cdashline{3-3}
			
			$(1,2,\frac{1}{2})$&
			&
			\hyperlink{vertex-6-ii}{V6-(ii)}&
			&
			&
			\\
			
			\cdashline{1-1}\cdashline{3-3}
			
			$(1,2,\frac{3}{2})$&
			&
			\hyperlink{vertex-6-vii}{V6-(vii)}&
			&
			&
			\\

			\hline
			\hline
			\multicolumn{6}{||c||}{$\mathcal{Q}_{HB}:\,(H^{\dagger}\,H)\,(B_{\mu\nu}\,B^{\mu\nu})$}\\
			\hline
			
			\textbf{Heavy fields}&
			\textbf{Diagram}&
			\textbf{Vertices}&
			\textbf{Heavy fields}&
			\textbf{Diagram}&
			\textbf{Vertices}\\
			\hline

			$(3,1,\frac{2}{3}),\,(3,3,\frac{2}{3})$&
			\multirow{8}{*}{\includegraphics[width=3.6cm, height=2.8cm]{smeft-phi2x2-6.pdf}}&
			\hyperlink{vertex-6-iii}{V6-(iii)}&
			\multirow{2}{*}{$(1,3,1),\,(1,1,1)$}&
			\multirow{4}{*}{\includegraphics[width=3.2cm, height=2.2cm]{smeft-phi2x2-3.pdf}}&
			\multirow{2}{*}{\hyperlink{vertex-1-i}{V1-(i)}}\\
			\cdashline{1-1}\cdashline{3-3}

			$(3,1,-\frac{1}{3}),\,(3,3,-\frac{1}{3})$&
			&
			\hyperlink{vertex-6-viii}{V6-(viii)}&
			&
			&
			\\
			\cdashline{1-1}\cdashline{3-3}\cdashline{4-4}\cdashline{6-6}
			
			$(3,2,\frac{7}{6})$&
			&
			\hyperlink{vertex-6-iv}{V6-(iv)}&
			\multirow{2}{*}{$(1,3,0),\,(1,1,0)$}&
			&
			\multirow{2}{*}{\hyperlink{vertex-1-ii}{V1-(ii)}}\\
			\cdashline{1-1}\cdashline{3-3}

			$(3,2,\frac{1}{6})$&
			&
			\hyperlink{vertex-6-v}{V6-(v)}, \hyperlink{vertex-6-ix}{V6-(ix)}&
			&
			&
			\\
			\cdashline{1-1}\cdashline{3-3}\cline{4-6}

			$(3,2,-\frac{5}{6})$&
			&
			\hyperlink{vertex-6-x}{V6-(x)}&
			\multirow{5}{*}{$(\{1,R_C\},\{1,R_L\},Y)$}&
			\multirow{5}{*}{\includegraphics[width=3.2cm, height=2.2cm]{smeft-phi2x2-7.pdf}}&
			\multirow{4}{*}{\hyperlink{vertex-1-ii}{V1-(ii)},}\\
			
			\cdashline{1-1}\cdashline{3-3}
			
			$(1,2,\frac{1}{2})$&
			&
			\hyperlink{vertex-6-ii}{V6-(ii)}&
			&
			&
			\multirow{4}{*}{\hyperlink{vertex-10-i}{V10-(i)}}\\
			\cdashline{1-1}\cdashline{3-3}
			
			$(1,2,\frac{3}{2})$&
			&
			\hyperlink{vertex-6-vii}{V6-(vii)}&
			&
			&
			\\
			\cdashline{1-1}\cdashline{3-3}
			
			$(1,3,0), (1,1,0)$&
			&
			\hyperlink{vertex-6-i}{V6-(i)}&
			&
			&
			\\
			\cdashline{1-1}\cdashline{3-3}

			$(1,3,1), (1,1,1)$&
			&
			\hyperlink{vertex-6-vi}{V6-(vi)}&
			&
			&
			\\
			\hline
			
	\end{tabular}}
\caption{Heavy field representations that are obtained by unfolding the $\phi^2X^2$ operators into non-trivial one-loop-level diagrams and the corresponding vertices.}
\label{table:smeft-phi2X2-1}
\end{table}
\clearpage

\begin{table}[!htb]
	\centering
	\renewcommand{\arraystretch}{2.2}
	{\scriptsize\begin{tabular}{||c|c|c||c|c|c||}
			\hline
			\multicolumn{6}{||c||}{$\mathcal{Q}_{HWB}:\,(H^{\dagger}\,\tau^I\,H)\,(W^{I}_{\mu\nu}\,B^{\mu\nu})$}\\
			\hline
			
			\textbf{Heavy fields}&
			\textbf{Diagram}&
			\textbf{Vertices}&
			\textbf{Heavy fields}&
			\textbf{Diagram}&
			\textbf{Vertices}\\
			\hline
			
			\multirow{2}{*}{$(1,3,1),\,(1,1,1)$}&
			\multirow{4}{*}{\includegraphics[width=3.2cm, height=2.2cm]{smeft-phi2x2-3.pdf}}&
			\multirow{2}{*}{\hyperlink{vertex-1-i}{V1-(i)}}&
			$(3,1,\frac{2}{3}),\,(3,3,\frac{2}{3})$&
			\multirow{4}{*}{\includegraphics[width=3.2cm, height=2.2cm]{smeft-phi2x2-6.pdf}}&
			\hyperlink{vertex-6-iii}{V6-(iii)}\\
			\cdashline{4-4}\cdashline{6-6}

			&
			&
			&
			$(3,1,-\frac{1}{3}),\,(3,3,-\frac{1}{3})$&
			&
			\hyperlink{vertex-6-viii}{V6-(viii)}\\
			\cdashline{1-1}\cdashline{3-3}\cdashline{4-4}\cdashline{6-6}
			
			\multirow{2}{*}{$(1,3,0),\,(1,1,0)$}&
			&
			\multirow{2}{*}{\hyperlink{vertex-1-ii}{V1-(ii)}}&
			$(1,3,0), (1,1,0)$&
			&
			\hyperlink{vertex-6-i}{V6-(i)}\\
			\cdashline{4-4}\cdashline{6-6}
			
			&
			&
			&
			$(1,3,1), (1,1,1)$&
			&
			\hyperlink{vertex-6-vi}{V6-(vi)}\\
			\hline

			$(3,2,\frac{7}{6})$&
			\multirow{5}{*}{\includegraphics[width=3.2cm, height=2.2cm]{smeft-phi2x2-5.pdf}}&
			\hyperlink{vertex-6-iv}{V6-(iv)}&
			\multirow{5}{*}{$(\{1,R_C\},R_L,Y)$}&
			\multirow{5}{*}{\includegraphics[width=3.2cm, height=2.2cm]{smeft-phi2x2-7.pdf}}&
			\multirow{3}{*}{\hyperlink{vertex-1-ii}{V1-(ii)},}\\
			\cdashline{1-1}\cdashline{3-3}
			
			$(3,2,\frac{1}{6})$&
			&
			\hyperlink{vertex-6-v}{V6-(v)}, \hyperlink{vertex-6-ix}{V6-(ix)}&
			&
			&
			\multirow{3}{*}{\hyperlink{vertex-10-i}{V10-(i)},}\\
			\cdashline{1-1}\cdashline{3-3}

			$(3,2,-\frac{5}{6})$&
			&
			\hyperlink{vertex-6-x}{V6-(x)}&
			&
			&
			\multirow{3}{*}{\hyperlink{vertex-10-ii}{V10-(ii)}}\\
			\cdashline{1-1}\cdashline{3-3}
			
			$(1,2,\frac{1}{2})$&
			&
			\hyperlink{vertex-6-ii}{V6-(ii)}&
			&
			&
			\\
			\cdashline{1-1}\cdashline{3-3}
			
			$(1,2,\frac{3}{2})$&
			&
			\hyperlink{vertex-6-vii}{V6-(vii)}&
			&
			&
			\\
			\hline			
	\end{tabular}}
	\caption{Table \ref{table:smeft-phi2X2-1} continued.}
	\label{table:smeft-phi2X2-2}
\end{table}

\subsection{\Large$\psi^4$}
	
Restricting ourselves to only baryon and lepton number conserving operators, we can subdivide the $\psi^4$ operators into several sub-categories based on the chirality of the fields constituting the individual operators. The representations of the heavy fields from which the origin of these operators can be retraced also vary based on the kind of fermion bilinears that can be identified within each operator.   

\subsubsection*{$(i)$ \underline{$(L\bar{L})\,(L\bar{L})$}}

This subclass contains operators constituted solely of the isospin doublet fermions $l$, $q$ and their conjugates, i.e., the operators - $\mathcal{Q}_{ll}$, $\mathcal{Q}_{lq}^{(1)}$, $\mathcal{Q}_{lq}^{(3)}$, $\mathcal{Q}_{qq}^{(1)}$, $\mathcal{Q}_{qq}^{(3)}$. In the Warsaw basis \cite{Grzadkowski:2010es}, these operators are expressed as the Lorentz contraction of two 4-vectors each of them being fermion bilinears of the form $\overline{\psi}\,\gamma_{\mu}\,\psi$. This also indicates that unfolding these operators must incorporate the SM vector bosons and heavy fields appear in the form of scalar or fermion loops coupling to these vector bosons.
  
The choice of the vector boson is based on whether the fermion bilinears in the particular operator transform as singlets or as triplets under weak isospin $SU(2)$. Therefore, $\mathcal{Q}_{ll}$, $\mathcal{Q}_{lq}^{(1)}$ and $\mathcal{Q}_{qq}^{(1)}$ involve $B_{\mu}$ whereas  $\mathcal{Q}_{lq}^{(3)}$ and $\mathcal{Q}_{qq}^{(3)}$ (whose constituent fermion bilinears have the form $\overline{\psi}\,\gamma_{\mu}\,\tau^I\,\psi$) involve $W^I_{\mu}$ in their respective unfolded diagrams. Fixing the vector boson ultimately fixes the representations in accordance with the results of Table~\ref{table:sm-vertices-3}. The presence of an SM vector boson at each vertex also enforces the conservation of fermion flavour.

One can note the absence of an operator $\mathcal{Q}_{ll}^{(3)}$ of the form $(\overline{l}_p\,\gamma^{\mu}\,\tau^I\,l_r)\,(\overline{l}_s\,\gamma_{\mu}\,\tau^I\,l_t)$ by examining the following equations:

\vspace{-0.6cm}
{\small\begin{eqnarray}
(\overline{l}_p\,\gamma^{\mu}\,\tau^I\,l_r)\,(\overline{l}_s\,\gamma_{\mu}\,\tau^I\,l_t) &=& (\overline{l}_p\,\gamma^{\mu}\,l_t)\,(\overline{l}_s\,\gamma_{\mu}\,l_r) - \frac{1}{2}(\overline{l}_p\,\gamma^{\mu}\,l_r)\,(\overline{l}_s\,\gamma_{\mu}\,l_t), \nonumber\\
(\overline{l}_p\,\gamma^{\mu}\,\tau^I\,l_r)\,(\overline{q}_s\,\gamma_{\mu}\,\tau^I\,q_t) &=& (\overline{l}_p\,\gamma^{\mu}\,q_t)\,(\overline{q}_s\,\gamma_{\mu}\,l_r) - \frac{1}{2}(\overline{l}_p\,\gamma^{\mu}\,l_r)\,(\overline{q}_s\,\gamma_{\mu}\,q_t). 
\end{eqnarray}}

Here, we have used the Fierz relations for the $SU(2)$ generators $(\tau^I)^i{}_j(\tau^I)^k{}_l = \delta^i{}_l\,\delta^k{}_j - \frac{1}{2}\delta^i{}_j\,\delta^k{}_l$. In the first equation, both terms on the right are identical with just the flavour indices shuffled. Thus, the first equation only relates two quantities so the operator basis can only include one of these. On the other hand, the second equation connects three different quantities, hence the operator basis can only include two of these. This explains why there are two operators $\mathcal{Q}_{lq}^{(1),(3)}$ with the same external states but only one operator  $\mathcal{Q}_{ll}$. And consequently, the absence of $W^{I}_{\mu}$ in the diagrams with $l$ and $\overline{l}$ as external lines on both vertices is explained. A similar argument explains the absence of an operator of the form  $(\overline{q}_p\,\gamma_{\mu}\,T^A\,q_r)(\overline{q}_s\,\gamma^{\mu}\,T^A\,q_t)$ and consequently the lack of a diagram involving $G^A_{\mu}$. Complete results for these operators have been presented in a condensed form in Table~\ref{table:smeft-psi4-LL}.

\begin{table}[!htb]
	\centering
	\renewcommand{\arraystretch}{2.3}
	{\scriptsize\begin{tabular}{||>{\centering}p{0.3cm}|>{\centering}p{0.3cm}||c|c|c||c|c|c||}
			
			\hline
			\multicolumn{8}{||c||}{$\mathcal{Q}_{\psi_1\psi_2}/\mathcal{Q}_{\psi_1\psi_2}^{(1)}:\, (\overline{\psi}_1\,\gamma^{\mu}\,\psi_1)\, (\overline{\psi}_2\,\gamma_{\mu}\,\psi_2)$}\\
			\hline
			
			{$\mathbf{\psi_1}$}&
			{$\mathbf{\psi_2}$}&
			\textbf{Heavy fields}&
			\textbf{Diagram}&
			\textbf{Vertices}&
			\textbf{Heavy fields}&
			\textbf{Diagram}&
			\textbf{Vertices}\\
			\hline
			
			$q$&
			$q$&
			\multirow{3}{*}{$(\{1,R_C\},\{1,R_L\},Y)$}&
			\multirow{3}{*}{\includegraphics[width=3cm,height=1.8cm]{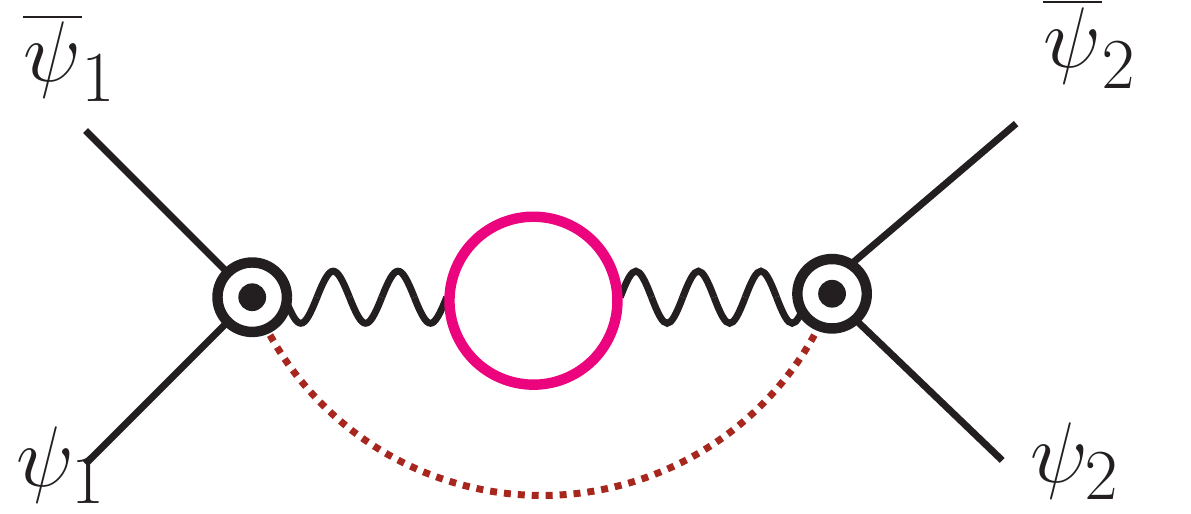}}&
			\multirow{3}{*}{\hyperlink{vertex-9-i}{V9-(i)}}&
			\multirow{3}{*}{$(\{1,R_C\},\{1,R_L\},Y)$}&
			\multirow{3}{*}{\includegraphics[width=3cm,height=1.8cm]{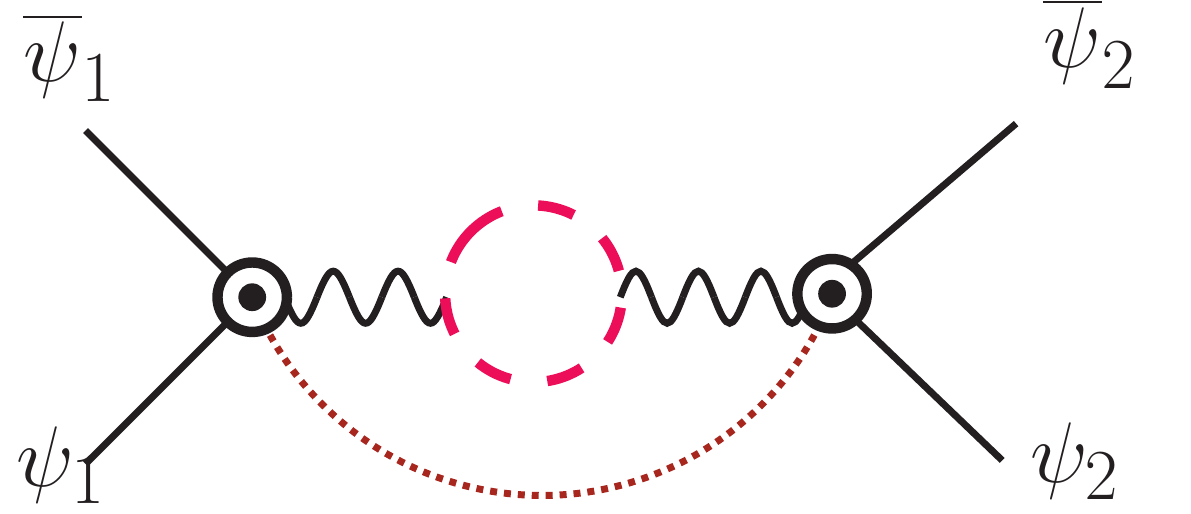}}&
			\multirow{3}{*}{\hyperlink{vertex-10-i}{V10-(i)}}\\
			
			\cdashline{1-2}
			
			$l$&
			$l$&
			&
			&
			&
			&
			&
			\\
			\cdashline{1-2}
			
			$q$&
			$l$&
			&
			&
			&
			&
			&
			\\
			\hline
			\hline
			\multicolumn{8}{||c||}{$\mathcal{Q}_{\psi_1\psi_2}^{(3)}:\, (\overline{\psi}_1\,\gamma^{\mu}\,\tau^I\,\psi_1)\, (\overline{\psi}_2\,\gamma_{\mu}\,\tau^I\,\psi_2)$}\\
			\hline
			
			{$\mathbf{\psi_1}$}&
			{$\mathbf{\psi_2}$}&
			\textbf{Heavy fields}&
			\textbf{Diagram}&
			\textbf{Vertices}&
			\textbf{Heavy fields}&
			\textbf{Diagram}&
			\textbf{Vertices}\\
			\hline
			
			\multirow{2}{*}{$q$}&
			\multirow{2}{*}{$l$}&
			\multirow{4}{*}{$(\{1,R_C\},R_L,\{0,Y\})$}&
			\multirow{4}{*}{\includegraphics[width=3cm,height=2cm]{psi4-eom1.pdf}}&
			\multirow{4}{*}{\hyperlink{vertex-9-ii}{V9-(ii)}}&
			\multirow{4}{*}{$(\{1,R_C\},R_L,\{0,Y\})$}&
			\multirow{4}{*}{\includegraphics[width=3cm,height=2cm]{psi4-eom2.pdf}}&
			\multirow{4}{*}{\hyperlink{vertex-10-ii}{V10-(ii)}}\\
			
			&&&&&&&\\
			\cdashline{1-2}
			
			\multirow{2}{*}{$q$}&
			\multirow{2}{*}{$q$}&
			&
			&
			&
			&
			&
			\\
			
			&
			&
			&
			&
			&
			&
			&
			\\
			\hline
	\end{tabular}}
	\caption{Heavy field representations that are obtained by unfolding the $\psi^4$ operators, composed of the left chiral SM fermions $q$ and $l$, into non-trivial one-loop-level diagrams and the corresponding vertices.}
	\label{table:smeft-psi4-LL}
\end{table}	

The form in which these operators appear within the SMEFT dimension-6 operator basis can only allow vector boson propagators in the unfolded diagrams but the ``external states" corresponding to these same operators can also be obtained from tree-level processes involving a scalar propagator. The corresponding results have been highlighted in Table~\ref{table:smeft-psi4-LL-2}.

\begin{table}[!htb]
	\centering
	\renewcommand{\arraystretch}{2.3}
	{\scriptsize\begin{tabular}{||>{\centering}p{1.3cm}|>{\centering}p{1.3cm}||c|c|c||}
			
			\hline
			\multicolumn{5}{||c||}{$\mathcal{Q}_{\psi_1\psi_2\psi_3\psi_4}:\, (\psi_1\,C\,\psi_2)\, (\overline{\psi}_3\,C\,\overline{\psi}_4)\,/\,(\psi_1\,C\,\tau^I\,\psi)_2\, (\overline{\psi}_3\,C\,\tau^I\,\overline{\psi}_4)$}\\
			\hline

			{$(\mathbf{\psi_1},\,\mathbf{\psi_2})$}&
			{$(\mathbf{\psi_3},\,\mathbf{\psi_4})$}&
			\textbf{Heavy fields}&
			\textbf{Diagram}&
			\textbf{Vertices}\\
			\hline

			$(l,\,l)$&
			$(l,\,l)$&
			$(1,1,1)$, $(1,3,1)$&
			\multirow{4}{*}{\hspace{0.4cm}\includegraphics[width=3.9cm,height=2.6cm]{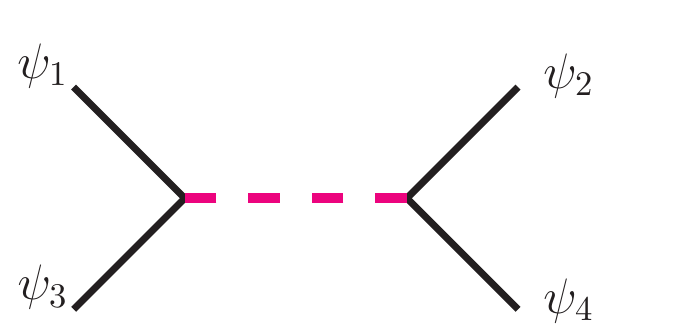}}&
			\hyperlink{vertex-5-ii}{V5-(ii)}\\

			\cdashline{1-3}\cdashline{5-5}
			
			$(l,\,q)$&
			$(l,\,q)$&
			$(3,1,-\frac{1}{3})$, $(3,3,-\frac{1}{3})$&
			&
			\hyperlink{vertex-5-ix}{V5-(ix)}\\
			\cdashline{1-3}\cdashline{5-5}
			
			\multirow{2}{*}{$(q,\,q)$}&
			\multirow{2}{*}{$(q,\,q)$}&
			$\,\,\,(3,1,-\frac{1}{3})$, $(3,3,-\frac{1}{3}),\,\,\,$&
			&
			\multirow{2}{*}{\hyperlink{vertex-5-v}{V5-(v)}}\\
			
			&
			&
			$(\overline{6},1,-\frac{1}{3})$, $(\overline{6},3,-\frac{1}{3})$&
			&
			\\
			\hline
	\end{tabular}}
	\caption{Heavy field representations that are obtained by unfolding the $\psi^4$ operators, composed of the left chiral SM fermions $q$ and $l$, into a tree-level diagram yielding non-zero contributions and the corresponding vertices that constitute the diagram for individual cases.}
	\label{table:smeft-psi4-LL-2}
\end{table}	

\subsubsection*{$(ii)$\underline{$(R\bar{R})\,(R\bar{R})$}}

This subclass contains operators constituted solely of the isospin singlet fermions $u$, $d$, $e$ and their conjugates, i.e., the operators - $\mathcal{Q}_{ee}$, $\mathcal{Q}_{uu}$, $\mathcal{Q}_{dd}$, $\mathcal{Q}_{eu}$, $\mathcal{Q}_{ed}$, $\mathcal{Q}_{ud}^{(1)}$, $\mathcal{Q}_{ud}^{(8)}$. These operators are also expressed as the Lorentz contraction of two 4-vectors each of them being fermion bilinears of the form $\overline{\psi}\,\gamma_{\mu}\,\psi$. Therefore their unfolding incorporate the SM vector bosons and heavy fields appear in the form of scalar or fermion loops coupling to these vector bosons. 

The diagrams corresponding to $\mathcal{Q}_{ud}^{(8)}$ incorporate $G^A_{\mu}$, while those for the rest of the operators incorporate $B_{\mu}$. Since all these fermions are isospin singlets, $W^I_{\mu}$ does not appear in any diagram. The results have been succinctly collected in Table~\ref{table:smeft-psi4-RR}.

\begin{table}[!htb]
	\centering
	\renewcommand{\arraystretch}{2.3}
	{\scriptsize\begin{tabular}{||>{\centering}p{0.3cm}|>{\centering}p{0.3cm}||c|c|c||c|c|c||}
			
			\hline
			\multicolumn{8}{||c||}{$\mathcal{Q}_{\psi_1\psi_2}/\mathcal{Q}_{\psi_1\psi_2}^{(1)}:\, (\overline{\psi}_1\,\gamma^{\mu}\,\psi_1)\, (\overline{\psi}_2\,\gamma_{\mu}\,\psi_2)$}\\
			\hline
			
			{$\mathbf{\psi_1}$}&
			{$\mathbf{\psi_2}$}&
			\textbf{Heavy fields}&
			\textbf{Diagram}&
			\textbf{Vertices}&
			\textbf{Heavy fields}&
			\textbf{Diagram}&
			\textbf{Vertices}\\
			\hline
			
			$e$&
			$e$&
			\multirow{6}{*}{$(\{1,R_C\},\{1,R_L\},Y)$}&
			\multirow{6}{*}{\includegraphics[width=3cm,height=2cm]{psi4-eom1.pdf}}&
			\multirow{6}{*}{\hyperlink{vertex-9-i}{V9-(i)}}&
			\multirow{6}{*}{$(\{1,R_C\},\{1,R_L\},Y)$}&
			\multirow{6}{*}{\includegraphics[width=3cm,height=2cm]{psi4-eom2.pdf}}&
			\multirow{6}{*}{\hyperlink{vertex-10-i}{V10-(i)}}\\
			\cdashline{1-2}
			
			$u$&
			$u$&
			&
			&
			&
			&
			&
			\\
			\cdashline{1-2}
			
			$d$&
			$d$&
			&
			&
			&
			&
			&
			\\
			\cdashline{1-2}
			
			$e$&
			$u$&
			&
			&
			&
			&
			&
			\\
			\cdashline{1-2}
			
			$e$&
			$d$&
			&
			&
			&
			&
			&
			\\
			\cdashline{1-2}
			
			$u$&
			$d$&
			&
			&
			&
			&
			&
			\\
			\hline 
			\hline
			\multicolumn{8}{||c||}{$\mathcal{Q}_{\psi_1\psi_2}^{(8)}:\, (\overline{\psi}_1\,\gamma^{\mu}\,T^A\,\psi_1)\, (\overline{\psi}_2\,\gamma_{\mu}\,T^A\,\psi_2)$}\\
			\hline
			
			{$\mathbf{\psi_1}$}&
			{$\mathbf{\psi_2}$}&
			\textbf{Heavy fields}&
			\textbf{Diagram}&
			\textbf{Vertices}&
			\textbf{Heavy fields}&
			\textbf{Diagram}&
			\textbf{Vertices}\\
			\hline
			
			\multirow{3}{*}{$u$}&
			\multirow{3}{*}{$d$}&
			\multirow{3}{*}{$(R_C,\{1,R_L\},\{0,Y\})$}&
			\multirow{3}{*}{\includegraphics[width=3cm,height=1.8cm]{psi4-eom1.pdf}}&
			\multirow{3}{*}{\hyperlink{vertex-9-iii}{V9-(iii)}}&
			\multirow{3}{*}{$(R_C,\{1,R_L\},\{0,Y\})$}&
			\multirow{3}{*}{\includegraphics[width=3cm,height=1.8cm]{psi4-eom2.pdf}}&
			\multirow{3}{*}{\hyperlink{vertex-10-iii}{V10-(iii)}}\\
			
			&&&&&&&\\

			&
			&
			&
			&
			&
			&
			&
			\\
			\hline
			\end{tabular}}
	\caption{Heavy field representations that are obtained by unfolding the $\psi^4$ operators, composed of the right chiral SM fermions $e$, $u$ and $d$, into non-trivial one-loop-level diagrams and the corresponding vertices.}
	\label{table:smeft-psi4-RR}
\end{table}	

Additionally, ``the external states" corresponding to these same operators can also be obtained from tree-level processes involving a scalar propagator. The corresponding results have been collected in Table~\ref{table:smeft-psi4-LL-2}.

\begin{table}[!htb]
	\centering
	\renewcommand{\arraystretch}{2.3}
	{\scriptsize\begin{tabular}{||>{\centering}p{1.3cm}|>{\centering}p{1.3cm}||c|c|c||}
			
			\hline
			\multicolumn{5}{||c||}{$\mathcal{Q}_{\psi_1\psi_2\psi_3\psi_4}:\, (\psi_1\,C\,\psi_2)\, (\overline{\psi}_3\,C\,\overline{\psi}_4)\,/\,(\psi_1\,C\,T^A\,\psi_2)\, (\overline{\psi}_3\,C\,T^A\,\overline{\psi}_4)$}\\
			\hline

			{$(\mathbf{\psi_1},\,\mathbf{\psi_2})$}&
			{$(\mathbf{\psi_3},\,\mathbf{\psi_4})$}&
			\textbf{Heavy fields}&
			\textbf{Diagram}&
			\textbf{Vertices}\\
			\hline
			
			$(e,\,e)$&
			$(e,\,e)$&
			$(1,1,2)$&
			\multirow{6}{*}{\hspace{0.4cm}\includegraphics[width=3.9cm,height=2.6cm]{smeft-psi4-6.pdf}}&
			\hyperlink{vertex-5-i}{V5-(i)}\\
			\cdashline{1-3}\cdashline{5-5}

			$(e,\,u)$&
			$(e,\,u)$&
			$(\overline{3},1,\frac{1}{3})$&
			&
			\hyperlink{vertex-5-xi}{V5-(xi)}\\
			\cdashline{1-3}\cdashline{5-5}
			
			$(e,\,d)$&
			$(e,\,d)$&
			$(\overline{3},1,\frac{4}{3})$&
			&
			\hyperlink{vertex-5-xii}{V5-(xii)}\\
			\cdashline{1-3}\cdashline{5-5}
			
			$(d,\,d)$&
			$(d,\,d)$&
			$(3,1,\frac{2}{3})$, $(\overline{6},1,\frac{2}{3})$&
			&
			\hyperlink{vertex-5-iii}{V5-(iii)}\\
			\cdashline{1-3}\cdashline{5-5}
			
			$(u,\,u)$&
			$(u,\,u)$&
			$(3,1,-\frac{4}{3})$, $(\overline{6},1,-\frac{4}{3})$&
			&
			\hyperlink{vertex-5-iv}{V5-(iv)}\\
			\cdashline{1-3}\cdashline{5-5}

			$(u,\,d)$&
			$(u,\,d)$&
			$\,\,\,(3,1,-\frac{1}{3})$, $(\overline{6},1,-\frac{1}{3})\,\,\,$&
			&
			\hyperlink{vertex-5-x}{V5-(x)}\\
			
			\hline
	\end{tabular}}
	\caption{Heavy field representations that are obtained by unfolding the $\psi^4$ operators, composed of the right chiral SM fermions $e$, $u$ and $d$, into a non-trivial tree-level diagram and the corresponding vertices for individual cases.}
	\label{table:smeft-psi4-RR-2}
\end{table}	
\newpage
\subsubsection*{$(iii)$\underline{$(L\bar{L})\,(R\bar{R})$}}	
There are two different ways of expressing the operators of this subclass as products of fermion bilinears:
\begin{enumerate}
	\item The first one is similar to the previous two cases with bilinears of the form $\overline{\psi}\,\gamma_{\mu}\,\psi$ and with the SM vector bosons in the diagrams. The diagrams corresponding to $\mathcal{Q}_{qu}^{(8)}$ and $\mathcal{Q}_{qd}^{(8)}$ incorporate $G^A_{\mu}$, while those for the rest of the operators incorporate $B_{\mu}$. Since in each operator two fermions are isospin singlets, $W^I_{\mu}$ does not appear in any diagram. Once again these diagrams have flavour conservation imposed at each vertex. The corresponding results are collected in Table~\ref{table:smeft-psi4-LR-1}.	
\begin{table}[!htb]
	\centering
	\renewcommand{\arraystretch}{2.2}
	{\scriptsize\begin{tabular}{||>{\centering}p{0.3cm}|>{\centering}p{0.3cm}||c|c|c||c|c|c||}
			
			\hline
			\multicolumn{8}{||c||}{$\mathcal{Q}_{\psi_1\psi_2}/\mathcal{Q}_{\psi_1\psi_2}^{(1)}:\, (\overline{\psi}_1\,\gamma^{\mu}\,\psi_1)\, (\overline{\psi}_2\,\gamma_{\mu}\,\psi_2)$}\\
			\hline
			
			{$\mathbf{\psi_1}$}&
			{$\mathbf{\psi_2}$}&
			\textbf{Heavy fields}&
			\textbf{Diagram}&
			\textbf{Vertices}&
			\textbf{Heavy fields}&
			\textbf{Diagram}&
			\textbf{Vertices}\\
			\hline
			
			$l$&
			$e$&
			\multirow{6}{*}{$(\{1,R_C\},\{1,R_L\},Y)$}&
			\multirow{6}{*}{\includegraphics[width=3cm,height=2.2cm]{psi4-eom1.pdf}}&
			\multirow{6}{*}{\hyperlink{vertex-9-i}{V9-(i)}}&
			\multirow{6}{*}{$(\{1,R_C\},\{1,R_L\},Y)$}&
			\multirow{6}{*}{\includegraphics[width=3cm,height=2.2cm]{psi4-eom2.pdf}}&
			\multirow{6}{*}{\hyperlink{vertex-10-i}{V10-(i)}}\\
			\cdashline{1-2}
			
			$l$&
			$u$&
			&
			&
			&
			&
			&
			\\
			\cdashline{1-2}
			
			$l$&
			$d$&
			&
			&
			&
			&
			&
			\\
			\cdashline{1-2}
			
			$q$&
			$e$&
			&
			&
			&
			&
			&
			\\
			\cdashline{1-2}
			
			$q$&
			$u$&
			&
			&
			&
			&
			&
			\\
			\cdashline{1-2}
			
			$q$&
			$d$&
			&
			&
			&
			&
			&
			\\
			
			\hline
			\hline
			\multicolumn{8}{||c||}{$\mathcal{Q}_{\psi_1\psi_2}^{(8)}:\, (\overline{\psi}_1\,\gamma^{\mu}\,T^A\,\psi_1)\, (\overline{\psi}_2\,\gamma_{\mu}\,T^A\,\psi_2)$}\\
			\hline
			
			{$\mathbf{\psi_1}$}&
			{$\mathbf{\psi_2}$}&
			\textbf{Heavy fields}&
			\textbf{Diagram}&
			\textbf{Vertices}&
			\textbf{Heavy fields}&
			\textbf{Diagram}&
			\textbf{Vertices}\\
			\hline
			
			\multirow{2}{*}{$q$}&
			\multirow{2}{*}{$u$}&
			\multirow{4}{*}{$(R_C,\{1,R_L\},\{0,Y\})$}&
			\multirow{4}{*}{\includegraphics[width=3cm,height=2cm]{psi4-eom1.pdf}}&
			\multirow{4}{*}{\hyperlink{vertex-9-iii}{V9-(iii)}}&
			\multirow{4}{*}{$(R_C,\{1,R_L\},\{0,Y\})$}&
			\multirow{4}{*}{\includegraphics[width=3cm,height=2cm]{psi4-eom2.pdf}}&
			\multirow{4}{*}{\hyperlink{vertex-10-iii}{V10-(iii)}}\\
			
			&&&&&&&\\
			\cdashline{1-2}
			
			\multirow{2}{*}{$q$}&
			\multirow{2}{*}{$d$}&
			&
			&
			&
			&
			&
			\\
			
			&
			&
			&
			&
			&
			&
			&
			\\
			\hline
	\end{tabular}}
	\caption{Heavy field representations that are obtained by unfolding the $\psi^4$ operators, composed of both left and the right chiral SM fermions, into non-trivial one-loop-level diagrams and the corresponding vertices. Each of the processes highlighted here are flavour conserving.}
	\label{table:smeft-psi4-LR-1}
\end{table}

	\item For operators of this subclass, instead of forming fermion bilinears of the form  $\overline{\psi}\,\gamma_{\mu}\,\psi$ which transform as 4-vectors under Lorentz transformations, we can construct Lorentz scalars of the form  $\overline{\psi}_1\,\psi_2$, where $\psi_1$ and $\psi_2$ have opposite chirality. As a result of this, in the unfolded diagram we can have the vertex $\overline{\psi}_1\,\psi_2\,\phi$ instead of $\overline{\psi}\,\gamma_{\mu}\,\psi\,V^{\mu}$. This brings into light a few interesting points:
	
\begin{itemize}
	\item The presence of a Yukawa like vertex allows for the possibility of flavour violation.
	
	\item This rearrangement of external legs also allows the appearance of certain heavy scalars at the tree-level itself.
	
	\item Even though operators of the form $(\overline{\psi}_1\,\psi_2)\,(\overline{\psi}_2\,\psi_1)$ may appear to be unique with respect to those of the form $(\overline{\psi}_1\,\gamma_{\mu}\,\psi_1)\,(\overline{\psi}_2\,\gamma^{\mu}\,\psi_2)$, they are actually interrelated through the Fierz relations of the $\gamma^{\mu}$-matrices or more appropriately\footnote{It must be kept in mind that when we are speaking of $\sigma^{\mu}$ and $\gamma^{\mu}$ in the same sentence, it is implied that we are working in the Weyl basis, where $\gamma^{\mu} = \begin{pmatrix}
		0 \ \ \sigma^{\mu}\\
		\overline{\sigma}^{\mu} \ \ 0
		\end{pmatrix}$ and for conveniently going between 2- and 4- component notation of fermions, we can understand a 4-component left-chiral fermion as one with the last two entries 0. Similarly, a 4-component right chiral fermion would have its first two entries as 0. } the relations of the $\sigma^{\mu}$ ($\equiv\, (\mathbf{1}_{2\times2},\,\vec{\sigma})$) as shown below:
 	
 	\vspace{-0.6cm}
	{\small\begin{eqnarray}\label{eq:gamma-fierz}
			(\sigma^\mu)_{\alpha\dot{\alpha}}\,(\sigma_\mu)_{\beta\dot{\beta}} 
			= 2\,\epsilon_{\alpha\beta}\,\epsilon_{\dot{\alpha}\dot{\beta}}, \hspace{0.7cm}
			(\sigma^\mu)_{\alpha\dot{\alpha}}\,(\overline{\sigma}_\mu)^{\dot{\beta}\beta} 
			=  2\,\delta^{\beta}_{\alpha}\, \delta^{\dot{\beta}}_{\dot{\alpha}}, \hspace{0.7cm}
			(\overline{\sigma}^\mu)^{\dot{\alpha}\alpha}\,(\overline{\sigma}_\mu)^{\dot{\beta}\beta} 
			=  2\,\epsilon^{\alpha \beta}\,\epsilon^{\dot{\alpha}\dot{\beta}}. 
	\end{eqnarray}}
	Using these we can show, for instance:
	
	\vspace{-0.6cm}
	{\small\begin{eqnarray} \label{eq:fierz-psi4-cls-1}
		(\overline{d}\,\gamma^\mu\, d)(\overline{q}\, \gamma_\mu\, q)
		&=& (\overline{d}^{\alpha}\,\sigma^{\mu}_{\alpha\dot{\alpha}}\, d^{\dot{\alpha}})\, (\overline{q}_{\dot{\beta}}\,  \overline{\sigma}^{\mu\dot{\beta}\beta}\, q_{\beta}) = 2\, (\overline{d}^\alpha \, q_{\beta}  \, \overline{q}_{\dot{\beta}} \, d^{\dot{\alpha}}) \, \delta^{\beta}_{\alpha}\, \delta^{\dot{\beta}}_{\dot{\alpha}} = 2\,(\overline{d}\,q)\,(\overline{q}\,d),\nonumber \\
		\label{eq:fierz-psi4-cls-2}
		 (\overline{d}\,\gamma^\mu\, T^A \,d)(\overline{q}\, \gamma_\mu\,T^A\, q)
		&=& (\overline{d}^{\alpha}\,\sigma^{\mu}_{\alpha\dot{\alpha}}\, T^A\, d^{\dot{\alpha}})\,(\overline{q}_{\dot{\beta}}\,T^A\, \overline{\sigma}^{\mu\dot{\beta}\beta}\,q_{\beta}) = 2\,\overline{d}^\alpha\,T^A\,q_{\beta}\,\overline{q}_{\dot{\beta}}\,T^A d^{\dot{\alpha}}\,\delta^{\beta}_{\alpha}\,\delta^{\dot{\beta}}_{\dot{\alpha}} \nonumber\\
		&=& 2\,(\overline{d}\,T^A\,q)\,(\overline{q}\, T^A\,d).
	\end{eqnarray}}
	Hence, we can easily modify the form of the operators while staying within the confines of the complete and independent Warsaw basis \cite{Grzadkowski:2010es}.
\end{itemize}
	In Table~\ref{table:smeft-psi4-LR-1}, the heavy field representations obtained were common for the operators $\mathcal{Q}_{le}$, $\mathcal{Q}_{lu}$, $\mathcal{Q}_{ld}$, $\mathcal{Q}_{qe}$, $\mathcal{Q}_{qu}^{(1)}$ and $\mathcal{Q}_{qd}^{(1)}$, and similarly for the operators $\mathcal{Q}_{qu}^{(8)}$ and $\mathcal{Q}_{qd}^{(8)}$ but after rearranging each operator in the manner suggested in Eq.~\ref{eq:fierz-psi4-cls-1}, we observe that there are heavy field representations which are common to some operators and other representations which are unique to certain operators. The specific results for each case have been catalogued in Tables~\ref{table:smeft-psi4-LR-2} and \ref{table:smeft-psi4-LR-3}. In a minimal setting, the following diagrams appear for each case:
	
	\begin{enumerate}
		\item Fig.~\ref{subfig:psi4-tree1} which is a tree level diagram containing a heavy scalar propagator.
		
		\item Fig.~\ref{subfig:psi4-loop1} with light-heavy mixing in the loop involving a single heavy fermion propagator along with the SM scalar and an SM fermion.
		
		\item Fig.~\ref{subfig:psi4-loop1} with light-heavy mixing in the loop involving heavy scalar propagators and light SM fermion propagators.
	\end{enumerate}
	
\begin{table}[!htb]
	\centering
	\renewcommand{\arraystretch}{2.2}
	{\scriptsize\begin{tabular}{||c|c|c||c|c|c||}
			\hline
			\multicolumn{3}{||c||}{$\mathcal{Q}_{le}:\,(\overline{l}_p\,\gamma^{\mu}\,l_r)(\overline{e}_s\,\gamma_{\mu}\,e_t)$}&
			\multicolumn{3}{c||}{$\mathcal{Q}_{ld}:\,(\overline{l}_p\,\gamma^{\mu}\,l_r)(\overline{d}_s\,\gamma_{\mu}\,d_t)$}\\
			
			\hline
			
			\textbf{Heavy fields}&
			\textbf{Diagram}&
			\textbf{Vertices}&
			\textbf{Heavy fields}&
			\textbf{Diagram}&
			\textbf{Vertices}\\
			\hline
			
			\multirow{4}{*}{$(1,2,\frac{1}{2})$}&
			\multirow{4}{*}{\includegraphics[width=3cm, height=1.8cm]{smeft-psi4-6.pdf}}&
			\multirow{4}{*}{\hyperlink{vertex-5-vi}{V5-(vi)}}&
			\multirow{4}{*}{$(3,2,\frac{1}{6})$}&
			\multirow{4}{*}{\includegraphics[width=3cm, height=1.8cm]{smeft-psi4-6.pdf}}&
			\multirow{4}{*}{\hyperlink{vertex-5-xvi}{V5-(xvi)}}\\

			&
			&
			&
			&
			&
			\\

			&
			&
			&
			&
			&
			\\

			&
			&
			&
			&
			&
			\\
			\hline
			
			$(1,3,0), (1,1,0)$&
			\multirow{4}{*}{\includegraphics[width=3cm, height=2cm]{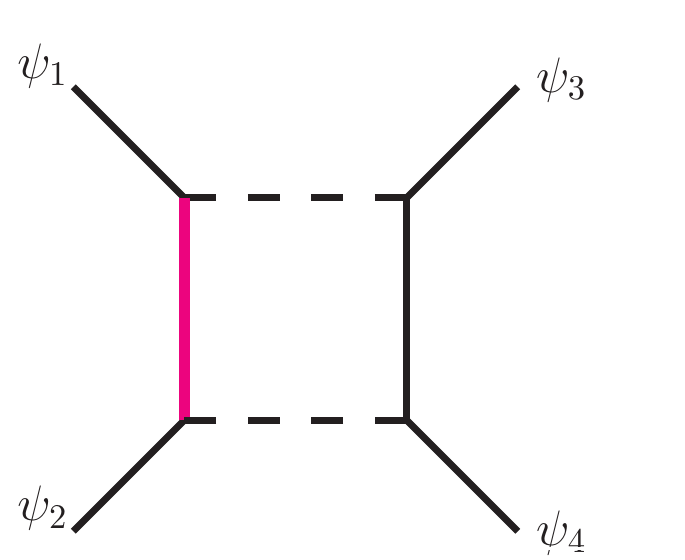}}&
			\hyperlink{vertex-6-i}{V6-(i)}&
			$(1,3,0), (1,1,0)$&
			\multirow{4}{*}{\includegraphics[width=3cm, height=2cm]{smeft-psi4-2.pdf}}&
			\hyperlink{vertex-6-i}{V6-(i)}\\
			\cdashline{1-1}\cdashline{3-3}\cdashline{4-4}\cdashline{6-6}
			
			$(1,3,1), (1,1,1)$&
			&
			\hyperlink{vertex-6-vi}{V6-(vi)}&
			$(1,3,1), (1,1,1)$&
			&
			\hyperlink{vertex-6-vi}{V6-(vi)}\\
			\cdashline{1-1}\cdashline{3-3}\cdashline{4-4}\cdashline{6-6}
			
			$(1,2,\frac{1}{2})$&
			&
			\hyperlink{vertex-6-ii}{V6-(ii)}&
			$(3,2,\frac{1}{6})$&
			&
			\hyperlink{vertex-6-v}{V6-(v)}\\
			\cdashline{1-1}\cdashline{3-3}\cdashline{4-4}\cdashline{6-6}
			
			$(1,2,\frac{3}{2})$&
			&
			\hyperlink{vertex-6-vii}{V6-(vii)}&
			$(3,2,-\frac{5}{6})$&
			&
			\hyperlink{vertex-6-x}{V6-(x)}\\
			\hline

			\multirow{2}{*}{$(3,2,\frac{7}{6})$}&
			\multirow{4}{*}{\includegraphics[width=3cm, height=2cm]{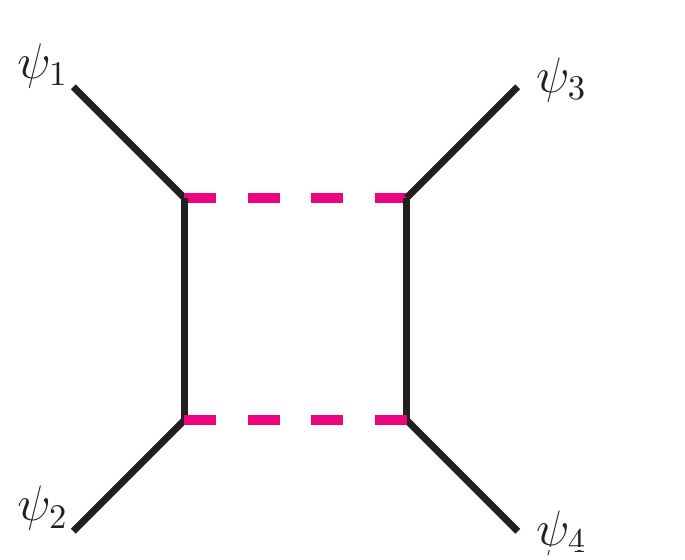}}&
			\multirow{2}{*}{\hyperlink{vertex-5-xiii}{V5-(xiii)}, \hyperlink{vertex-5-xiv}{V5-(xiv)}}&
			\multirow{4}{*}{$(3,1,-\frac{1}{3})$}&
			\multirow{4}{*}{\includegraphics[width=3cm, height=2cm]{smeft-psi4-4.pdf}}&
			\multirow{4}{*}{\hyperlink{vertex-5-ix}{V5-(ix)}, \hyperlink{vertex-5-x}{V5-(x)}}\\

			&
			&
			&
			&
			&
			\\
			\cdashline{1-1}\cdashline{3-3}
			
			\multirow{2}{*}{$(3,1,-\frac{1}{3})$}&
			&
			\multirow{2}{*}{\hyperlink{vertex-5-ix}{V5-(ix)}, \hyperlink{vertex-5-xi}{V5-(xi)}}&
			&
			&
			\\

			&
			&
			&
			&
			&
			\\
			\hline
			\hline
			\multicolumn{3}{||c||}{$\mathcal{Q}_{lu}:\,(\overline{l}_p\,\gamma^{\mu}\,l_r)(\overline{u}_s\,\gamma_{\mu}\,u_t)$}&
			\multicolumn{3}{c||}{$\mathcal{Q}_{qe}:\,(\overline{q}_p\,\gamma^{\mu}\,q_r)(\overline{e}_s\,\gamma_{\mu}\,e_t)$}\\
			\hline
			
			\textbf{Heavy fields}&
			\textbf{Diagram}&
			\textbf{Vertices}&
			\textbf{Heavy fields}&
			\textbf{Diagram}&
			\textbf{Vertices}\\
			\hline
			
			\multirow{4}{*}{$(3,2,\frac{7}{6})$}&
			\multirow{4}{*}{\includegraphics[width=3cm, height=1.8cm]{smeft-psi4-6.pdf}}&
			\multirow{4}{*}{\hyperlink{vertex-5-xiv}{V5-(xiv)}}&
			\multirow{4}{*}{$(3,2,\frac{7}{6})$}&
			\multirow{4}{*}{\includegraphics[width=3cm, height=1.8cm]{smeft-psi4-6.pdf}}&
			\multirow{4}{*}{\hyperlink{vertex-5-xiv}{V5-(xiv)}}\\

			&
			&
			&
			&
			&
			\\

			&
			&
			&
			&
			&
			\\

			&
			&
			&
			&
			&
			\\
			\hline

			$(1,3,0), (1,1,0)$&
			\multirow{4}{*}{\includegraphics[width=3cm, height=2cm]{smeft-psi4-2.pdf}}&
			\hyperlink{vertex-6-i}{V6-(i)}&
			$(3,1,\frac{2}{3}), (3,3,\frac{2}{3})$&
			\multirow{4}{*}{\includegraphics[width=3cm, height=2cm]{smeft-psi4-2.pdf}}&
			\hyperlink{vertex-6-iii}{V6-(iii)}\\
			\cdashline{1-1}\cdashline{3-3}\cdashline{4-4}\cdashline{6-6}
			
			$(1,3,1), (1,1,1)$&
			&
			\hyperlink{vertex-6-vi}{V6-(vi)}&
			$(3,1,-\frac{1}{3}), (3,3,-\frac{1}{3})$&
			&
			\hyperlink{vertex-6-viii}{V6-(viii)}\\
			\cdashline{1-1}\cdashline{3-3}\cdashline{4-4}\cdashline{6-6}
			
			$(3,2,\frac{7}{6})$&
			&
			\hyperlink{vertex-6-iv}{V6-(iv)}&
			$(1,2,\frac{1}{2})$&
			&
			\hyperlink{vertex-6-ii}{V6-(ii)}\\
			\cdashline{1-1}\cdashline{3-3}\cdashline{4-4}\cdashline{6-6}
			
			$(3,2,\frac{1}{6})$&
			&
			\hyperlink{vertex-6-ix}{V6-(ix)}&
			$(1,2,\frac{3}{2})$&
			&
			\hyperlink{vertex-6-vii}{V6-(vii)}\\
			\hline

			\multirow{4}{*}{$(3,1,-\frac{1}{3})$}&
			\multirow{4}{*}{\includegraphics[width=3cm, height=2cm]{smeft-psi4-4.pdf}}&
			\multirow{2}{*}{\hyperlink{vertex-5-ix}{V5-(ix)}, \hyperlink{vertex-5-x}{V5-(x)}}&
			\multirow{4}{*}{$(3,1,-\frac{1}{3})$}&
			\multirow{4}{*}{\includegraphics[width=3cm, height=2cm]{smeft-psi4-4.pdf}}&
			\multirow{2}{*}{\hyperlink{vertex-5-ix}{V5-(ix)}, \hyperlink{vertex-5-xi}{V5-(xi)}}\\
			
			&
			&
			&
			&
			&
			\\
			\cdashline{3-3}\cdashline{6-6}
			
			&
			&
			\multirow{2}{*}{\hyperlink{vertex-5-ix}{V5-(ix)}, \hyperlink{vertex-5-xi}{V5-(xi)}}&
			&
			&
			\multirow{2}{*}{\hyperlink{vertex-5-v}{V5-(v)}, \hyperlink{vertex-5-xi}{V5-(xi)}}\\

			&
			&
			&
			&
			&
			\\
			\hline
			
	\end{tabular}}
	\caption{Heavy field representations that are obtained by unfolding the $\psi^4$ operators, composed of both left and right chiral SM fermions, into non-trivial tree- and(or) one-loop-level diagrams and the corresponding vertices. These diagrams are obtained after the form of the operators have been modified using Fierz identities. The processes highlighted here can allow for flavour violation.}
	\label{table:smeft-psi4-LR-2}
\end{table}
\clearpage

\begin{table}[!htb]
	\centering
	\renewcommand{\arraystretch}{2.2}
	{\scriptsize\begin{tabular}{||c|c|c||c|c|c||}
			\hline
			\multicolumn{3}{||c||}{$\mathcal{Q}_{qu}^{(1)}:\,(\overline{q}_p\,\gamma^{\mu}\,q_r)(\overline{u}_s\,\gamma_{\mu}\,u_t)$}&
			\multicolumn{3}{c||}{$\mathcal{Q}_{qd}^{(1)}:\,(\overline{q}_p\,\gamma^{\mu}\,q_r)(\overline{d}_s\,\gamma_{\mu}\,d_t)$}\\
			\hline
			
			\textbf{Heavy fields}&
			\textbf{Diagram}&
			\textbf{Vertices}&
			\textbf{Heavy fields}&
			\textbf{Diagram}&
			\textbf{Vertices}\\
			\hline
			
			\multirow{4}{*}{$(1,2,\frac{1}{2})$}&
			\multirow{4}{*}{\includegraphics[width=3cm, height=1.8cm]{smeft-psi4-6.pdf}}&
			\multirow{4}{*}{\hyperlink{vertex-5-viii}{V5-(viii)}}&
			\multirow{4}{*}{$(1,2,\frac{1}{2})$}&
			\multirow{4}{*}{\includegraphics[width=3cm, height=1.8cm]{smeft-psi4-6.pdf}}&
			\multirow{4}{*}{\hyperlink{vertex-5-vii}{V5-(vii)}}\\

			&
			&
			&
			&
			&
			\\

			&
			&
			&
			&
			&
			\\
			
			&
			&
			&
			&
			&
			\\
			\hline

			$(3,1,\frac{2}{3}), (3,3,\frac{2}{3})$&
			\multirow{4}{*}{\includegraphics[width=3cm, height=2cm]{smeft-psi4-2.pdf}}&
			\hyperlink{vertex-6-iii}{V6-(iii)}&
			$(3,1,\frac{2}{3}), (3,3,\frac{2}{3})$&
			\multirow{4}{*}{\includegraphics[width=3cm, height=2cm]{smeft-psi4-2.pdf}}&
			\hyperlink{vertex-6-iii}{V6-(iii)}\\
			\cdashline{1-1}\cdashline{3-3}\cdashline{4-4}\cdashline{6-6}
			
			$(3,1,-\frac{1}{3}), (3,3,-\frac{1}{3})$&
			&
			\hyperlink{vertex-6-viii}{V6-(viii)}&
			$(3,1,-\frac{1}{3}), (3,3,-\frac{1}{3})$&
			&
			\hyperlink{vertex-6-viii}{V6-(viii)}\\
			\cdashline{1-1}\cdashline{3-3}\cdashline{4-4}\cdashline{6-6}
			
			$(3,2,\frac{7}{6})$&
			&
			\hyperlink{vertex-6-iv}{V6-(iv)}&
			$(3,2,\frac{1}{6})$&
			&
			\hyperlink{vertex-6-v}{V6-(v)}\\
			\cdashline{1-1}\cdashline{3-3}\cdashline{4-4}\cdashline{6-6}
			
			$(3,2,\frac{1}{6})$&
			&
			\hyperlink{vertex-6-ix}{V6-(ix)}&
			$(3,2,-\frac{5}{6})$&
			&
			\hyperlink{vertex-6-x}{V6-(x)}\\
			\hline
			
			\multirow{4}{*}{$(3,1,-\frac{1}{3})$}&
			\multirow{6}{*}{\includegraphics[width=3cm, height=2cm]{smeft-psi4-4.pdf}}&
			\hyperlink{vertex-5-ix}{V5-(ix)}, \hyperlink{vertex-5-x}{V5-(x)}&
			\multirow{3}{*}{$(3,1,-\frac{1}{3})$}&
			\multirow{6}{*}{\includegraphics[width=3cm, height=2cm]{smeft-psi4-4.pdf}}&
			\hyperlink{vertex-5-ix}{V5-(ix)}, \hyperlink{vertex-5-x}{V5-(x)}\\
			\cdashline{3-3}
			
			&
			&
			\hyperlink{vertex-5-v}{V5-(v)}, \hyperlink{vertex-5-x}{V5-(x)}&
			&
			&
			\hyperlink{vertex-5-v}{V5-(v)}, \hyperlink{vertex-5-x}{V5-(x)}\\
			\cdashline{3-3}
			
			&
			&
			\hyperlink{vertex-5-ix}{V5-(ix)}, \hyperlink{vertex-5-xi}{V5-(xi)}&
			&
			&
			\\
			\cdashline{3-3}\cdashline{4-4}\cdashline{6-6}
			
			&
			&
			\hyperlink{vertex-5-v}{V5-(v)}, \hyperlink{vertex-5-xi}{V5-(xi)}&
			\multirow{3}{*}{$(6,1,\frac{1}{3})$}&
			&
			\multirow{3}{*}{\hyperlink{vertex-5-v}{V5-(v)}, \hyperlink{vertex-5-x}{V5-(x)}}\\
			\cdashline{1-1}\cdashline{3-3}
			
			$(6,1,\frac{1}{3})$&
			&
			\hyperlink{vertex-5-v}{V5-(v)}, \hyperlink{vertex-5-x}{V5-(x)}&
			&
			&
			\\
			
			\cdashline{1-1}\cdashline{3-3}
			
			$(3,2,\frac{7}{6})$&
			&
			\hyperlink{vertex-5-xiii}{V5-(xiii)}, \hyperlink{vertex-5-xiv}{V5-(xiv)}&
			&
			&
			\\

			\hline
			\hline
			\multicolumn{3}{||c||}{$\mathcal{Q}_{qu}^{(8)}:\,(\overline{q}_p\,\gamma^{\mu}\,T^A\,q_r)(\overline{u}_s\,\gamma_{\mu}\,T^A\,u_t)$}&
			\multicolumn{3}{c||}{$\mathcal{Q}_{qd}^{(8)}:\,(\overline{q}_p\,\gamma^{\mu}\,T^A\,q_r)(\overline{d}_s\,\gamma_{\mu}\,T^A\,d_t)$}\\
			\hline
			
			\textbf{Heavy fields}&
			\textbf{Diagram}&
			\textbf{Vertices}&
			\textbf{Heavy fields}&
			\textbf{Diagram}&
			\textbf{Vertices}\\
			\hline
			
			\multirow{4}{*}{$(8,2,\frac{1}{2})$}&
			\multirow{4}{*}{\includegraphics[width=3cm, height=1.8cm]{smeft-psi4-6.pdf}}&
			\multirow{4}{*}{\hyperlink{vertex-5-viii}{V5-(viii)}}&
			\multirow{4}{*}{$(8,2,\frac{1}{2})$}&
			\multirow{4}{*}{\includegraphics[width=3cm, height=1.8cm]{smeft-psi4-6.pdf}}&
			\multirow{4}{*}{\hyperlink{vertex-5-vii}{V5-(vii)}}\\

			&
			&
			&
			&
			&
			\\

			&
			&
			&
			&
			&
			\\

			&
			&
			&
			&
			&
			\\
			\hline
			
	\end{tabular}}
	\caption{Table \ref{table:smeft-psi4-LR-2} continued.}
	\label{table:smeft-psi4-LR-3}
\end{table}

\end{enumerate}

\subsubsection*{$(iv)$ \underline{$(L\bar{R})\,(R\bar{L})$, $(\bar{L}R)\,(\bar{L}R)$}}	

The first of these two subclasses contains a single operator $\mathcal{Q}_{ledq}$ and it allows heavy scalars with specific quantum numbers to appear through tree as well as loop diagrams.

The other subclass contains four operators $\mathcal{Q}^{(1)}_{quqd}$, $\mathcal{Q}^{(8)}_{quqd}$, $\mathcal{Q}^{(1)}_{lequ}$, $\mathcal{Q}^{(3)}_{lequ}$. The first three of these admit heavy scalars through tree-level processes. Different representations are obtained by permuting the external legs suitably. $\mathcal{Q}^{(3)}_{lequ}$ is built of non-trivial tensor structures and therefore requires special attention. The results corresponding to $\mathcal{Q}_{ledq}$, $\mathcal{Q}^{(1)}_{quqd}$, $\mathcal{Q}^{(8)}_{quqd}$ and $\mathcal{Q}^{(1)}_{lequ}$ have been collected in Table~\ref{table:smeft-psi4-LR-4}.
	\begin{table}[!htb]
	\centering
	\renewcommand{\arraystretch}{2.2}
	{\scriptsize\begin{tabular}{|c|c|c|}
			\hline
			\hline
			\multicolumn{3}{|c|}{$\mathcal{Q}_{ledq}:\,(\overline{l}_p^j\,e_r)(\overline{d}_s\,q_{tj})$}\\
			\hline
			
			\textbf{Heavy fields}&
			\textbf{Diagram}&
			\textbf{Vertices}\\
			\hline
			
			\multirow{4}{*}{$(1,2,\frac{1}{2})$}&
			\multirow{4}{*}{\includegraphics[width=3cm, height=1.8cm]{smeft-psi4-6.pdf}}&
			\multirow{4}{*}{\hyperlink{vertex-5-vi}{V5-(vi)},  \hyperlink{vertex-5-vii}{V5-(vii)}}\\

			&
			&
			\\

			&
			&
			\\

			&
			&
			\\
			\cline{1-3}

			\multirow{4}{*}{$(3,1,-\frac{1}{3})$}&
			\multirow{4}{*}{\includegraphics[width=3cm, height=1.8cm]{smeft-psi4-4.pdf}}&
			\multirow{4}{*}{\hyperlink{vertex-5-v}{V5-(v)}, \hyperlink{vertex-5-ix}{V5-(ix)}, \hyperlink{vertex-5-x}{V5-(x)}, \hyperlink{vertex-5-xi}{V5-(xi)}}\\
			
			&
			&
			\\
			
			&
			&
			\\

			&
			&
			\\
			\hline
			\multicolumn{3}{|c|}{$\mathcal{Q}_{quqd}^{(1)}:\, (\overline{q}_p^j\,u_r)\epsilon_{jk}(\overline{q}_s^k\,d_t)$}\\
			\hline

			\textbf{Heavy fields}&
			\textbf{Diagram}&
			\textbf{Vertices}\\
			\hline

			\multirow{2}{*}{$(1,2,\frac{1}{2})$}&
			\multirow{3}{*}{\includegraphics[width=3cm, height=1.8cm]{smeft-psi4-6.pdf}}&
			\multirow{2}{*}{\hyperlink{vertex-5-vii}{V5-(vii)},  \hyperlink{vertex-5-viii}{V5-(viii)}}\\

			&
			&
			\\
			\cdashline{1-1}\cdashline{3-3}

			\multirow{2}{*}{$(3,1,-\frac{1}{3}),\,(\overline{6},1,-\frac{1}{3}) $}&
			&
			\multirow{2}{*}{\hyperlink{vertex-5-v}{V5-(v)},  \hyperlink{vertex-5-x}{V5-(x)}}\\

			&
			&
			\\
			\hline
			\hline
			\multicolumn{3}{|c|}{$\mathcal{Q}_{quqd}^{(8)}:\, (\overline{q}_p^j\,T^A\,u_r)\epsilon_{jk}(\overline{q}_s^k\,T^A\,d_t)$}\\
			\hline
			
			\textbf{Heavy fields}&
			\textbf{Diagram}&
			\textbf{Vertices}\\
			\hline
			
			\multirow{3}{*}{$(8,2,\frac{1}{2})$}&
			\multirow{2}{*}{\includegraphics[width=3cm, height=1.8cm]{smeft-psi4-6.pdf}}&
			\multirow{3}{*}{\hyperlink{vertex-5-vii}{V5-(vii), \hyperlink{vertex-5-viii}{V5-(viii)}}}\\

			&
			&
			\\
			
			&
			&
			\\
			
			\hline
			\hline
			\multicolumn{3}{|c|}{$\mathcal{Q}_{lequ}^{(1)}:\, (\overline{l}_p^j\,e_r)\,\epsilon_{jk}\,(\overline{q}_s^k\,u_t)\,/\, \textcolor{OliveGreen}{(\overline{l}_p^j\,u_{r\alpha})\,\epsilon_{jk}\,(\overline{q}_s^{k\alpha}\,e_t)}$}\\
			\hline
			
			\textbf{Heavy fields}&
			\textbf{Diagram}&
			\textbf{Vertices}\\
			\hline

			$(1,2,\frac{1}{2})$&
			\multirow{2}{*}{\includegraphics[width=3cm, height=1.8cm]{smeft-psi4-6.pdf}}&
			\hyperlink{vertex-5-vi}{V5-(vi)},  \hyperlink{vertex-5-viii}{V5-(viii)}\\
			\cdashline{1-1}\cdashline{3-3}

			$(3,1,-\frac{1}{3})$&
			&
			\hyperlink{vertex-5-ix}{V5-(ix)},  \hyperlink{vertex-5-xi}{V5-(xi)}\\
			\cdashline{1-1}\cdashline{3-3}

			$(3,2,\frac{7}{6})$&
			&
			\hyperlink{vertex-5-xiii}{V5-(xiii)},  \hyperlink{vertex-5-xiv}{V5-(xiv)}\\
			
			\hline
	\end{tabular}}
	\caption{Heavy field representations that are obtained by unfolding the $\psi^4$ operators, composed of both left and right chiral SM fermions, into non-trivial tree- and(or) one-loop-level diagrams and the corresponding vertices. These operators contain atleast 3 unique fields as their constituents and they do not permit a vector boson propagators when we restrict to minimal scenarios. The operator highlighted in green color is constituted of the same external states as a SMEFT operator but leads to unique heavy field representations, see Eq.~\eqref{eq:fierz-psi4-cls-3}.}
	\label{table:smeft-psi4-LR-4}
\end{table}
\clearpage
\noindent
One can observe that $\mathcal{Q}^{(3)}_{lequ}\,\equiv\, (\overline{l}^j_p\,\sigma^{\mu\nu}\,e_r)\,\epsilon_{jk}\, (\overline{q}^k_s\,\sigma_{\mu\nu}\,u_t)$ contains the same external states as $\mathcal{Q}^{(1)}_{lequ}\,\equiv\, (\overline{l}^j_p\,e_r)\,\epsilon_{jk}\, (\overline{q}^k_s\,u_t)$ and the operator $(\overline{l}^j_p\,u_r)\,\epsilon_{jk}\, (\overline{q}^k_s\,e_t)$. Also, $\mathcal{Q}^{(3)}_{lequ}$ is related to these 2 operators as shown below:

{\small\begin{eqnarray}\label{eq:fierz-psi4-cls-3}
	& (\bar{l}^j\,\sigma_{\mu\nu}\,e)\,\epsilon_{jk}\,(\bar{q}^k\,\sigma^{\mu\nu}\,u)
	& = [ (\bar{l}^j)^{\alpha}\, (\sigma_{\mu\nu})_{\alpha}^{\beta}\, (e)_{\beta}]\,\epsilon_{jk}\, [(\bar{q}^k)^{\rho}\,(\sigma^{\mu\nu})_{\rho}^{\theta}\, (u)_{\theta}] \nonumber  \\
	& & = [(\bar{l}^j)^{\alpha}\,(\sigma_{\mu})_{\alpha \dot{\beta}}\, (\overline{\sigma}_{\nu})^{\dot{\beta} \beta}\, (e)_{\beta}] \,\epsilon_{jk}\,[(\bar{q}^k)^{\rho}\,(\sigma^{\mu})_{\rho \dot{\theta}}\,(\overline{\sigma}^{\nu})^{\dot{\theta} \theta}\, (u)_{\theta}] \nonumber \\
	& & = 4\,(\bar{l}^j\, e)\,\epsilon_{jk}\,(\bar{q}^k\, u) - 8\,(\bar{l}^j\, u)\,\epsilon_{jk}\,(\bar{q}^k\, e) \,.
	\end{eqnarray}}

\noindent Here, we have used $(\sigma^{\mu\nu})_{\alpha}^{\beta} = (\sigma^{\mu})_{\alpha\dot{\beta}} (\overline{\sigma}^{\nu})^{\dot{\beta}\beta}$, $(\overline{\sigma}_{\mu\nu})_{\dot{\alpha}}^{\dot{\beta}} = (\overline{\sigma}_{\mu})^{\dot{\beta}\beta}(\sigma_{\nu})_{\beta\dot{\alpha}}$ in addition to Eq.~\eqref{eq:gamma-fierz}. The last diagram in Table~\ref{table:smeft-psi4-LR-4} describes the unfolding of both $\mathcal{Q}^{(1)}_{lequ}$ and $(\overline{l}^j_p\,e_r)\,\epsilon_{jk}\, (\overline{q}^k_s\,u_t)$. To highlight this fact we have written the latter operator structure in colored text in Table~\ref{table:smeft-psi4-LR-4}. Therefore, instead of unfolding $\mathcal{Q}^{(3)}_{lequ}$ explicitly, we can indirectly relate it to the heavy fields that lead to those two operators. However, if we choose to forego our conditions of minimality then, this operator can be unfolded into a multi-loop diagram with multiple heavy field propagators within some of the loops. This is discussed more properly in Section \ref{subsec:multiloop}. 

\subsection{\Large$X^3$}

This class contains only two operators - $\mathcal{Q}_G$ and $\mathcal{Q}_W$. Heavy scalars, as well as heavy fermions, appear through one-loop-level diagrams, and the minimal choice corresponds to Figs.~\ref{subfig:x3-loop1} and \ref{subfig:x3-loop5}. $\mathcal{Q}_G$ only encapsulates color non-singlets and $\mathcal{Q}_W$ only isospin non-singlet representations, with no restrictions on the other quantum numbers in each case. These results have been summarized in Table~\ref{table:smeft-X3}.

	\begin{table}[!htb]
		\centering
		\renewcommand{\arraystretch}{2.2}
		{\scriptsize\begin{tabular}{||c|c|c||c|c|c||}
				\hline
				\multicolumn{3}{||c||}{$\mathcal{Q}_G:\,f^{ABC}\,G^{A\nu}_{\mu}\,G^{B\rho}_{\nu}\,G^{C\mu}_{\rho}$}&
				\multicolumn{3}{|c|}{$\mathcal{Q}_W:\,\epsilon^{IJK}\,W^{I\nu}_{\mu}\,W^{J\rho}_{\nu}\,W^{K\mu}_{\rho}$}\\
				\hline
				
				\textbf{Heavy fields}&
				\textbf{Diagram}&
				\textbf{Vertices}&
				\textbf{Heavy fields}&
				\textbf{Diagram}&
				\textbf{Vertices}\\
				\hline

				\multirow{4}{*}{$(R_C,\{1,R_L\},\{0,Y\})$}&
				\multirow{4}{*}{\includegraphics[width=2.8cm, height=2cm]{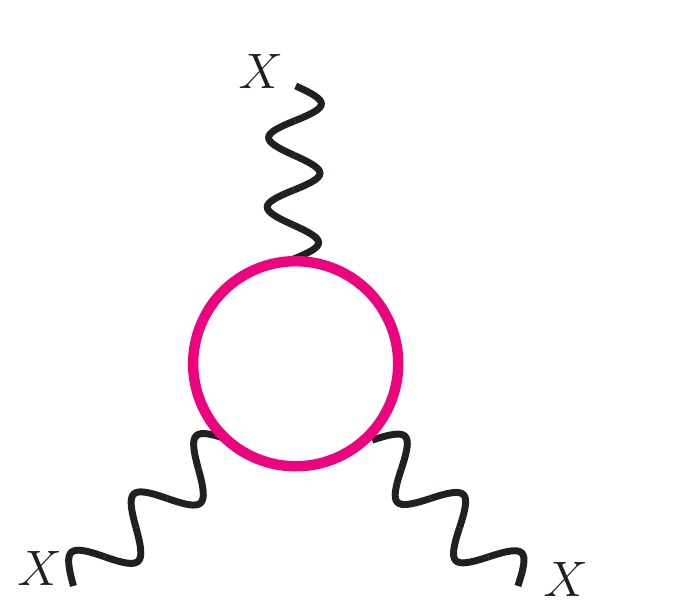}}&
				\multirow{4}{*}{\hyperlink{vertex-9-iii}{V9-(iii)}}&
				\multirow{4}{*}{$(\{1,R_C\},R_L,\{0,Y\})$}&
				\multirow{4}{*}{\includegraphics[width=2.8cm, height=2cm]{smeft-x3-1.pdf}}&
				\multirow{4}{*}{\hyperlink{vertex-9-ii}{V9-(ii)}}\\

				&
				&
				&
				&
				&
				\\
				
				&
				&
				&
				&
				&
				\\
				
				&
				&
				&
				&
				&
				\\
				
				\hline

				\multirow{4}{*}{$(R_C,\{1,R_L\},\{0,Y\})$}&
				\multirow{4}{*}{\includegraphics[width=2.8cm, height=2cm]{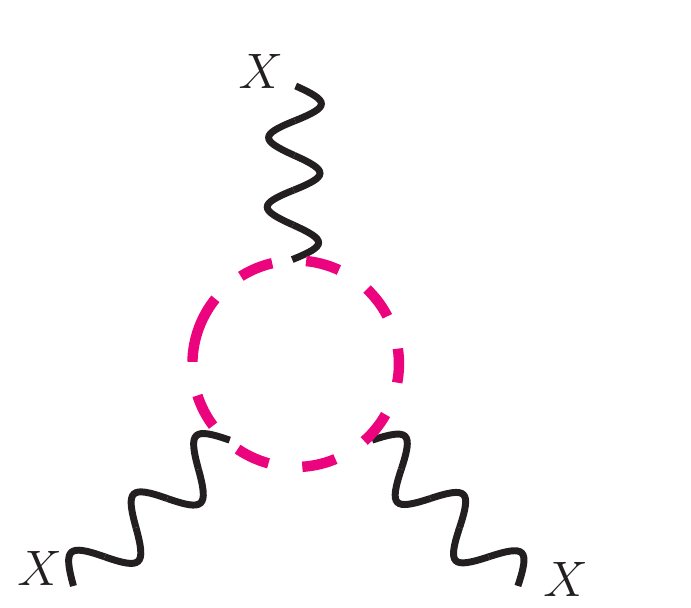}}&
				\multirow{4}{*}{\hyperlink{vertex-10-iii}{V10-(iii)}}&
				\multirow{4}{*}{$(\{1,R_C\},R_L,\{0,Y\})$}&
				\multirow{4}{*}{\includegraphics[width=2.8cm, height=2cm]{smeft-x3-2.pdf}}&
				\multirow{4}{*}{\hyperlink{vertex-10-ii}{V10-(ii)}}\\

				&
				&
				&
				&
				&
				\\

				&
				&
				&
				&
				&
				\\

				&
				&
				&
				&
				&
				\\
				\hline
		\end{tabular}}
		\caption{Heavy field representations that are obtained by unfolding the $X^3$ operators into non-trivial one-loop-level diagrams and the corresponding vertices.}
		\label{table:smeft-X3}
	\end{table}
	\clearpage
	\section{Departure from minimality}\label{sec:non-minimal-cases}
	
	While one can always study scenarios that depart radically from the notion of minimality described here, we have discussed a few cases that are necessary for the sake of completeness in the context of SMEFT.
	
	\subsection{Diagrams with multiple fundamental vertices}
	As already discussed previously, the most straightforward departure from our notion of minimality occurs if we include diagrams such as Fig.~\ref{subfig:phi6-tree2}, which includes both trilinear and quartic scalar vertices.
	
	\subsection{Heavy-heavy mixing in the loops}
	
	To depict the simultaneous appearance of multiple heavy fields in the loop, we consider operators of the $\psi^2\phi X$ class. For this class, no heavy field emerges through tree- or ``simple" one-loop-level diagrams when the light fields are fixed to be the SM degrees of freedom. We have listed the general representations allowed for each operator in Table~\ref{table:multi-heavy}. These must satisfy equations corresponding to the conservation of symmetries at each vertex.
	
	\begin{table}[!htb]
		\centering
		\renewcommand{\arraystretch}{2.0}
		{\scriptsize\begin{tabular}{|c|c|c|c|}
				
				\hline
				
				\textbf{Operator}&
				\textbf{Heavy fields}&
				\textbf{Diagram}&
				\textbf{Vertices involved}\\
				\hline
				$(\bar{l}_p\,\sigma^{\mu\nu}\,e_r)\,H\,B_{\mu\nu}$&
				$(\{1,R_{C_1}\},\{1,R_{L_1}\},Y_1)_{\Psi}\, \oplus$&
				\multirow{8}{*}{\includegraphics[width=3.5cm,height=3.5cm,trim = 0 0 30 0]{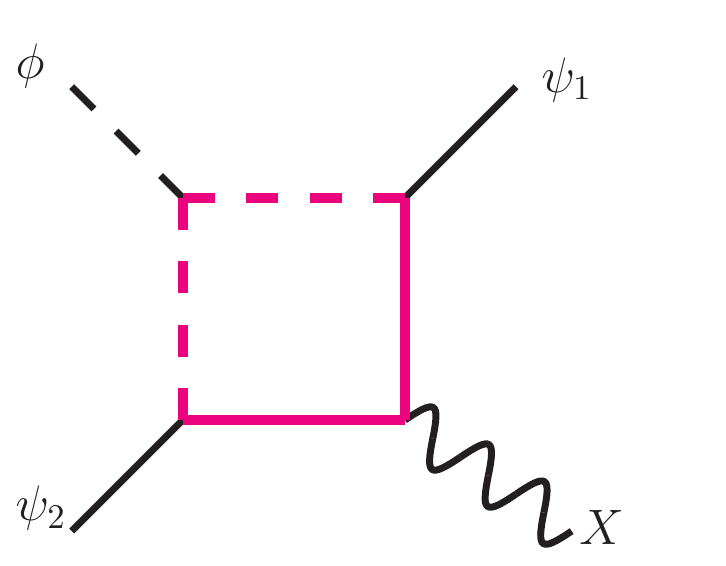}}&
				\hyperlink{vertex-2}{V2}, \hyperlink{vertex-7-i}{V7-(i)}, \hyperlink{vertex-7-ii}{V7-(ii)}, \hyperlink{vertex-9-i}{V9-(i)}\\
				\cdashline{1-1}\cdashline{4-4}
				
				$(\bar{q}_p\,\sigma^{\mu\nu}\,d_r)\,H\,B_{\mu\nu}$&
				$(R_{C_2},R_{L_2},Y_2)_{\Phi}\, \oplus\, (R_{C_3},R_{L_3},Y_3)_{\Phi}$&&
				\hyperlink{vertex-2}{V2}, \hyperlink{vertex-7-iii}{V7-(iii)}, \hyperlink{vertex-7-v}{V7-(v)}, \hyperlink{vertex-9-i}{V9-(i)}\\
				\cdashline{1-1}\cdashline{4-4}
				$(\bar{q}_p\,\sigma^{\mu\nu}\,u_r)\,\tilde{H}\,B_{\mu\nu}$&&&
				\hyperlink{vertex-2}{V2}, \hyperlink{vertex-7-iv}{V7-(iv)}, \hyperlink{vertex-7-v}{V7-(v)}, \hyperlink{vertex-9-i}{V9-(i)}\\
				\cdashline{1-2}\cdashline{4-4}
				
				$(\bar{l}_p\,\sigma^{\mu\nu}\,e_r)\,\tau^I\,H\,W^I_{\mu\nu}$&
				$(\{1,R_{C_1}\},R_{L_1},\{0,Y_1\})_{\Psi}\, \oplus$&&
				\hyperlink{vertex-2}{V2}, \hyperlink{vertex-7-i}{V7-(i)}, \hyperlink{vertex-7-ii}{V7-(ii)}, \hyperlink{vertex-9-ii}{V9-(ii)}\\
				\cdashline{1-1}\cdashline{4-4}
				$(\bar{q}_p\,\sigma^{\mu\nu}\,d_r)\,\tau^I\,H\,W^I_{\mu\nu}$&
				$(R_{C_2},R_{L_2},Y_2)_{\Phi}\, \oplus\, (R_{C_3},R_{L_3},Y_3)_{\Phi}$&&
				\hyperlink{vertex-2}{V2}, \hyperlink{vertex-7-iii}{V7-(iii)}, \hyperlink{vertex-7-v}{V7-(v)}, \hyperlink{vertex-9-ii}{V9-(ii)}\\
				\cdashline{1-1}\cdashline{4-4}
				$(\bar{q}_p\,\sigma^{\mu\nu}\,u_r)\,\tau^I\,\tilde{H}\,W^I_{\mu\nu}$&&&
				\hyperlink{vertex-2}{V2}, \hyperlink{vertex-7-iv}{V7-(iv)}, \hyperlink{vertex-7-v}{V7-(v)}, \hyperlink{vertex-9-ii}{V9-(ii)}\\
				\cdashline{1-2}\cdashline{4-4}
				
				$(\bar{q}_p\,\sigma^{\mu\nu}\,T^A\,d_r)\,H\,G^A_{\mu\nu}$&
				$(R_{C_1},\{1,R_{L_1}\},\{0,Y_1\})_{\Psi}\, \oplus$&&
				\hyperlink{vertex-2}{V2}, \hyperlink{vertex-7-iii}{V7-(iii)}, \hyperlink{vertex-7-v}{V7-(v)}, \hyperlink{vertex-9-iii}{V9-(iii)}\\
				\cdashline{1-1}\cdashline{4-4}
				$(\bar{q}_p\,\sigma^{\mu\nu}\,T^A\,u_r)\,\tilde{H}\,G^A_{\mu\nu}$&
				$(R_{C_2},R_{L_2},Y_2)_{\Phi}\, \oplus\, (R_{C_3},R_{L_3},Y_3)_{\Phi}$&&
				\hyperlink{vertex-2}{V2}, \hyperlink{vertex-7-iv}{V7-(iv)}, \hyperlink{vertex-7-v}{V7-(v)}, \hyperlink{vertex-9-iii}{V9-(iii)}\\
				\hline
		\end{tabular}}
		\caption{Combinations of heavy field representations that are required to unfold the $\psi^2\phi\,X$ operators into non-trivial one-loop-level diagrams and the corresponding vertices for individual cases.}
		\label{table:multi-heavy}
	\end{table}

\subsection{Multi-loop diagrams}\label{subsec:multiloop}
	
As discussed previously, to confine ourselves within a minimal analysis we can relate $\mathcal{Q}^{(3)}_{lequ}$ indirectly with the heavy fields embedded in the diagrams corresponding to  $\mathcal{Q}^{(1)}_{lequ}$ and $(\overline{l}^j_p\,e_r)\,\epsilon_{jk}\, (\overline{q}^k_s\,u_t)$. But, if we relax the restriction of being minimal we can unfold $\mathcal{Q}^{(3)}_{lequ}$ into a three loop diagram having heavy-heavy-light mixing between two of the loops. The explicit combinations of heavy field representations for different permutations of the external legs have been described in Table~\ref{table:smeft-psi4-tensor-op}.  
\clearpage

\begin{table}[!htb]
	\centering
	\renewcommand{\arraystretch}{2.0}
	{\scriptsize\begin{tabular}{|c|c|c|c|c|c|c|c|}
			
			\hline
			\multicolumn{8}{|c|}{$\mathcal{Q}^{(3)}_{lequ}:\, (\overline{l}^j_p\,\sigma^{\mu\nu}\,e_r)\,\epsilon_{jk}\, (\overline{q}^k_s\,\sigma_{\mu\nu}\,u_t)$}\\
			\hline

			$ (\psi_1,\psi_2) $&
			\multicolumn{2}{c|}{\textbf{loop1}  }&
			\multicolumn{2}{c|}{\textbf{loop2} }&
			\multirow{2}{*}{ $X$ }&
			\multirow{2}{*}{ \textbf{Corresponding diagram} }&
			\multirow{2}{*}{ \textbf{Vertices involved} } \\
			\cline{2-5}

			$ (\psi_3,\psi_4) $&
			$ \Phi_1 $ &
			$ \Psi_1 $ &
			$ \Phi_2 $ &
			$ \Psi_2 $ &
			&
			&
			\\
			\hline

			\multirow{3}{*}{ $ (\overline{l},e) $} &
			$ (1,1,0)$ &
			\multirow{2}{*}{$(1,2,\text{-}\frac{1}{2}) $} &
			$ (1,1,0)$ &
			$(3,1,\frac{2}{3}) $ &
			$ B $ &
			\multirow{7}{*}{\includegraphics[width=5.2cm,height=3.5cm]{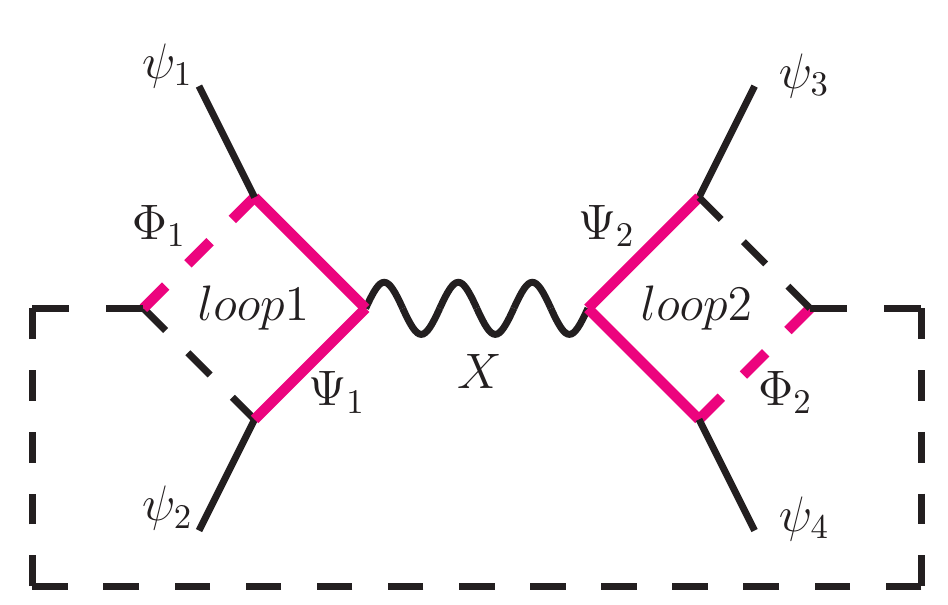}}&
			\multirow{2}{*}{ \hyperlink{vertex-1-ii}{V1-(ii)}, \hyperlink{vertex-6-ii}{V6-(ii)}, \hyperlink{vertex-6-iii}{V6-(iii)} }\\
			\cdashline{4-6} 
			
			\multirow{3}{*}{ $ (\overline{q},u) $ }&
			$ (1,3,0) $ &
			&
			$ (1,3,0)$ &
			$(3,3,\frac{2}{3}) $ &
			$ B/W $&
			&
			\\
			\cdashline{2-6} \cdashline{8-8}
			
			&
			$ (1,1,1)$ &
			\multirow{2}{*}{$(1,2,\frac{3}{2}) $ } &
			$ (1,1,1)$ &
			$(\bar{3},1,\frac{1}{3}) $ &
			$ B $&
			&
			\multirow{2}{*}{ \hyperlink{vertex-1-i}{V1-(i)}, \hyperlink{vertex-6-vii}{V6-(vii)}, \hyperlink{vertex-6-viii}{V6-(viii)} }\\
			\cdashline{4-6}
			
			&
			$ (1,3,1) $ &
			&
			$ (1,3,1)$ &
			$(\bar{3},3,\frac{1}{3}) $ &
			$ B/W $&
			&
			\\

			\cline{1-6} \cline{8-8}
			
			\multirow{2}{*}{ $ (e,\overline{l}) $ } &
			$ (1,1,0)$ &
			$(1,1,1) $ &
			$ (1,1,0)$ &
			$(\bar{3},2,\frac{1}{6}) $ &
			$ B/W $&
			&
			\hyperlink{vertex-1-ii}{V1-(ii)}, \hyperlink{vertex-6-vi}{V6-(vi)}, \hyperlink{vertex-6-ix}{V6-(ix)} \\
			
			\cdashline{2-6} \cdashline{8-8}
			
			\multirow{2}{*}{$ (u,\overline{q}) $ } &
			\multirow{2}{*}{$ (1,3,1) $} &
			\multirow{2}{*}{$ (1,3,0) $} &
			$ (1,1,1)$ &
			\multirow{2}{*}{$(3,2,\frac{5}{6}) $ }&
			\multirow{2}{*}{$ W $ }&
			&
			\multirow{2}{*}{ \hyperlink{vertex-1-i}{V1-(i)}, \hyperlink{vertex-6-i}{V6-(i)}, \hyperlink{vertex-6-iv}{V6-(iv)} } \\

			&
			&
			&
			$ (1,3,1)$ &
			&
			&
			&
			\\

			\hline

	\end{tabular}}
	\caption{Combinations of heavy field representations that appear within the loops when the operator $\mathcal{Q}^{(3)}_{lequ}$ is unfolded into a multi-loop diagram. The different orientations of the external legs and the choice of the intermediate vector boson have been considered separately and the vertices. have been listed for individual cases.}
	\label{table:smeft-psi4-tensor-op}
\end{table}	

%	\clearpage
	\section{Operator driven BSM construction: Validation and illustration through examples} \label{sec:section5}

	\subsection{Validating the diagrammatic method}
	The analysis thus far has centred on unfolding the dimension-6 effective operators of SMEFT diagrammatically and obtaining BSM fields that can offer non-zero contributions to them. Here, we have demonstrated the consistency of our results with those of a top-down analysis. Focussing on a single heavy scalar, here the lepto-quark $\Theta$ with quantum numbers $(3,2,\frac{1}{6})$, we have first collected the effective operators that have been found to receive contributions from it based on our diagrammatic approach, in Table~\ref{table:slq-example} below.
	
\begin{table}[!htb]
	\centering
	\renewcommand{\arraystretch}{2.0}
	{\scriptsize\begin{tabular}{||c|c||c|c||c|c||c|c||}
			\hline
			\textbf{Operator} & \textbf{Table} &\textbf{Operator} &\textbf{Table} & \textbf{Operator} & \textbf{Table} &\textbf{Operator} & \textbf{Table} \\
			\hline \hline
			$ \mathcal{Q}_{H} $ & \ref{table:smeft-phi6} &$ \mathcal{Q}_{H\square},\,\mathcal{Q}_{H\mathcal{D}} $ &\ref{table:smeft-phi4D2-1} & 
			$ \mathcal{Q}_{eH}, \mathcal{Q}_{uH}, \mathcal{Q}_{dH}$ &\ref{table:smeft-eom} & 
			$ \mathcal{Q}_{HWB} $ &
			\ref{table:smeft-phi2X2-2}\\
			\hline

			$ \mathcal{Q}_{Hl}^{(1)}, \mathcal{Q}_{He}, \mathcal{Q}_{Hl}^{(3)},$ & \multirow{2}{*}{\ref{table:smeft-psi2phi2D}} &
			$ \mathcal{Q}_{qq}^{(1)},\mathcal{Q}_{lq}^{(1)}, \mathcal{Q}_{ll} $& \multirow{2}{*}{\ref{table:smeft-psi4-LL}} &
			$ \mathcal{Q}_{ee}, \mathcal{Q}_{uu}, \mathcal{Q}_{dd} $& \multirow{2}{*}{\ref{table:smeft-psi4-RR}} & 
			$ \mathcal{Q}_{le}, \mathcal{Q}_{lu}, \mathcal{Q}_{qe}, $ &
			\multirow{2}{*}{\ref{table:smeft-psi4-LR-1}} \\

			$\mathcal{Q}_{Hq}^{(1)}, \mathcal{Q}_{Hu}, \mathcal{Q}_{Hd}, \mathcal{Q}_{Hq}^{(3)} $& &
			$ \mathcal{Q}_{lq}^{(3)}, \mathcal{Q}_{qq}^{(3)} $&
			&
			$ \mathcal{Q}_{eu}, \mathcal{Q}_{ed}, \mathcal{Q}_{ud}^{(1)}, \mathcal{Q}_{ud}^{(8)} $&
			&
			$ \mathcal{Q}_{qu}^{(1)}, \mathcal{Q}_{qu}^{(8)}, \mathcal{Q}_{qd}^{(1)}, \mathcal{Q}_{qd}^{(8)} $ &
			\\
			\hline

			$ \mathcal{Q}_{G}, \mathcal{Q}_{W} $ &
			\ref{table:smeft-X3} &
			$ \mathcal{Q}_{ld} $ &
			\ref{table:smeft-psi4-LR-3} &
			$ \mathcal{Q}_{HG}, \mathcal{Q}_{HW}, \mathcal{Q}_{HB}$ &
			\ref{table:smeft-phi2X2-1} &
			&
			\\
			
			\hline

	\end{tabular}}
	\caption{List of dimension-6 SMEFT operators which receive non-zero contributions from the scalar lepto-quark $(3,2,\frac{1}{6})$ based on the diagrammatic unfolding of the operators. Also listed are the tables where the unfolding has explicitly been demonstrated for each operator.}
	\label{table:slq-example}
\end{table}
	
	To test the validity of our method and our results, we first build the most general UV Lagrangian for a model where the SM field content is extended by a lepto-quark scalar $ \Theta $ with gauge quantum numbers $(3,2,\frac{1}{6})$ under the SM gauge groups using \texttt{GrIP} \cite{Banerjee:2020bym}, which is shown in the covariant form below:
	\begin{align}
	\mathcal{L}_\Theta =& \mathcal{L}_{SM} + \lvert\mathcal{D}_{\mu} \, \Theta\rvert^2 - m^2_{\Theta} \lvert\Theta\rvert^2 - \eta_1 \, (H^{\dagger}H) (\Theta^{\dagger} \Theta) - \eta_2 \,(\Theta^{\dagger}\tau^I \Theta)(H^{\dagger} \tau^I H)\nonumber\\
	&-\lambda_1\ (\Theta^{\dagger} \Theta)^2 - \lambda_2 (\Theta^{\dagger}\tau^I \Theta)^2
	- y_{{}_\Theta}^{pr}\, ( \epsilon_{ij}\; \Theta^{\alpha i}\; \overline{d}_{p\alpha}\; l^{j}_{r} + h.c.).
	\end{align}
	Here, $ \mathcal{L}_{SM} $ represents the Standard Model Lagrangian.
	We have used \texttt{CoDEx}\cite{Bakshi:2018ics} to integrate-out the heavy scalar $\Theta$ and determine the SMEFT dimension-6 effective operators. The output from \texttt{CoDEx} exhibit non-zero Wilson coefficients for the following operators at the matching scale:\\
	\\
	$\mathcal{Q}_{HD}, \mathcal{Q}_{ll}, \mathcal{Q}_{Hu}, \mathcal{Q}_{Hd}, \mathcal{Q}_{He}, \mathcal{Q}_{Hq}^{(1)}, \mathcal{Q}_{Hl}^{(1)}, \mathcal{Q}_{Hl}^{(3)}, \mathcal{Q}_{Hq}^{(3)}, \mathcal{Q}_{HWB}, \mathcal{Q}_{H\square}, \mathcal{Q}_{HB}, \mathcal{Q}_{HW},  \mathcal{Q}_H, \mathcal{Q}_{G}, \mathcal{Q}_{HG}, \mathcal{Q}_{eH}, \mathcal{Q}_{uH}, \mathcal{Q}_{dH},$
	
	$\mathcal{Q}_{qq}^{(1)}, \mathcal{Q}_{qq}^{(3)}, \mathcal{Q}_{uu}, \mathcal{Q}_{dd}, \mathcal{Q}_{ud}^{(1)}, \mathcal{Q}_{lq}^{(1)}, \mathcal{Q}_{ee}, \mathcal{Q}_{eu}, \mathcal{Q}_{ed}, \mathcal{Q}_{le},  \mathcal{Q}_{lu}, \mathcal{Q}_{ld}, \mathcal{Q}_{qe}, \mathcal{Q}_{qu}^{(1)}, \mathcal{Q}_{qd}^{(1)},  \mathcal{Q}_{lq}^{(3)}, \mathcal{Q}_{W}, \mathcal{Q}_{ud}^{(8)},  \mathcal{Q}_{qd}^{(8)}, \mathcal{Q}_{qu}^{(8)}. $\\
	\\
	These operators are completely in agreement with the results of Table~\ref{table:slq-example}. This shows that results obtained using the techniques outlined in this paper are consistent with the results available in the literature \cite{Bakshi:2020eyg,Gherardi:2020det}.

	\subsection{The minimal  extension of the SM as root of CP even D6 SMEFT operators}
	As the first application of our method, we have discussed how one can extend the SM degrees of freedom in \textit{the most minimal way} so as to explain the origin of as many dimension-6 $CP$ and $B$-, $L$- conserving operators as possible. Out of the 59 operators\footnote{All these conserve baryon and lepton numbers.} that constitute a complete and independent basis for SMEFT at dimension-6 only 52 are found to be $CP$ conserving.
	
	Since our aim is to build \textit{minimal extensions} of the SM, based on the list of heavy field representations that have already been tabulated corresponding to every single operator, one must first identify the representations which are common to the largest subset of operators. We have found that the scalar field $\Theta_{(3,2,\frac{1}{6})}$ provides non-zero contributions to 39 operators, while $\mathcal{H}_{(1,2,\frac{1}{2})}$ (which constitutes the Two Higgs Doublet Model \cite{Anisha:2019nzx,Crivellin:2016ihg,Karmakar:2019vnq}) leads to 37 operators. The latter of the two is more \textit{minimal} on account of it being an $SU(3)$ singlet. Therefore we include it as the first extension to our degrees of freedom. Next, one can see that the operators involving gluons - \textcolor{RubineRed}{$\mathcal{Q}_{G}$, $\mathcal{Q}_{HG}$} and those that are products of fermion bilinears transforming as $SU(3)$ octets - \textcolor{RubineRed}{$\mathcal{Q}^{(8)}_{ud}, \mathcal{Q}^{(8)}_{qd}, \mathcal{Q}^{(8)}_{qu}$} can only be generated by scalars with non-trivial $SU(3)$ quantum numbers. This motivates us to include a second heavy field. We have chosen $\varphi_{(3,1,-\frac{1}{3})}$ which being a color triplet $SU(2)$ singlet so as to ensure minimality. 
	
	In Table~\ref{table:2hdm+slq-example}, we have listed all 42 operators that receive non-zero contributions from a color-singlet, isospin-doublet scalar $\mathcal{H}$ and(or) a color-triplet, isospin-singlet scalar $\varphi$. The entries in black colour are only generated by $\mathcal{H}$, those in pink are generated by $\varphi$ alone, while the ones in green receive contributions from both these fields.  For easy reference, adjacent to each operator (or a collection of operators) we have also specified the table where it has been shown how unfolding the operator yields $\mathcal{H}$ and(or) $\varphi$. 
	
	\begin{table}[!htb]
		\centering
		\renewcommand{\arraystretch}{2.0}
		{\scriptsize\begin{tabular}{||c|c||c|c||c|c||c|c||}
				\hline
				\textbf{Operator} & \textbf{Table} &\textbf{Operator} &\textbf{Table} & \textbf{Operator} & \textbf{Table} &\textbf{Operator} & \textbf{Table} \\
				\hline \hline
				\textcolor{ForestGreen}{$ \mathcal{Q}_{H} $} & \ref{table:smeft-phi6} &
				\textcolor{ForestGreen}{$ \mathcal{Q}_{H\square},\,\mathcal{Q}_{H\mathcal{D}} $} &
				\ref{table:smeft-phi4D2-1} & 
				$ \mathcal{Q}_{eH}, \textcolor{ForestGreen}{\mathcal{Q}_{uH}, \mathcal{Q}_{dH}} $ &\ref{table:smeft-psi2phi3} & 
				$ \textcolor{ForestGreen}{\mathcal{Q}^{(1)}_{quqd},\,\mathcal{Q}^{(1)}_{lequ},\,\mathcal{Q}_{ledq} } $&
				\ref{table:smeft-psi4-LR-4} \\
				\hline

				$ \textcolor{ForestGreen}{\mathcal{Q}_{Hl}^{(1)}, \mathcal{Q}_{He}}, \mathcal{Q}_{Hl}^{(3)},$ & \multirow{2}{*}{\ref{table:smeft-psi2phi2D}} &
				$\textcolor{ForestGreen}{ \mathcal{Q}_{qq}^{(1)},\mathcal{Q}_{lq}^{(1)}, \mathcal{Q}_{ll}} $
				& \multirow{2}{*}{\ref{table:smeft-psi4-LL}} &
				$ \textcolor{ForestGreen}{\mathcal{Q}_{ee}, \mathcal{Q}_{uu}, \mathcal{Q}_{dd}} $& \multirow{2}{*}{\ref{table:smeft-psi4-RR}} & 
				$ \textcolor{ForestGreen}{\mathcal{Q}_{le}, \mathcal{Q}_{lu}, \mathcal{Q}_{qe},} $ &
				\multirow{2}{*}{\ref{table:smeft-psi4-LR-1}} \\

				$\textcolor{ForestGreen}{\mathcal{Q}_{Hq}^{(1)}, \mathcal{Q}_{Hu}, \mathcal{Q}_{Hd}}, \mathcal{Q}_{Hq}^{(3)} $& &
				$ \mathcal{Q}_{lq}^{(3)}, \mathcal{Q}_{qq}^{(3)} $&
				&
				$ \textcolor{ForestGreen}{\mathcal{Q}_{eu}, \mathcal{Q}_{ed}, \mathcal{Q}_{ud}^{(1)}}, \textcolor{RubineRed}{\mathcal{Q}_{ud}^{(8)}} $&
				&
				$ \textcolor{ForestGreen}{\mathcal{Q}_{qu}^{(1)}, \mathcal{Q}_{qd}^{(1)},} \textcolor{RubineRed}{\mathcal{Q}^{(8)}_{qd},\,\mathcal{Q}^{(8)}_{qu}} $ &
				\\
				\hline

				$ \mathcal{Q}_{W},\textcolor{RubineRed}{ \mathcal{Q}_{G} } $ &
				\ref{table:smeft-X3} &
				$ \textcolor{ForestGreen}{ \mathcal{Q}_{ld} }$ &
				\ref{table:smeft-psi4-LR-1}, \ref{table:smeft-psi4-LR-2} &
				$  \textcolor{ForestGreen}{\mathcal{Q}_{HW}}, \mathcal{Q}_{HB}, \textcolor{RubineRed}{ \mathcal{Q}_{HG}} $ &
				\ref{table:smeft-phi2X2-1} &
				$ \mathcal{Q}_{HWB} $ &
				\ref{table:smeft-phi2X2-2}\\
				
				\hline

		\end{tabular}}
		\caption{SMEFT dimension-6 effective operators unfolded by minimal extension. The operators in black are generated by the $\mathcal{H}_{(1,2,\frac{1}{2})}$ extension alone. Further extension by the $\varphi_{(3,1,-\frac{1}{3})} $ generates the extra operators highlighted in pink color. The ones shown in green are common to both these new fields. }
		\label{table:2hdm+slq-example}
	\end{table}

	\noindent
	
	Of the remaining 10 operators, 8 belong to the $\psi^2\phi X$ class and these can only be generated using models containing both scalar and fermion extensions, i.e., we are required to forego the notion of minimality, see Table~\ref{table:multi-heavy}, as long as one only considers 1-particle-irreducible (1PI) diagrams. The last two operators are $\mathcal{Q}^{(8)}_{quqd},\, \mathcal{Q}^{(3)}_{lequ}$ which belong to the $\psi^4$ class and require special attention. The first of these $\mathcal{Q}^{(8)}_{quqd}$ can be directly generated by only a color-octet scalar extension or indirectly by a color-sextet scalar if one takes Fierz identities into account. Similarly, $\mathcal{Q}^{(3)}_{lequ}$ can be shown to receive contributions from $\mathcal{H}$ and $\varphi$ after employing certain Fierz identities, see Eq.~\eqref{eq:fierz-psi4-cls-3} and Table~\ref{table:smeft-psi4-LR-4}. But its unfolding can only be conducted at the level of multi-loop diagrams, see Table~\ref{table:smeft-psi4-tensor-op}. 
	
	\subsection{Role of observables on the choice of BSMs}

The analyses based on SMEFT start with the construction of a UV Lagrangian, followed by integrating out heavy fields to obtain effective operators and ultimately connecting those operators to well-defined observables. The prevalence of a large number of candidate scenarios that appear to be distinct makes conducting comparative analyses and verifying claims of one model being more vital than others, an arduous task.  

This is where our techniques truly shine. Based on the particular observable we are interested in, we can readily obtain the set of relevant operators. Then, using the results of this paper as a dictionary, i.e., by looking at the unfolding of these operators catalogued here, one can easily arrive at the list of heavy field representations which are actually relevant for such observables. To illustrate this point, we have considered the operators contributing to the electroweak precision observables at leading order (EWPO-LO), listed below:
\begin{eqnarray}
\{ \mathcal{Q}_{ll},\, \mathcal{Q}_{H\mathcal{D}},\, \mathcal{Q}_{HWB},\, \mathcal{Q}_{Hq}^{(3)},\, \mathcal{Q}_{Hl}^{(3)},\, \mathcal{Q}_{Hq}^{(1)},\, \mathcal{Q}_{Hl}^{(1)},\, \mathcal{Q}_{He},\, \mathcal{Q}_{Hu},\, \mathcal{Q}_{Hd} \}. \nonumber
\end{eqnarray}

\noindent
We have followed two separate routes, first, where the SM extension is a scalar and second where it is a fermion. The procedure has been elucidated in Table~\ref{table:sm-extension}. 

\begin{table}[h]
	\centering
	\renewcommand{\arraystretch}{2.0}
	{\scriptsize\begin{tabular}{|c|c|c|}
			\hline
			\multicolumn{3}{|c|}{\textbf{Scalar extension}}\\
			
			\hline
			
			\textbf{Operators} &
			\textbf{Heavy Fields} &
			\textbf{Table} \\
			\hline
			
			$ \{\mathcal{Q}_{H\mathcal{D}}\} $ &
			$ (\{1,R_C\},\{1,R_L\},\{0,Y\}) $ &
			\ref{table:smeft-phi4D2-1} \\
			
			$ + $ &
			$ \downarrow $&
			\\
			
			$ \{\mathcal{Q}_{ll},\, \mathcal{Q}_{Hq}^{(1)},\, \mathcal{Q}_{Hl}^{(1)}, \mathcal{Q}_{He},\, \mathcal{Q}_{Hu},\, \mathcal{Q}_{Hd}\} $  &
			$ (\{1,R_C\},\{1,R_L\},Y) $ &
			\ref{table:smeft-psi2phi2D}, \ref{table:smeft-psi4-LL}\\
			
			+ &
			$ \downarrow $&
			\\
			
			$ \{\mathcal{Q}_{Hq}^{(3)},\, \mathcal{Q}_{Hl}^{(3)},\, \mathcal{Q}_{HWB} \} $ &
			$(\{1,R_C\},R_L,Y)$ &
			\ref{table:smeft-psi2phi2D}, \ref{table:smeft-phi2X2-2} \\
			
			\hline

	\end{tabular}}

	{\scriptsize\begin{tabular}{|c|c|c|}
			\hline
			\multicolumn{3}{|c|}{\textbf{Fermion extension}} \\
			
			\hline
			
			\textbf{Operators} &
			\textbf{Heavy Fields} &
			\textbf{Table} \\
			\hline
			
			$ \{\mathcal{Q}_{ll},\, \mathcal{Q}_{Hq}^{(1)},\, \mathcal{Q}_{Hl}^{(1)}, \mathcal{Q}_{He},\, \mathcal{Q}_{Hu},\, \mathcal{Q}_{Hd} \} $ &
			$ (\{1,R_C\},\{1,R_L\},Y) $ & 
			\ref{table:smeft-psi2phi2D}, \ref{table:smeft-psi4-LL} \\
			
			+ &
			$ \downarrow $ &
			\\
			
			$ \{\mathcal{Q}_{Hq}^{(3)},\, \mathcal{Q}_{Hl}^{(3)} \} $&
			$ (\{1,R_C\},R_L,Y) $ &
			\ref{table:smeft-psi2phi2D}\\

			+ &
			$ \downarrow $ &
			\\
			
			\multirow{2}{*}{$ \{ \mathcal{Q}_{HWB} \} $ }&
			$(R_C, R_L, Y)  \in \{(3,3,-\frac{1}{3}),\,(3,3,\frac{2}{3}),$ $(1,2,\frac{3}{2}),\,(1,2,\frac{1}{2}),$&
			\multirow{2}{*}{\ref{table:smeft-phi2X2-2} }\\
			
			&
			$ (1,3,1), \,(3,2,\frac{7}{6}),\,(3,2,\frac{1}{6}),\,(3,2,-\frac{5}{6}) \}$ &
			\\
			
			+ &
			$ \downarrow $ &
			\\
			
			\multirow{2}{*}{$ \{\mathcal{Q}_{H\mathcal{D}}\} $ }&
			Necessitates the inclusion of a second heavy fermion $(R_{C^\prime}, R_{L^{\prime}}, Y^{\prime}) $&
			\multirow{2}{*}{\ref{table:smeft-phi4D2-1} (last entry)}\\
			
			&
			such that $(R_{C}, R_{L}, Y) \oplus (R_{C^\prime}, R_{L^{\prime}}, Y^{\prime}) = (1,2,\pm\frac{1}{2})$, see \hyperlink{vertex-8}{V8}&
			\\
			\hline

	\end{tabular}}
	\caption{Restrictions imposed on the heavy field representation as we take into account more and more operators relevant to electroweak precision observables. The ``+" sign indicates that more operators are being added and ``$\downarrow$" shows how the permitted representations get modified gradually. For each step we have also referred to the Tables where the relevant diagrams have been catalogued.}
	\label{table:sm-extension}
\end{table}

For the case of scalar extensions, the operator $\mathcal{Q}_{H\mathcal{D}}$ allows all possible representations (trivial as well as non-trivial). Taking the operators $\{ \mathcal{Q}_{ll},\, \mathcal{Q}_{Hq}^{(1)},\, \mathcal{Q}_{Hl}^{(1)},\, \mathcal{Q}_{He},\, \mathcal{Q}_{Hu},\, \mathcal{Q}_{Hd} \}$  into account it is evident that only fields with non-zero hypercharge can be included. Finally, bringing $\mathcal{Q}_{HWB},\, \mathcal{Q}_{Hq}^{(3)},$ and $\mathcal{Q}_{Hl}^{(3)}$ into the picture we get another constraint that the $SU(2)$ quantum number must also be non-singlet.
For the fermion case the procedure takes a different shape. Starting with $\{ \mathcal{Q}_{ll},\, \mathcal{Q}_{Hq}^{(1)},\, \mathcal{Q}_{Hl}^{(1)},\, \mathcal{Q}_{He},\, \mathcal{Q}_{Hu},\, \mathcal{Q}_{Hd} \}$, restriction is set on the $U(1)$ charge and including $ \mathcal{Q}_{Hq}^{(3)},\, \mathcal{Q}_{Hl}^{(3)}$ constrains the $SU(2)$ quantum number. Then, adding $\mathcal{Q}_{HWB}$ we get well-defined heavy fermion representations as shown in Table~\ref{table:sm-extension}. Lastly, adding $\mathcal{Q}_{H\mathcal{D}}$ to the list suggests that we require more than a single fermion extension of the SM corresponding to each of the eight representations listed before.

Thus we have seen how, by progressively enlarging the operator set, we can be led to well-defined heavy field representations that actually contribute to the observables under consideration.

%\newpage
	\section{Conclusion and Remarks}
	
	In this work, we have highlighted the principles and salient features of an operator driven prescription for UV model building. Starting from the bottom-up extension of a low energy theory, we have described how to catalogue heavy field representations that provide non-zero contributions to specific processes and observables of the low-energy theory. The main procedure involves the identification of Lorentz invariant renormalizable vertices followed by fixing some of the legs of those vertices to be the low energy quantum fields and investigating the heavy field representations (or their combinations in some cases) which can be assigned to the remaining leg(s) while ensuring the conservation of the internal symmetries of the theory. Thereafter, it has been described how the higher mass dimension effective operators of the low energy theory can be ``unfolded" into tree- and loop-level processes using these vertices involving light as well as heavy fields. 
	
	Following a general discussion using schematic diagrams, we have made our ideas concrete by using the example of the Standard Model as the low energy theory and by exhibiting how to unfold each operator belonging to the dimension-6 basis of SMEFT. We have kept our analysis minimal by only considering such UV theories that do not extend the SM gauge group. Also, the notion of minimality is reflected in the choice of tree and one-level diagrams included for specific cases. We have briefly described how non-minimal scenarios can be approached. Using appropriate examples, we have demonstrated how our results agree with those obtained using the conventional top-down approach of EFT. 
	
	The discussion underlines the economy as well as the systematic nature of our approach where the most significant heavy field extension(s) can be arrived at by examining the effective operators of the low energy theory alone. This is in contrast to the strenuous endeavour of building numerous plausible UV Lagrangians, integrating out the heavy fields to obtain subsets of the low energy effective Lagrangian and conducting multiple comparative analyses to adjudge the significance of the candidate UV models.    
	
	\section{Acknowledgement}
	
	The work of JC, SDB and SUR is supported by
	the Science and Engineering Research Board, Government of India, under the agreements SERB/PHY/2016348 (Early Career Research Award) and SERB/PHY/2019501
	(MATRICS) and Initiation Research Grant, agreement number IITK/PHY/2015077, by
	IIT Kanpur. SP is supported by the MHRD, Government of India, under the Prime Minister's Research Fellows (PMRF) Scheme, 2020. M.S. is supported by the STFC under grant ST/P001246/1.
	
	\appendix
	
	\section{The Standard Model field content}\label{sec:app}
	
	\begin{table}[h!]
		\centering
		\renewcommand{\arraystretch}{1.7}
		{\scriptsize\begin{tabular}{|c|c|c|c|c|c|c|}
				\hline
				\textbf{Field} & \textbf{$SU(3)_C$} & \textbf{$SU(2)_{L}$}&\textbf{$U(1)_{Y}$}&Baryon No.&Lepton No.&Spin\\
				\hline
				$H$    &1&2&1/2&0&0&0\\
				$q^p_L$       &3&2&1/6&1/3&0&1/2\\
				$u^p_R$     &3&1&2/3&1/3&0&1/2\\
				$d^p_R$     &3&1&-1/3&1/3&0&1/2\\
				$l^p_L$       &1&2&-1/2&0&-1&1/2\\
				$e^p_R$     &1&1&-1&0&-1&1/2\\
				\hline
				$G^A_{\mu}$ &8&1&0&0&0&1\\
				$W^I_{\mu}$ &1&3&0&0&0&1\\
				$B_{\mu}$   &1&1&0&0&0&1\\
				\hline
		\end{tabular}}
		\caption{\small Standard Model: Gauge and global quantum numbers and spins of the fields. Here, $A = 1,2,\cdots,8$; $I = 1,2,3;\,\, p = 1,2,3$ and $\mu = 0,1,2,3$ refer to the $SU(3), SU(2)$, flavour and Lorentz indices respectively.} 
		\label{table:SM-fields}
	\end{table}

	\section*{}
	\providecommand{\href}[2]{#2}
	\addcontentsline*{toc}{section}{}
	\bibliographystyle{JHEP}
	\bibliography{SMEFT_Roots}
	
\end{document}